\documentclass[aps,prd,superscriptaddress,twocolumn,floatfix,showpacs]{revtex4}
\newcommand{\beq}{\begin{equation}}
\newcommand{\eeq}{\end{equation}}
\newcommand{\beqn}{\begin{eqnarray}}
\newcommand{\eeqn}{\end{eqnarray}}

\newcommand{\A}{A}
\newcommand{\B}{B}
\newcommand{\D}{C}
\newcommand{\F}{D}
\newcommand{\G}{E}
\newcommand{\ve}[1]{\mbox{\boldmath $#1$}}
\usepackage{graphicx}
\usepackage{epsf}
\begin{document}
\title{General Relativistic Simulations of Black Hole-Neutron
  Star Mergers: \\ Effects of black-hole spin}

\author{Zachariah B. Etienne}
\email{zetienne@illinois.edu}
\affiliation{Department of Physics, University of Illinois at
  Urbana-Champaign, Urbana, IL 61801}
\author{Yuk Tung Liu}
\affiliation{Department of Physics, University of Illinois at
  Urbana-Champaign, Urbana, IL 61801}
\author{Stuart~L.~Shapiro}
\altaffiliation{Also at Department of Astronomy and NCSA, University of
  Illinois at Urbana-Champaign, Urbana, IL 61801}
\affiliation{Department of Physics, University of Illinois at
  Urbana-Champaign, Urbana, IL 61801}
\author{Thomas W. Baumgarte}
\altaffiliation{Also at Department of Physics, University of Illinois at
  Urbana-Champaign, Urbana, IL 61801}
\affiliation{Department of Physics and Astronomy, Bowdoin College,
  Brunswick, ME 04011}

\begin{abstract}
Black hole-neutron star (BHNS) binary mergers are candidate engines 
for generating both short-hard gamma-ray bursts (SGRBs) and detectable
gravitational waves.  Using our most recent conformal thin-sandwich
BHNS initial data and our fully general relativistic hydrodynamics
code, which is now AMR-capable, we are able to efficiently and
accurately simulate these binaries from large separations
through inspiral, merger, and ringdown. We evolve the metric using the BSSN
formulation with the standard moving puncture gauge conditions and handle the
hydrodynamics with a high-resolution shock-capturing scheme.  We explore the
effects of BH spin (aligned and anti-aligned with the orbital angular
momentum) by evolving three sets of initial data with BH:NS mass ratio
$q=3$: the data sets are nearly identical, except the BH spin is
varied between $a/M_{\rm BH}=-0.5$ (anti-aligned), 0.0, and 0.75. The
number of orbits before merger increases with $a/M_{\rm BH}$, as
expected. We also study the nonspinning BH case in more detail,
varying $q$ between 1, 3, and 5.  We calculate gravitational waveforms
for the cases we simulate and compare them to binary black-hole
waveforms.  Only a small disk ($<0.01M_\odot$) forms for the
anti-aligned spin case ($a/M_{\rm BH}=-0.5$) and for the most extreme
mass ratio case ($q=5$).  By contrast, a massive ($M_{\rm disk}\approx
0.2M_\odot$), hot disk forms in the rapidly spinning ($a/M_{\rm
  BH}=0.75$) aligned BH case. Such a disk could drive a SGRB, 
possibly by, e.g.,
producing a copious flux of neutrino-antineutino pairs.
\end{abstract}


\maketitle

\section{Introduction}

Mergers of black hole-neutron star (BHNS) binaries are expected
to be among the most promising sources of gravitational waves
detectable by ground-based laser interferometers like
LIGO~\cite{LIGO1,LIGO2}, VIRGO~\cite{VIRGO1,VIRGO2}, GEO~\cite{GEO},
TAMA~\cite{TAMA1,TAMA2}, and AIGO~\cite{aigo}, as well as by the
proposed space-based interferometers LISA~\cite{LISA} and
DECIGO~\cite{DECIGO}. Theoretical models indicate that a neutron
star-neutron star (NSNS)~\cite{HMNS1,HMNS2,ShiUIUC,STU1,STU2,ST,SST}
or black hole-neutron star (BHNS)~\cite{FBSTR,FBST,SU1,SU2,SST,ST07}
binary merger may result in a hot, massive disk around a black hole, whose
temperatures and densities could be sufficient to trigger a short-hard
gamma-ray burst (SGRB).  Indeed, SGRBs have been repeatedly associated
with galaxies with extremely low star formation rates (see~\cite{SGRB_local}
and references therein for a review), indicating
that the source is likely to involve an evolved population, rather
than main sequence stars. 
The number of detectable BHNS mergers in the observable universe is still an
open question.  The uncertainties arise from some aspects of
population synthesis calculations that are only partially understood.
The estimated event rate of BHNS mergers observable by an
Advanced LIGO detector typically falls in the range ${\cal R}\sim
1-100~{\rm yr}^{-1}$~\cite{KBKOW}.

Motivated by the significance of BHNS binaries for both GRB
and gravitational-wave physics, many simulations of BHNS systems
have been performed over the past years. In most dynamical simulations
of BHNS binaries to date, the
self-gravity of the neutron star (NS) and/or the tidal gravity of the
black hole (BH) are
treated in a Newtonian or post-Newtonian framework (see, e.g.,
\cite{Lee00,RSW,Rosswog,Koba,RKLRasio}; see also~\cite{LRA}, who
provide fully relativistic head-on collision simulations).

Our first BHNS merger simulations were performed in an approximate
relativistic framework~\cite{FBSTR,FBST}, which assumed conformal
flatness of the spatial metric throughout the evolution
(see~\cite{Isen,WMM}).  In particular, we evolved extreme mass ratio
($q=M_{\rm BH}/M_{\rm NS}$$\gg$1) initial data developed earlier by
our group~\cite{BSS,TBFS05}.  Though this simulation technique only
allows for crude estimates, we found that mergers of irrotational BHNS
binaries may lead to disks with masses up to 0.3 $M_{\odot}$, with
sufficient heating to emit the neutrino fluxes required to launch an
SGRB.

To date the only self-consistent, general relativistic,
dynamical simulations of BHNS inspiral and
coalescence have been those of Shibata {\it et al.}~\cite{SU1,SU2,ST07,yst08}, 
Etienne {\it et al.}~\cite{eflstb08} (hereafter Paper~I), and 
Duez {\it et al.}~\cite{dfkpst08}.
In many earlier calculations, especially those with initial mass ratios
$q\lesssim 3$, significant disks are formed
after the NS is disrupted, and, for very stiff nuclear
equations of state (EOSs), the core of the NS may survive
the initial mass transfer episode and remain bound.  These findings
contrast with some semi-analytic relativistic arguments, which suggest
that disks with appreciable masses are difficult to form via the
merger of BHNS binaries~\cite{MCMiller}.

In their first set of fully relativistic BHNS simulations, Shibata and
Ury\=u found disk masses in the range of 0.1 -- 0.3 $M_{\odot}$ for
corotating neutron stars~\cite{SU1,SU2}. Subsequently, Shibata and
Taniguchi (\cite{ST07}, hereafter ST) found smaller disk masses of
0.04 -- 0.16 $M_{\odot}$ for irrotational neutron
stars, which are more realistic.  
However, our own fully general relativistic simulations 
of irrotational BHNSs
in Paper~I suggest that the disk mass is no more than 0.04
$M_{\odot}$.  Such a small disk mass, if true for generic cases,
might disfavor BHNS mergers as possible central engines for SGRBs.

Recently, Yamamoto, Shibata, and Taniguchi presented a new code which
implements an adaptive mesh refinement (AMR)
algorithm~\cite{yst08}. In one of their test simulations of BHNS
binaries, they use the same initial data as in ST, but they find that
the disk mass is less than 0.001 $M_{\odot}$. This result disagrees
with that in ST, but is consistent with our result in Paper~I. They
attribute the discrepancy to different grid structures and
computational methods.

Most recently, Duez {\it et al.} developed a new fully general
relativistic hydrodynamics code~\cite{dfkpst08} that evolves
the generalized harmonic formulation of Einstein's equations using a
pseudospectral method, and the equations of hydrodynamics via
shock-capturing, finite-difference techniques. They use this code to
perform a simulation of a single equal-mass BHNS and find that the
disk mass is less than 0.03$M_{\odot}$, which agrees with our result
in Paper~I.

As we report in this paper, our code has been modified to use the
moving-box Carpet~\cite{Carpet} AMR infrastructure. With AMR, our
code is now capable of performing BHNS simulations with resolutions
equivalent to Paper~I at only about $\sim1\%$ the total computational
cost. This enables us to simulate strong-field regions at higher
resolution while simultaneously extending our grid's outer boundaries
much farther into the wavezone for more accurate wave extraction.
We also evolve initial data at larger binary separations than in Paper
I.  Some of the BHNS initial data evolved in this paper are also new, as
the BHs are now spinning. All of our initial data are constructed in
the conformal thin-sandwich (CTS) formalism. The BH spin is
incorporated by imposing boundary conditions on the BH horizon as
in~\cite{CP04,Caudill}.  The initial NS is irrotational and is modeled by an
$n=1$ polytropic EOS.

In this paper, we present a new set of BHNS simulations that probes
how the initial BH spin and the binary mass ratio affect the
dynamics and outcome of the merger.  In particular, we study mergers
of nonspinning BHNS binaries with mass ratios $q=1$, $3$, and $5$. For
the mass ratio $q=3$, we study three cases in which the BH spin parameter
($\tilde{a}\equiv J_{\rm BH}/M_{\rm BH}^2$) is $-0.5$
(counter-rotating), 0 and 0.75.  As in Paper~I, we focus on binaries
in which the NS tidally disrupts before reaching the
innermost stable circular orbit (ISCO) and plunging into the
BH. These systems are the most likely to yield a significant disk
after merger. We thus do not evolve extreme mass-ratio binaries 
in this paper.

Starting with approximately the same initial orbital angular velocity $\Omega$,
we find that when the BH is spinning, the merger time is delayed
when the BH spin aligns with the orbital angular momentum, and
is shortened when the BH spin is anti-aligned with the orbital angular
momentum.  For example, evolutions of initial data with $M\Omega
\approx 0.033$ [$M$ is the initial Arnowitt-Deser-Misner (ADM) mass of
  the system],  require about 4.25 orbits before merger for
$\tilde{a}=-0.5$, but 6.5 orbits before merger for
$\tilde{a} = 0.75$.  This finding is expected, 
since for large, aligned spins there is more angular momentum that
must be radiated away to bring the binary to merger.  This
result is consistent with similar findings in the case of BHBHs (see
e.g.~\cite{clz06}).

After the merger, we find a remnant disk with rest mass between $\lesssim
0.01M_\odot$ and $\approx 0.2M_\odot$ (assuming the NS has a rest
mass of $M_0=1.4M_\odot$). When the initial BH is nonspinning, the disk
mass is $\approx 0.06M_\odot$ for $q=3$, $\approx 0.03M_\odot$ for
$q=1$ and tiny ($\lesssim 0.01M_\odot$) for $q=5$. For a fixed mass ratio
$q=3$, the disk mass is $\approx 0.2 M_\odot$ for $\tilde{a}=0.75$ and
tiny ($\lesssim 0.01M_\odot$) for $\tilde{a}=-0.5$. The increase in disk mass
with increasing BH spin agrees qualitatively with the
semi-relativistic, smoothed particle hydrodynamics (SPH) simulations of
large mass-ratio ($q\approx 10$) BHNS mergers reported
in~\cite{RKLRasio}.

We find that the remnant disks are hot. A rough estimate suggests that
the temperature in the disks is $T\sim 5\times 10^{10}$K. The
results of BH disk simulations in~\cite{SRJ} suggest that the disk
could generate a gamma-ray energy $E\sim 10^{47}$--$10^{50}$erg
from neutrino-antineutrino annihilation, which is promising for
BHNS mergers as plausible central engines for SGRBs.

We compute gravitational waveforms and the effective wave strain in
the frequency domain for our mergers.  As in Paper~I, we find
measurable differences between BHNS waveforms and those produced by
BHBH mergers within the Advanced LIGO band.  These differences appear
at frequencies at which the NS is tidally disrupted and accreted by
the black hole. Extracting the masses and spins of the objects from
the inspiral data, the merger waveforms could provide information about the
radius of the NS, which in turn may be used to constrain the NS
equation of state.

In contrast with Paper~I, we find substantial disks in two of our
nonspinning BH cases.  The reason is that in Paper~I, we imposed
limits on pressure to prevent spurious heating/cooling in
the low-density artificial ``atmosphere''.  We find that imposing
these artificial limits spuriously removes angular momentum,
especially during the NS tidal disruption phase, thereby suppressing
disk formation (see Section~\ref{sec:Pceiling} for further details).

As we cautioned in Paper~I, our findings are fundamentally limited by
the physics we do {\it not} model in these preliminary simulations
(e.g., neutrino transport, magnetic fields, etc), as well as by 
our choice of a simple $\Gamma$-law EOS
to model the (cold and hot) nuclear EOS.
Disk masses may depend sensitively on the initial structure of the neutron
star, especially the low-density outer regions, which is determined
by the adopted EOS.  So all BHNS merger
results should be viewed in light of these caveats.  
Our main focus has been to establish that reliable simulations 
in full general relativity can now be launched to address these 
important issues.

This paper is organized as follows.  In Secs.~\ref{sec:basic_eqns} and
\ref{sec:numerical}, we briefly outline the basic equations and their
specific implementation in our general relativistic, hydrodynamics
scheme. Here we also provide an overview of our initial data, 
gauge conditions, matter evolution equations, and diagnostics.  
Sec.~\ref{sec:code_tests} presents code tests designed to validate our new AMR-based scheme.
In Sec.~\ref{sec:results}, we review the results of our BHNS merger
simulations.  In Sec.~\ref{sec:discussion} we summarize our findings
and comment on future directions.

\section{Basic Equations}
\label{sec:basic_eqns}

The formulation and numerical scheme for our simulations are basically
the same as those already reported in~\cite{DLSS,eflstb08}, to which
the reader may refer for details.  Here we introduce our notation,
summarize our method, and point out the latest changes to our
numerical technique.  Geometrized units ($G = c = 1$) are adopted,
except where stated explicitly.  Greek indices denote all four
spacetime dimensions (0, 1, 2, and 3), and Latin indices imply spatial
parts only (1, 2, and 3).

We use the 3+1 formulation of general relativity and decompose
the metric into the following form:
\beq
  ds^2 = -\alpha^2 dt^2
+ \gamma_{ij} (dx^i + \beta^i dt) (dx^j + \beta^j dt) \ .
\eeq
The fundamental variables for the metric evolution are the spatial
three-metric $\gamma_{ij}$ and extrinsic curvature $K_{ij}$. We adopt
the Baumgarte-Shapiro-Shibata-Nakamura (BSSN) 
formalism~\cite{SN,BS} in which 
the evolution variables are the conformal exponent $\phi
\equiv \ln (\gamma)/12$, the conformal 3-metric $\tilde
\gamma_{ij}=e^{-4\phi}\gamma_{ij}$, three auxiliary functions
$\tilde{\Gamma}^i \equiv -\tilde \gamma^{ij}{}_{,j}$, the trace of
the extrinsic curvature $K$, and the tracefree part of the conformal extrinsic
curvature $\tilde A_{ij} \equiv e^{-4\phi}(K_{ij}-\gamma_{ij} K/3)$.
Here, $\gamma={\rm det}(\gamma_{ij})$. The full spacetime metric $g_{\mu \nu}$
is related to the three-metric $\gamma_{\mu \nu}$ by $\gamma_{\mu \nu}
= g_{\mu \nu} + n_{\mu} n_{\nu}$, where the future-directed, timelike
unit vector $n^{\mu}$ normal to the time slice can be written in terms
of the lapse $\alpha$ and shift $\beta^i$ as $n^{\mu} = \alpha^{-1}
(1,-\beta^i)$. The evolution equations of these BSSN variables are 
given by Eqs.~(9)--(13) in Paper~I. 

It has been suggested that 
evolving $\chi=e^{-4\phi}$ or $W=e^{-2\phi}$ instead of $\phi$ 
gives more accurate results in BHBH simulations (see 
e.g.~\cite{bghhst08,FAU_BBH,rit08}). We have tried all three 
techniques. We find that in the presence of hydrodynamic matter, 
the evolution of the variable $\phi$ yields a more stable evolution 
inside the apparent horizon (AH) near the puncture, leading to better 
rest-mass conservation (see Sec.~\ref{sec:num_metric_hydro}). We therefore
adopt $\phi$-evolution in all of our BHNS simulations. 

It has been suggested that Kreiss-Oliger dissipation is sometimes
useful to add in the BSSN evolution equations outside the AH to reduce
high-frequency numerical noise associated with AMR refinement
interfaces~\cite{goddard06}.  We have also tried this technique, and
found that the dissipation results in slightly smaller Hamiltonian and
momentum constraint violations.  We typically do not use it in our
BHNS simulations.  However, in one case (Case~\G, see
Sec.~\ref{sec:results}), we found that the presence of hydrodynamic
matter causes the conformal related metric $\tilde{\gamma}_{ij}$ to
lose positive definiteness near the puncture, which is
unphysical. This behavior eventually causes the code to crash.  We
find that adding Kreiss-Oliger dissipation, coupled with a high but
finite pressure ceiling deep inside the AH, solves this
problem.

We adopt standard puncture gauge conditions: an advective
``1+log'' slicing condition for the lapse and a 
``Gamma-freezing'' condition for the shift~\cite{GodGauge}. 
Thus we have 
\beqn
  \partial_0 \alpha &=& -2\alpha K  \ , \label{eq:1+log} \\ 
  \partial_0 \beta^i &=& (3/4) B^i \ , \\ 
  \partial_0 B^i &=& \partial_0 \tilde{\Gamma}^i - \eta B^i \ ,
\label{puncturegauge}
\eeqn
where $\partial_0 \equiv \partial_t - \beta^j \partial_j$. The
$\eta$ parameter is set to $\approx 0.55/M$ at the beginning of all
simulations, where $M$ is the initial ADM mass of the BHNS binary.  If
the value for $\eta$ is fixed at this value throughout the simulation,
both Hamiltonian and momentum constraint violation increase to
marginally unacceptable levels ($\gtrsim10\%$, as measured by Eqs.~(40) 
and (41) in Paper~I) outside the BH during the merger phase,
especially in the high BH spin cases.  By simply doubling
$\eta$ to $\approx 1.1/M$ early in the inspiral phase (specifically,
at $t\approx55M$) and fixing it at that value significantly
reduces constraint violations outside the BH during merger.
However, late in the merger stage of Case~\B, we find that
Hamiltonian constraint violation increases rapidly 
($\gtrsim10\%$) outside the BH, unless we enable Kreiss-Oliger
dissipation as well.

The fundamental matter variables are the rest-mass density 
$\rho_0$, specific internal energy $\epsilon$, pressure $P$, and 
four-velocity $u^{\mu}$. We adopt a $\Gamma$-law EOS 
$P=(\Gamma-1)\rho_0 \epsilon$ with $\Gamma=2$, which reduces to 
an $n=1$ polytropic law for the initial (cold) neutron star matter. 
The stress-energy tensor is given by 
\beq
  T_{\mu \nu} = \rho_0 h u_\mu u_\mu + P g_{\mu \nu} \ ,
\eeq
where $h=1+\epsilon+P/\rho_0$ is the specific enthalpy. 
In our numerical implementation of the hydrodynamics 
equations, we evolve the ``conservative'' variables 
$\rho_*$, $\tilde{S}_i$, and $\tilde{\tau}$. They are 
defined as 
\beqn
&&\rho_* \equiv - \sqrt{\gamma}\, \rho_0 n_{\mu} u^{\mu} \ ,
\label{eq:rhos} \\
&& \tilde{S}_i \equiv -  \sqrt{\gamma}\, T_{\mu \nu}n^{\mu} \gamma^{\nu}_{~i}
\ , \\
&& \tilde{\tau} \equiv  \sqrt{\gamma}\, T_{\mu \nu}n^{\mu} n^{\nu} - \rho_* \ .
\label{eq:S0} 
\eeqn
The evolution equations for these variables are given by Eqs.~(21)--(24) 
in Paper~I.

\section{Numerical Methods}
\label{sec:numerical}

\subsection{Initial data}

Our initial data are constructed by solving Einstein's constraint
equations in the CTS formalism, which allows
us to impose an approximate helical Killing vector by setting the time
derivatives of the conformally related metric $\tilde{\gamma}_{ij}$ 
to zero. We model the NS
as an irrotational $n=1$ polytrope, and impose the black hole
equilibrium boundary conditions of Cook and Pfeiffer~\cite{CP04}
on the black hole horizon. 
Details of this method can be found in~\cite{TBFS07b}.  Our
initial data here differ from those described in \cite{TBFS07b} only
in terms of the black hole spin, which can be specified by virtue of a
free (vectorial) parameter $\Omega_{\rm r}^i$ that appears in the
boundary conditions of~\cite{CP04}.  In \cite{TBFS07b} we adopted  
the method in~\cite{Caudill} by iterating
over this parameter until the quasilocal spin $J_{\rm BH}$ of the
black hole vanishes, making both binary components irrotational.  Here we
also consider rotating black hole configurations, for which we iterate
over $\Omega_{\rm r}^i$ until the black hole spin equals certain
specified values.  We focus on black hole spins that are aligned or
anti-aligned with the orbital angular momentum, and consider $\tilde a
\equiv J_{\rm BH}/M_{\rm BH}^2 = 0.75$ (aligned), $0.0$ (nonspinning), 
and $-0.5$ (anti-aligned).  

The initial data are calculated using the {\tt Lorene} spectral methods numerical
libraries \cite{web:Lorene}. 
To map these spectral configurations onto our
non-spectral simulation grid, we first construct our numerical grid
and record the positions of each point in physical coordinates.  Then
we evaluate the field and hydrodynamic quantities at these points
based on their spectral coefficients~\cite{ft01}.  Finally, the
excised BH region is filled with constraint-violating initial data, using the ``smooth junk''
technique we developed and validated in~\cite{EFLSB} (see also 
\cite{Turducken,Turducken2}).   
In particular, we extrapolate all initial data quantities from the BH
exterior into the interior with a 7th order polynomial, using a uniform
stencil spacing of $\Delta r \approx 0.3 r_{\rm AH}$. 

All of the NSs considered in this paper have a compaction of
${\cal C}=M_{\rm NS}/R_{\rm NS} = 0.145$, where
$M_{\rm NS}$ is the ADM mass and $R_{\rm NS}$ is the 
(circumferential) radius of the NS in isolation. Since we model 
the NS with an $n=1$ 
($\Gamma=2$) polytropic EOS, the rest mass of the NS, $M_0$, scales with the 
polytropic constant $\kappa$ as $M_0 \propto \kappa^{1/2}$. 
For a NS with compaction ${\cal C}=0.145$, we find the ADM mass 
for the isolated NS to be $M_{\rm NS}=1.30M_\odot (M_0/1.4M_\odot)$, 
with an isotropic radius $R_{\rm iso}=11.2{\rm km} (M_0/1.4M_\odot)$ 
and circumferential (Schwarzschild) radius of 
$R_{\rm NS}=13.2{\rm km} (M_0/1.4M_\odot)$. The maximum 
rest-mass density of this NS is $\rho_{0,\rm max} = 9\times 
10^{14} \mbox{g cm}^{-3} (1.4M_\odot/M_0)^2$. 

\subsection{Evolution of the metric and hydrodynamics}
\label{sec:num_metric_hydro}

We evolve the BSSN equations
with fourth-order accurate, centered finite-differencing stencils,
except on shift advection terms, where we use fourth-order accurate
upwind stencils.  We apply Sommerfeld outgoing wave boundary
conditions to all BSSN fields, as in Paper~I.  Our code is embedded in
the Cactus parallelization framework~\cite{Cactus}, and our
fourth-order Runge-Kutta timestepping is managed by the {\tt MoL}
(Method of Lines) thorn, with a Courant-Friedrichs-Lewy (CFL) factor
set to 0.45 in all BHNS simulations.  We use the
Carpet~\cite{Carpet} infrastructure to implement the moving-box
adaptive mesh refinement. In all AMR simulations presented here, we
use second-order temporal prolongation, coupled with fifth-order
spatial prolongation. The apparent horizon (AH) of the
black hole is computed with the {\tt AHFinderDirect} Cactus
thorn~\cite{ahfinderdirect}.

We write the general relativistic hydrodynamics equations in
conservative form. They are evolved by a high-resolution
shock-capturing (HRSC) technique~\cite{DLSS} that employs the
monotonized central (MC) reconstruction scheme~\cite{vL77} coupled to
the Harten, Lax, and van Leer (HLL) approximate Riemann solver~\cite{HLL}. 
The adopted hydrodynamic scheme is second-order accurate for smooth 
flows, and first-order accurate when discontinuities (e.g.\ shocks) 
arise. To stabilize our hydrodynamic scheme in regions where there is no
matter, we maintain a tenuous atmosphere on our grid, with a density
floor $\rho_{\rm atm}$ set equal to $10^{-10}$ times the initial
maximum density on our grid. The initial atmospheric pressure
$P_{\rm atm}$ is set equal to the cold
polytropic value $P_{\rm atm} = \kappa \rho_{\rm atm}^{\Gamma}$,
where $\kappa$ is the polytropic constant at $t=0$.  Throughout the
evolution, we impose limits on the atmospheric pressure to prevent
spurious heating and negative values of the internal energy
$\epsilon$. Specifically, we require $P_{\rm min}\leq P \leq P_{\rm max}$, 
where $P_{\rm max}=10 \kappa \rho_0^\Gamma$ and $P_{\rm min}=\kappa 
\rho_0^\Gamma/2$. 
Whenever $P$ exceeds $P_{\rm max}$ or drops below $P_{\rm min}$, we 
reset $P$ to $P_{\rm max}$ or $P_{\rm min}$, respectively.  Applying
these limits everywhere on our grid would artificially
sap the angular momentum in the tidally disrupted NS, allowing
matter to fall spuriously into the BH and thereby suppressing 
disk formation (see Sec.~\ref{sec:results}). To effectively 
eliminate this spurious
angular momentum loss, we impose these pressure limits only 
in regions where the rest-mass density remains very low ($\rho_0 < 100
\rho_{\rm atm}$) or deep inside the AH, where $e^{6\phi} > \psi^6_{\rm
  max}$ (see Sec.~\ref{sec:Pceiling} for more details).

At each timestep, 
we need to recover the ``primitive variables'' 
$\rho_0$, $P$, and $v^i$ from the ``conservative'' variables 
$\rho_*$, $\tilde{\tau}$, and $\tilde{S}_i$. We perform the 
inversion as specified in Eqs.~(57)--(62) of~\cite{DLSS}, but with a
slightly modified analytic quartic solver from the GNU Scientific
Library that outputs only the real roots.  We use the same technique as
in Paper~I to ensure that the values of $\tilde{S}_i$ and
$\tilde{\tau}$ yield physically valid primitive variables,
except we reset $\tilde{\tau}$ to
$10^{-10}\tilde{\tau}_{0,{\rm max}}$ (where
$\tilde{\tau}_{0,{\rm max}}$ is the maximum value of $\tilde{\tau}$
initially) when either $\tilde{S}_i$ or $\tilde{\tau}$ is 
unphysical [i.e., violate one of the inequalities~(34)~or~(35)
in Paper~I]. The restrictions usually apply only to the region near
the puncture or in the low-density atmosphere.

\subsection{Diagnostics}
\label{sec:diagnostics}

During the evolution, we monitor the Hamiltonian and momentum
constraints calculated by Eqs.~(40)--(43) of Paper~I.  Gravitational
waves are extracted using both the Regge-Wheeler-Zerilli-Moncrief
formalism and the Newman-Penrose Weyl scalar $\psi_4$. The extraction
technique is summarized in Sec.~IIIF of Paper~I. 

We also monitor the ADM mass and angular momentum of the system, which
can be calculated by surface integrals at a large distance (Eqs.~(37)
and (39) of Paper~I). However, with our AMR grid, the resolution is
rather low in regions very far from the binary, which causes our
angular momentum measurement to suffer from numerical noise.  
This problem can be overcome by using Gauss's law to split the 
distant surface integral into a sum of integrals over an inner
surface and the volume between the inner and distant surfaces.
The reasons are twofold: (1) the inner surface integrals are 
computed more accurately, as the inner surface resides in the region where
numerical resolution is higher, and (2) most of the contribution in
the volume integral is from the strong-field domain, which is also in the 
inner (high-resolution) region.  The equations we use to
calculate ADM mass and angular momentum with minimal numerical noise
are as follows (see e.g.\ Appendix~A in~\cite{YBS} for a derivation): 
\beqn
  M&=& \int_V d^3x \left(\psi^5\rho + {1\over16\pi}\psi^5 \tilde{A}_{ij}
\tilde{A}^{ij} - {1\over16\pi}\tilde{\Gamma}^{ijk}\tilde{\Gamma}_{jik} \right.
\ \ \label{eq:M_sur_vol} \\
&& \left. + {1-\psi\over16\pi}\tilde{R} - {1\over24\pi}\psi^5K^2\right) \cr
 && + {1\over 2\pi} \oint_S
\left( \frac{1}{8}\tilde{\Gamma}^i-\tilde{\gamma}^{ij} \partial_j \psi
\right) d\Sigma_i \ , \nonumber
\eeqn
\beqn
  J_i &=& {1\over8\pi} \epsilon_{ij}{}^n\int_V d^3x
           \bigl[\psi^6(\tilde{A}^j{}_n + {2\over3}x^j\partial_nK 
\label{eq:J_sur_vol} \\
         && - {1\over2} x^j\tilde{A}_{km}\partial_n\tilde{\gamma}^{km}) 
         + 8\pi x^j S_n\bigr] \cr
           && + {1\over8\pi} \epsilon_{ij}{}^n \oint_S
           \psi^6 x^j \tilde{A}^m{}_n d\Sigma_m \ . \nonumber
\eeqn
Here $S$ is a surface surrounding the BH, $V$ is the volume between the 
inner surface $S$ and a distant surface, $\psi = e^\phi$, 
$\rho=n_\mu n_\nu T^{\mu \nu}$, $S_i = -n_\mu \gamma_{i \nu} T^{\mu \nu}$, 
$\tilde{R}$ is the Ricci scalar associated with $\tilde{\gamma}_{ij}$, 
and $\tilde{\Gamma}_{ijk}$ are Christoffel symbols associated with 
$\tilde{\gamma}_{ij}$. 

When hydrodynamic matter is evolved on a fixed uniform grid, our
hydrodynamic scheme guarantees that the rest mass $M_0$ is conserved
to machine roundoff error.  This strict conservation is no longer maintained
in an AMR grid, where spatial and temporal prolongation is performed
at the refinement boundaries.  Hence we also monitor the
rest mass
\beq
  M_0 = \int \rho_* d^3x
\label{eq:m0}
\eeq
during the evolution. Rest-mass conservation is also violated whenever 
$\rho_0$ is reset to the atmosphere value. This usually happens only in the 
very low-density atmosphere or deep inside the AH where accuracy
is not maintained.  The low-density regions do not affect rest-mass 
conservation significantly. However, this is not the case near
the puncture,  where the rest-mass density can be very high,
especially after the NS is tidally disrupted and matter falls into the
BH and accumulates near the puncture.  The evolution near the puncture
is sensitive to the choice of conformal variable used in the BSSN
evolution, as discussed in Sec.~\ref{sec:basic_eqns}. We find that the
$\phi$ evolution leads to better rest-mass conservation than $\chi$ or
$W$ evolutions.

We measure the thermal energy generated by shocks via the quantity
$K\equiv P/P_{\rm cold}$, where $P_{\rm cold}=\kappa \rho_0^\Gamma$ is
the pressure associated with the cold EOS that characterizes the
initial matter. The specific internal energy can be decomposed into a
cold part and a thermal part: $\epsilon = \epsilon_{\rm cold} +
\epsilon_{\rm th}$ with 
\begin{equation}
\epsilon_{\rm cold} = -\int P_{\rm cold}
d(1/\rho_0) = \frac{\kappa}{\Gamma - 1} \rho_0^{\Gamma-1}\ .
\end{equation}
Hence the relationship between $K$ and $\epsilon_{\rm th}$ is 
\beqn
\epsilon_{\rm th} & = & \epsilon - \epsilon_{\rm cold} =
\frac{1}{\Gamma - 1} \frac{P}{\rho_0} - \frac{\kappa}{\Gamma - 1}
\rho_0^{\Gamma-1} \nonumber \\
& = & (K - 1) \epsilon_{\rm cold} \ .
\eeqn
For shock-heated gas, we always have $K>1$ (see 
Appendix~\ref{app:shock-heating}).

\section{AMR Code Tests}
\label{sec:code_tests}

Our general relativistic magnetohydrodynamic (GRMHD) code has been 
thoroughly tested by passing a robust suite of tests. These tests include
maintaining stable rotating stars in stationary equilibrium, reproducing
the exact Oppenheimer-Snyder solution for collapse to a black hole,
and reproducing analytic solutions for MHD shocks, nonlinear
MHD wave propagation, magnetized Bondi accretion, and MHD waves induced
by linear gravitational waves~\cite{DLSS,dlss05b}. Our code has also been
compared with Shibata \& Sekiguchi's GRMHD code~\cite{SS05} by
performing simulations of identical magnetized hypermassive
neutron stars~\cite{dlsss06a,dlsss06b} and of the magnetorotational
collapse of identical stellar
cores~\cite{ShiUIUC}. We obtain good agreement between these two independent
codes. Our code has also been used to simulate the collapse of very
massive, magnetized, rotating stars to black holes~\cite{lss07};
merging BHBH binaries~\cite{EFLSB}, BHNS binaries (Paper~I), magnetized NSNS
binaries~\cite{lset08}; and relativistic hydrodynamic matter in the
presence of puncture black holes~\cite{FBEST}.  Recently, our code has
been generalized to incorporate (optically thick) radiation transport
and its feedback on MHD fluids~\cite{flls08}. 

All of the above-mentioned tests and simulations were performed 
on grids with uniform spacing. In some of the simulations, we 
utilized the multiple-transition fisheye transformation~\cite{RIT2} 
so that a uniform computational grid spacing 
corresponds to physical coordinates with spatially varying
resolution. Recently, we have modified our code 
so that we can use the moving-box AMR infrastructure provided by
Carpet~\cite{Carpet}. To test our new code, we have performed 
shock-tube tests and 3+1 simulations of linear gravitational waves, single 
stationary and boosted puncture BHs, puncture BHBH binaries, 
stable, rapidly and differentially rotating
relativistic stars, and relativistic Bondi accretion onto a
Schwarzschild BH.  
All of our 3+1 AMR code tests were performed assuming equatorial
symmetry (i.e., symmetry about the $z=0$ orbital plane),  which we assume
in all BHNS evolutions presented in this paper.
In many of the tests we have analytic solutions with which to compare. 
For the BHBH tests, 
we compare our results with the numerical results reported in the literature. 
We have confirmed that our new code can evolve these 
systems reliably with AMR. Below we briefly report results from both our BHBH 
and differentially rotating relativistic star AMR code tests.

\subsection{Binary black holes}
 
In this subsection we summarize an important testbed for calibrating
the vacuum sector of our code: BHBH merger simulations.  We provide
details on the setup of our initial data and numerical grid, as well
as on the adopted numerical techniques and simulation results, in the
tables and figures of Appendix~\ref{app:BHBH}. Here we point out some
highlights.

We first simulate an equal-mass BHBH system. We use puncture initial
data~\cite{BeiO94,BeiO96,BB97} with initial binary separation
$D=9.89M$, where $M$ is the ADM mass of the binary. The initial
``bare'' mass and momentum of each puncture are chosen according
to~\cite{tb04} to make the orbit quasicircular (see also
\cite{Bau00}).  This initial configuration is close to the R4 run
presented in~\cite{goddard06,goddard07}. We use 8 refinement levels
centered at each BH and place the outer boundary at $323.4M$.
The resolution in the innermost refinement level is $M/32$, so that
there are about 15 grid points across the diameter of each
black hole's AH initially.  This number rapidly
increases during the early inspiral to a fixed value of about 24 grid
points due to our gauge choice, and increases again to about 40 grid
points, post-merger.
We find that the binary inspiral orbit, gravitational waveform, energy and 
angular momentum emitted by gravitational waves, and the final 
black hole's spin parameter are very similar to those reported by others
in~\cite{goddard06,goddard07}. 

To further validate our numerical results, we compute the quantities
\beqn
  \delta E &=& (M-M_f-\Delta E_{\rm GW})/M \ , \label{eq:deltae} \\ 
  \delta J &=& (J-J_f-\Delta J_{\rm GW})/J \ , \label{eq:deltaj}
\eeqn
where $J$ is the ADM angular momentum of the initial binary, 
$M_f$ and $J_f$ are the ADM mass and angular momentum 
of the final (merged) system, and $\Delta E_{\rm GW}$ and
$\Delta J_{\rm GW}$ are the energy and angular momentum carried 
off by
gravitational waves. We find that both $M_f$ and $J_f$ are very 
close to $M_{\rm BH}$ and $J_{\rm BH}$, the mass and angular momentum 
of the final BH.
We compute $M_{\rm BH}$ and $J_{\rm BH}$ from the
irreducible mass $M_{\rm irr}$ and the ratio of proper polar to
equatorial circumference $C_r$ of the final black hole's
horizon. Specifically, we first use Eq.~(5.3) of~\cite{bhspin} to
solve for the dimensionless spin $\tilde{a}\equiv J_{\rm BH}/M_{\rm
  BH}^2$:
\beq
  C_r=\frac{1+\sqrt{1-\tilde{a}^2}}{\pi}
E\left(-\frac{\tilde{a}^2}{(1+\sqrt{1-\tilde{a}^2})^2}\right) \ ,
\label{eq:spinbh} 
\eeq 
where $E(x)$ is the complete elliptic integral of the second kind. 
We next compute $M_{\rm BH}$ by $M_{\rm BH} = (M_{\rm irr}/\tilde{a})
\sqrt{2(1-\sqrt{1-\tilde{a}^2})}$ and $J_{\rm BH}$ by 
$J_{\rm BH} = \tilde{a} M_{\rm BH}^2$. These formulae are derived 
for a Kerr spacetime. They are applicable to the merged BH as the
spacetime approaches Kerr at late times. Gravitational waves are
extracted via the Newman-Penrose scalar $\psi_4$.  
The quantities $\Delta E_{\rm GW}$ and $\Delta J_{\rm GW}$ are computed by 
integrating Eqs.~(50) and (52) of Paper~I. Conservation of energy and 
angular momentum demands that $\delta E=0=\delta J$. Hence these parameters 
are good indicators of the accuracy of the simulations, whenever a
significant amount of energy and angular momentum are radiated during 
the evolution. We found $\delta E \approx 4\times 10^{-4}$ and 
$\delta J \approx 4 \times 10^{-3}$ in our equal-mass BHBH
simulation.

We next perform an unequal-mass BHBH simulation. The initial 
data are generated by the {\tt TwoPunctures} code~\cite{abt04} with 
initial binary (coordinate) separation $D\approx7M$. 
The approximate initial momentum of each BH for a quasicircular
configuration is calculated by the 3PN expression using
Eq.~(65) of~\cite{bghhst08}. The ``bare'' masses of the punctures 
are adjusted so that the irreducible mass ratio of the BHs equals
3. The initial configuration approximates one of the cases
studied in~\cite{bcgshhb07}. We use 9 and 10 refinement levels centered 
at the larger and smaller BH, respectively, with an outer boundary at
$413.9M$.  The resolution around the large and small BH is
$M/50$ and $M/100$, respectively, so that there are initially
about 36 and 22 grid points across the diameter of the larger and
smaller black holes' AH, respectively.  We find that 
our computed values of $\Delta E_{\rm GW}$ and $\Delta J_{\rm GW}$ agree with 
Table~I of~\cite{bcgshhb07} to within 4\%, and that the kick velocity 
agrees with Fig.~2 of~\cite{JenaQ}. We note that exact agreement is 
not expected as our initial configurations are slightly different. 
Finally, we find excellent energy and angular momentum conservation:
$\delta E \approx -2\times 10^{-4}$ and $\delta J \approx 9\times
10^{-3}$.

These tests demonstrate our code's ability to evolve dynamical vacuum 
spacetimes containing moving BHs. We next study our code's 
ability to handle hydrodynamics in a strong-field spacetime with
moving AMR boxes. 

\subsection{Rotating relativistic star}

\begin{figure*}
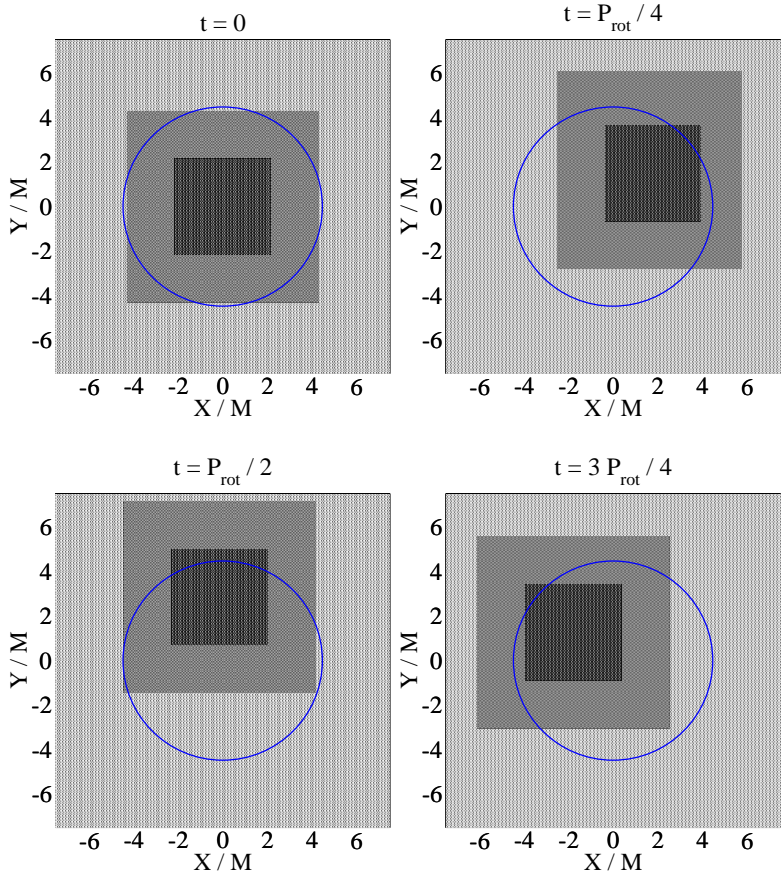

\vspace{-4mm}
\begin{center}
\epsfxsize=2.0in
\leavevmode
\epsffile{rns_t000.eps}
\epsfxsize=2.0in
\leavevmode
\epsffile{rns_t025.eps} \\
\vspace{5mm}
\epsfxsize=2.0in
\leavevmode
\epsffile{rns_t050.eps}
\epsfxsize=2.0in
\leavevmode
\epsffile{rns_t075.eps}
\caption{The moving AMR grid in the rotating relativistic star test. Shown
here are the snapshots of the two innermost refinement boxes 
(the nested squares) and part of the third refinement level in 
the equatorial plane. The circle denotes the boundary of the star.}
\label{fig:rns_grid_setup}
\end{center}
\end{figure*}

In our BHNS simulations, the NS is placed inside a high
resolution refinement box initially. However, when it is tidally
disrupted by the black hole, the NS matter spreads into other
refinement boxes, with different resolutions.  The purpose of this test
is to demonstrate the ability of our AMR code to maintain the
equilibrium of a stable, differentially rotating relativistic star
crossing refinement boundaries during the evolution, and verify 
second-order convergence as the resolution is improved.

The star considered in this test is the same hypermassive neutron star 
studied in~\cite{bss00}, which is modeled by an $n=1$ polytropic EOS.
The mass of the star is 1.7 times greater than the maximum mass of 
a {\it nonrotating} neutron star with the same EOS, but has been shown to be 
dynamically stable in previous simulations~\cite{bss00,dmsb03,DLSS}. 
The star is highly flattened because of its rapid rotation. It has an 
equatorial (coordinate) radius $R_e \approx 4.5M$, and a polar radius 
$R_p\approx 1.2M$. The rapid rotation also causes a torus-like density 
distribution in which the maximum density occurs off-center. 
The central rotation period of the star is $P_{\rm rot}=38.4M$. 

Our AMR grid contains four refinement levels that move in time. 
The side-lengths of the 
refinement boxes in the equatorial plane are $4.29M$, $8.69M$, 
$22.9M$, and $45.8M$. Since we impose equatorial symmetry, the 
lengths in the $z$-direction are half of the sizes stated above. 
The coarsest grid is fixed and the outer boundary is placed 
at $x$, $y =\pm 46M$ and $z=43M$. We initially center the star 
at the origin, and move the center of the four refinement 
boxes according to the following prescription: 
\beq
  x_c = A_x \sin \omega t \ \ , \ \ y_c = A_y (1-\cos \omega t) \ \ , 
\ \ z_c = 0 \ ,
\eeq
where $A_x$, $A_y$ and $\omega$ are constants. We set the amplitudes 
$A_x=1.79M$, $A_y=1.43M$ and the angular frequency of the moving center 
$\omega = 0.168/M$ ($2\pi/\omega = 0.98P_{\rm rot}$). 
With these parameters, the bulk of the star crosses 
the innermost refinement boxes during the evolution 
(see Fig.~\ref{fig:rns_grid_setup}). The grid spacing doubles at each 
successive refinement level. We evolve the star with three different 
resolutions, which we label low, medium and high resolution runs. 
The grid spacing in the innermost refinement level is $M/11.2$,
$M/16.7$, and $M/22.2$ for the low, medium and high resolution runs,
respectively.

\begin{figure}
\epsfxsize=3.4in
\leavevmode
\epsffile{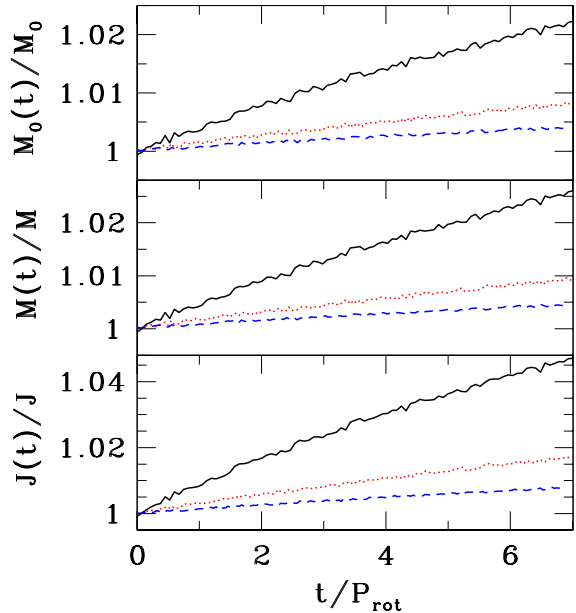}
\caption{Evolution of rest mass $M_0$, ADM mass $M$, and angular 
momentum $J$ of a rapidly rotating star using a moving AMR grid with 
low (black solid lines), medium (red dotted lines) and high (blue 
dashed lines) resolution.}
\label{fig:rns-diagnostics}
\end{figure}

Figure~\ref{fig:rns-diagnostics} shows the evolution of the rest mass
$M_0$, ADM mass $M$, and angular momentum $J$ of the star,
calculated by Eqs.~(\ref{eq:M_sur_vol})--(\ref{eq:m0}). Since there is
no black hole in the spacetime, we calculate only the volume integrals
of $M$ and $J$ over the entire spatial grid. As mentioned in
Sec.~\ref{sec:diagnostics}, strict rest-mass conservation is not
expected when using an AMR grid.  However, the deviations in
$M_0$, $M$ and $J$ converge to zero at slightly higher than second
order with increasing resolution.  In addition, the amplitudes of
the gravitational radiation decrease to zero at second order with increasing
resolution. Gravitational radiation does not contribute significantly
to the nonconservation of $M$ and $J$ observed in 
Fig.~\ref{fig:rns-diagnostics}; $\Delta E_{\rm GW}/M \lesssim 10^{-3}$ and
$\Delta J_{\rm GW}/J \lesssim 10^{-3}$ for all three runs.

\begin{figure}
\epsfxsize=3.4in
\leavevmode
\epsffile{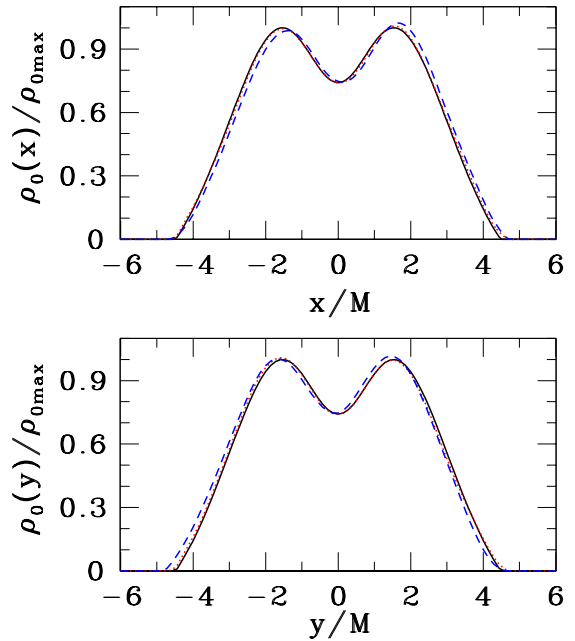}
\caption{Density profiles in the equatorial plane along the $x$-axis
  (upper graph) and $y$-axis (lower graph) for the high resolution
  run.  The rest-mass density $\rho_0$ is normalized to the initial
  maximum density, $\rho_{0,\rm max}$. Solid (black), dotted (red) and
  dashed (blue) lines show the profiles at times $t=0$, $3.43P_{\rm
  rot}$ and $6.86P_{\rm rot}$, respectively.}
\label{fig:rns-profile}
\end{figure}

Figure~\ref{fig:rns-profile} demonstrates that the equilibrium 
profile of the rest-mass density is well-maintained
in the high resolution simulation, apart from a slight drift
of the star's center.

\section{Results}
\label{sec:results}

\subsection{Overview}

\begin{table}
\caption{Initial BHNS configurations. Here, the mass ratio
  is defined as $q\equiv M_{\rm BH}/M_{NS}$, where $M_{\rm BH}$ and
  $M_{NS}$ are the ADM masses of the black hole and neutron star at
  infinite separation, $M$ is the total ADM mass of the binary
  system, $D_0$ the initial binary coordinate separation,
  $\tilde{a}=J_{\rm BH}/M_{\rm BH}^2$ is the spin parameter of the
  black hole (always aligned or anti-aligned with the orbital angular
  momentum), and $\Omega$ is the initial orbital angular
  velocity. All neutron stars have the same nondimensional rest
  (baryon) mass $\bar{M}_0\equiv M_0/\kappa^{1/2}=0.15$, which is
  $\approx 83\%$ the maximum rest mass of a nonrotating NS with the
  same $n=1$ polytropic EOS. 
}
\begin{tabular}{ccccccc}
\hline
\hline
Case & $\tilde{a}$ & $q$ & & $D_0/M$ & $J/M^2$ & $M\Omega$ \\
\hline
 \A      &  0.00 & 3.0 & &  8.81 & 0.702 & 0.0333 \\
 \A-MSep &  0.00 & 3.0 & &  7.17 & 0.668 & 0.0441 \\
 \A-SSep &  0.00 & 3.0 & &  5.41 & 0.638 & 0.0623 \\
 \B      &  0.75 & 3.0 & &  8.81 & 1.096 & 0.0328 \\
 \B-SSep &  0.75 & 3.0 & &  5.53 & 1.011 & 0.0594 \\
 \D      &$-$0.50\ \ \  & 3.0 & &  8.81 & 0.443 & 0.0338 \\
 \F      &  0.00 & 5.0 & &  8.79 & 0.518 & 0.0333 \\
 \G      &  0.00 & 1.0 & &  8.61 & 0.938 & 0.0347 \\
\hline
\hline
\end{tabular}
\label{table:initconf}
\end{table}

\begin{table*}
\caption{Grid configurations used in our simulations. Here, $N_{\rm
    AH}$ denotes the number of grid points covering the diameter of
    the (spherical) apparent horizon initially, and $N_{\rm NS}$
    denotes the number of grid points covering the smallest diameter
    of the neutron star initially.}
\begin{tabular}{cccccc}
  \hline \hline
  Case & $M\Omega$ & Grid Hierarchy (in units of $M$)$^{(a)}$
  & Max.~resolution & $N_{\rm AH}$ & $N_{\rm NS}$ \\
  \hline \hline
  \A      & 0.0333 & (196.7, 98.35, 49.18, 24.59, 12.29, 6.147, 3.073,
  1.414 [1.660]) & $M/32.5$ & 41 & 85 \\
  \A-MSep & 0.0441 & (197.0, 98.49, 49.24, 24.62, 12.31, 6.156, 3.078,
  1.416 [1.662]) & $M/32.5$ & 40 & 82 \\
  \A-SSep & 0.0623 & (197.3, 98.67, 49.33, 24.67, 12.33, 6.167, 3.083,
  1.418 [1.665]) & $M/32.4$ & 40 & 76 \\
  \B      & 0.0328 & (210.2, 92.49, 46.24, 23.12, 11.56, 5.780, 2.890,
  1.445 [1.642], 0.7554 [N/A])  & $M/60.9$ & 56 & 80 \\
  \B-SSep-HRes & 0.0594 & (197.6, 98.79, 49.40, 24.70, 12.35, 6.174,
  3.087, 1.544, 0.7718 [N/A]) & $M/64.8$ & 59 & 78 \\
  \B-SSep-MRes & 0.0594 & (197.6, 98.79, 49.40, 24.70, 12.35, 6.174,
  3.087, 1.544, 0.7718 [N/A]) & $M/47.9$ & 43 & 57 \\
  \B-SSep-LRes & 0.0594 & (197.6, 98.79, 49.40, 24.70, 12.35, 6.174,
  3.087, 1.544, 0.7718 [N/A]) & $M/41.5$ & 37 & 50 \\
  \D      & 0.0338 & (196.6, 98.31, 49.16, 24.58, 12.29, 6.145, 3.072,
  1.413 [1.659]) & $M/32.5$ & 37 & 85 \\
  \F      & 0.0333 & (196.3, 98.12, 49.06, 24.53, 12.26, 4.415, 2.208,
  1.104) & $M/48.9$ & 69 & 84 \\
  \G      & 0.0347 & (217.6, 108.8, 54.40, 27.20, 13.60, 6.801, 3.400,
  1.496 [N/A], 0.7349 [N/A]) & $M/58.8$ & 47 & 78 \\
  \hline \hline
\end{tabular}

\begin{flushleft}
$^{(a)}$ There are two sets of nested refinement boxes: one centered
  on the NS and one on the BH.  This column specifies the half side length
  of the refinement boxes centered on both the BH and NS. When
  the side length around the NS is different, we specify the NS half side 
  length in square brackets.  If there is no corresponding NS
  refinement box (as is the case when the NS is significantly larger
  than the BH), we write [N/A] for that box.
\end{flushleft}
\label{table:GridStructure}
\end{table*}

\begin{figure*}
\vspace{-4mm}
\begin{center}
\epsfxsize=3.5in
\leavevmode
\epsffile{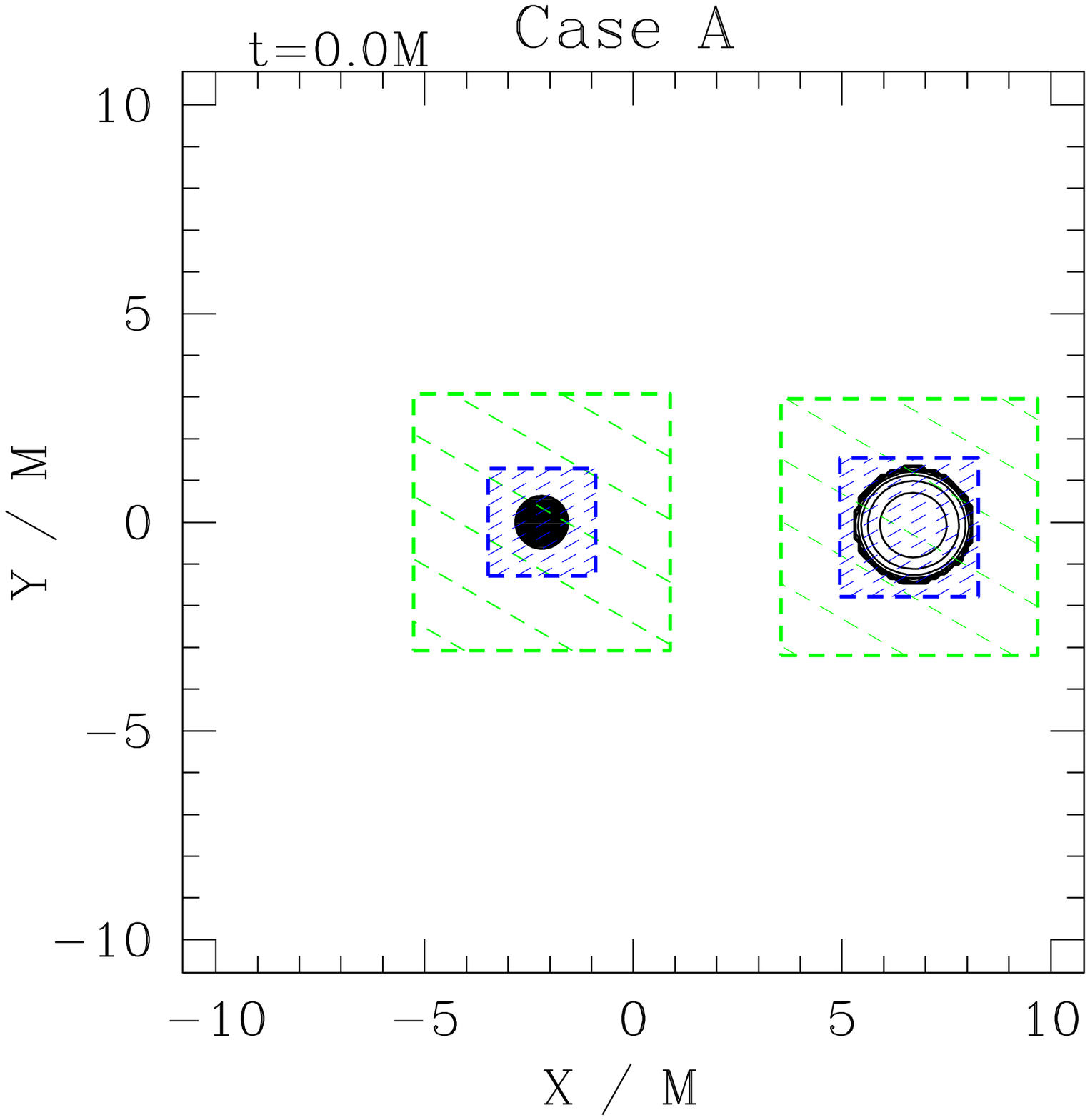}
\epsfxsize=3.5in
\leavevmode
\epsffile{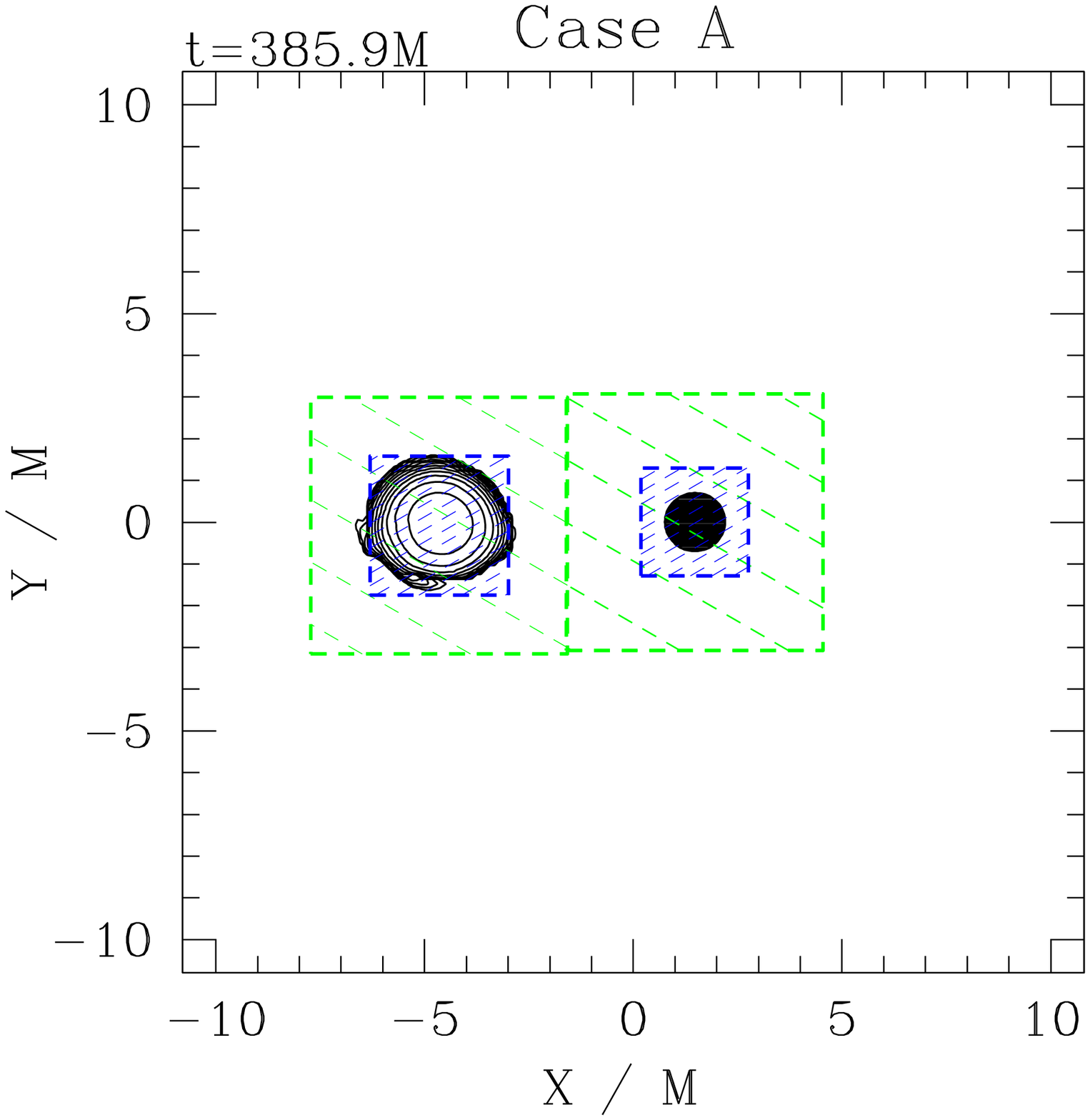}\\
\epsfxsize=3.5in
\leavevmode
\epsffile{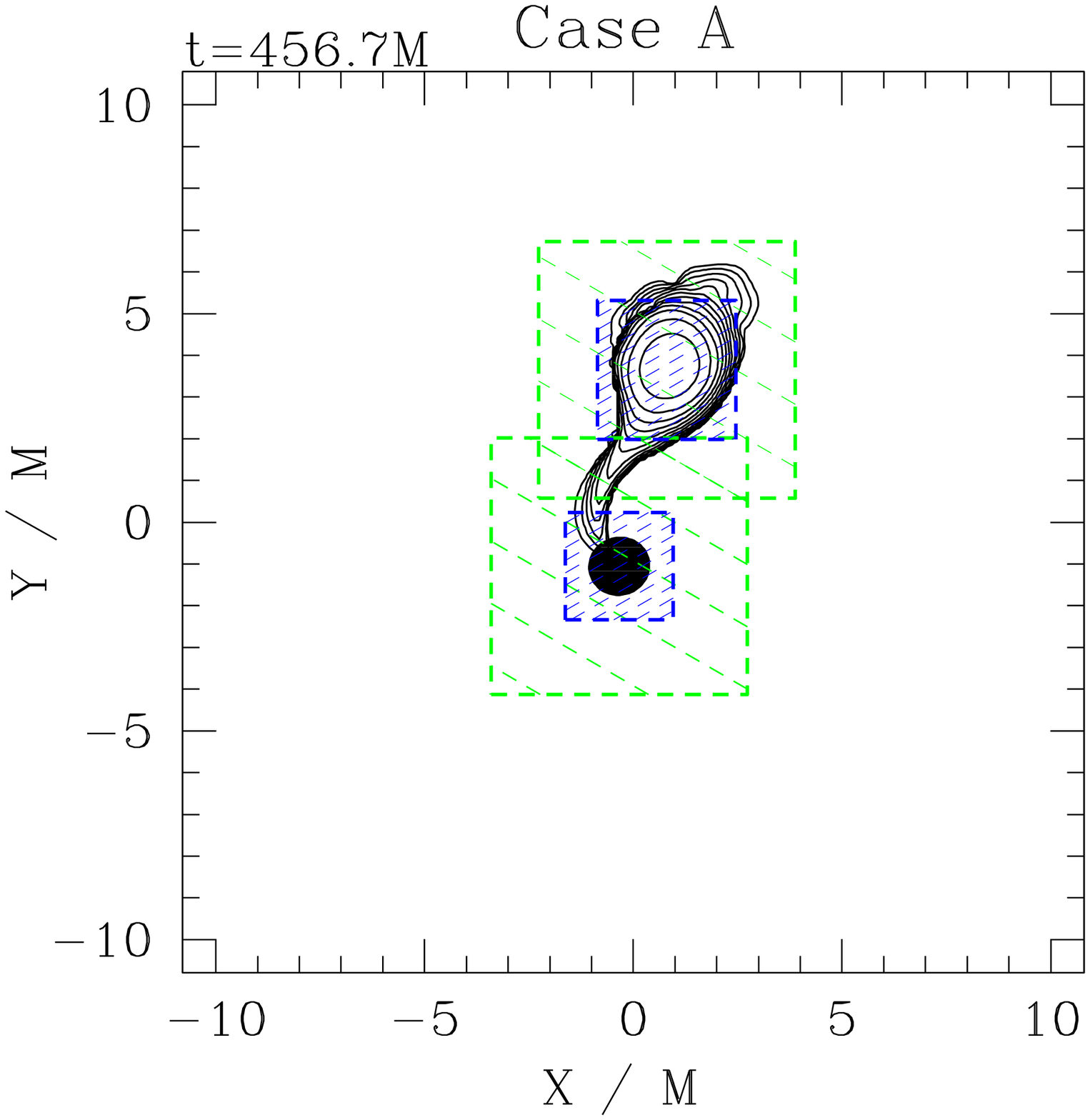}
\epsfxsize=3.5in
\leavevmode
\epsffile{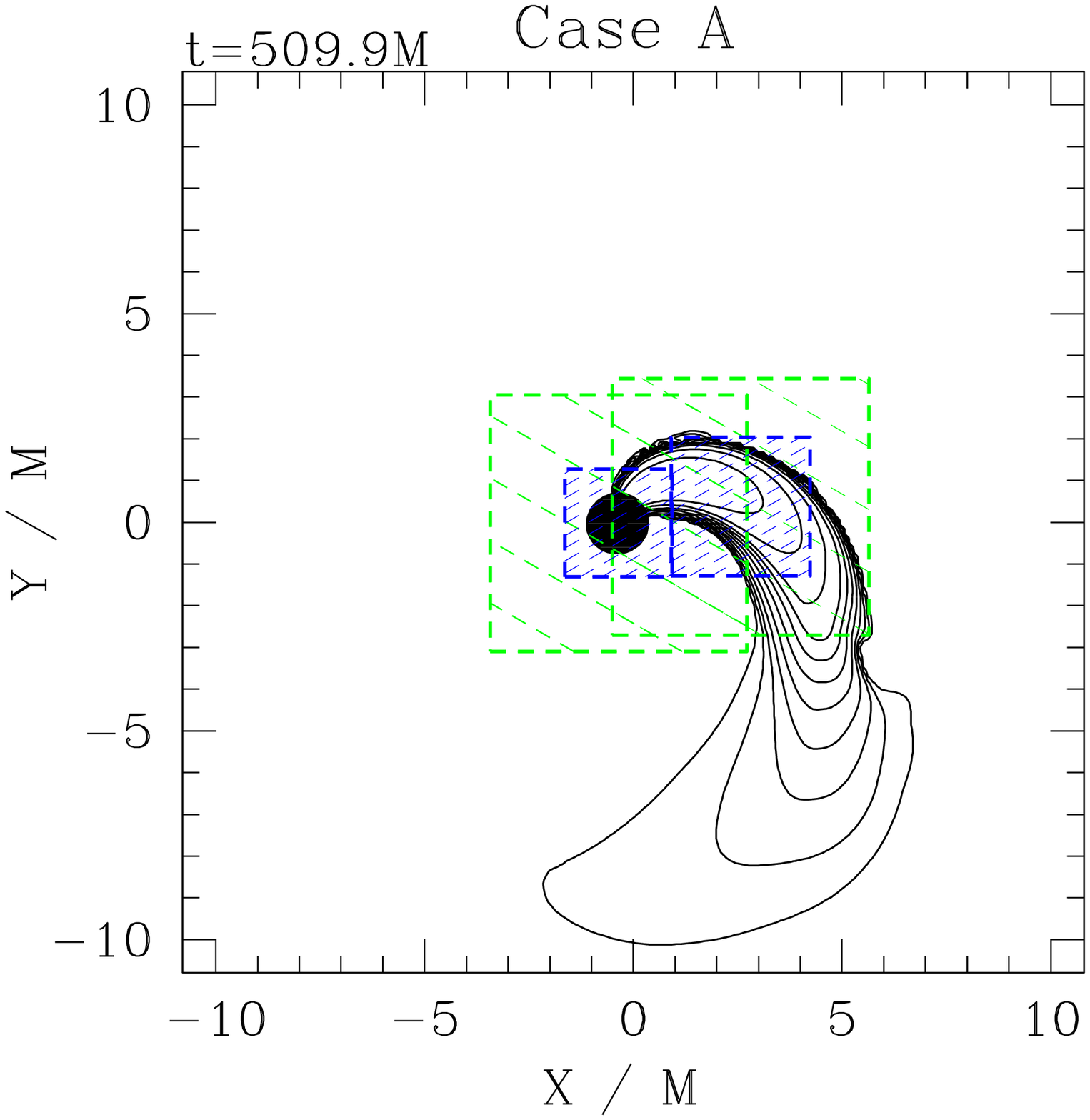}
\caption{AMR grid setup during a typical BHNS simulation, as observed in
  the equatorial plane. Only the two innermost refinement boxes are
  shown for clarity. The black region denotes the apparent horizon of
  the black hole. Neutron star density contours are drawn 
  according to $\rho_0 = \rho_{0,{\rm max}} 10^{-0.38j-0.04}$ 
  ($j$=0, 1,... 12). The maximum rest-mass density of the initial NS 
  is $\kappa \rho_{0,{\rm max}}=0.126$, or
  $\rho_{0,{\rm max}}=9\times 10^{14} (1.4M_\odot/M_0)^2$g cm$^{-3}$.}
\label{fig:refbox}
\end{center}
\end{figure*}

\begin{table*}
\caption{BHNS simulation results. Here $M_{\rm disk}$ is the
  rest mass of the material outside the AH at the end of the simulation,
  $\tilde{a}_f$ is the value of $\tilde{a}$ computed by solving
  Eq.~(\ref{eq:spinbh}) at late times.  The total energy and angular
  momentum carried off by the gravitational radiation are given by $\Delta E_{\rm
  GW}$ and $\Delta J_{\rm GW}$, respectively. $v_{\rm kick}$ is the kick velocity 
  due to recoil.
  $N_{\rm orbits}$ specifies the
  number of orbits in the inspiral phase before merger.
  $\delta E$ and $\delta J$ measure the violation of energy and
  angular momentum conservation, as defined in Eqs.~(\ref{eq:deltae})
  and (\ref{eq:deltaj}), respectively.}
\begin{tabular}{ccccccccc}
\hline
\hline
Case & $M_{\rm disk}/M_0$ & $\tilde{a}_f$ & $\Delta
E_{\rm{GW}}/M$ & $\Delta J_{\rm{GW}}/J$ & $v_{\rm kick}$ (km/s) &
$N_{\rm{orbits}}$ & $\delta E$ & $\delta J$ \\
\hline
 \A      &  3.9\% & 0.559 & 0.93\% & 17.4\% & 33 & 4.5 & $-0.02\%$ &
2.2\% \\
 \A-MSep &  3.8\% & 0.560 & 0.81\% & 13.2\% & 33 & 2.5 & $-0.02\%$ &
1.9\% \\
 \A-SSep &  2.9\% & 0.557 & 0.45\% & 11.4\% & 33 & 1.75 & $-0.02\%$ &
 $-0.36\%$ \\
 \B      & 15\% & 0.881 & 0.92\% & 13.3\% & 22 & 6.5 & $-0.04\%$ &
2.8\% \\
 \B-SSep & 14.3\% & 0.882 & 0.60\% &  6.4\% & 79 & 2.0 & $-0.02\%$ &
2.6\% \\
 \D      &  0.8\% & 0.331 & 0.98\% & 24.4\% & 49 & 3.25 & $-0.01\%$ &
1.3\% \\
 \F      &  0.8\% & 0.418 & 0.98\% & 19.2\% & 73 & 6.25 & $-0.02\%$ &
1.9\% \\
 \G      &  2.3\% & 0.851 & 0.35\% &  7.2\% & 17 & 2.25 & $-0.01\%$ &
0.9\% \\
\hline
\hline
\end{tabular}
\label{table:results}
\end{table*}

We perform a number of BHNS simulations with varying initial 
configurations and numerical resolutions.  Table~\ref{table:initconf}
provides an overview of our chosen initial configurations, and 
Table~\ref{table:GridStructure} specifies the AMR grid structure used
in each case.  
The first three cases in Table~\ref{table:initconf} 
(Cases~\A, \A-MSep and \A-SSep) correspond to the same 
BHNS binary system tracked from different initial separations 
along a quasiequilibrium sequence.
The objective of these simulations is to demonstrate that the fate 
of the binaries, and their merger waveforms, 
are insensitive to the initial binary separation at which we begin the 
simulation, provided it is sufficiently large.
For Case~\B-SSep, we evolve the
system with three different resolutions (see
Table~\ref{table:GridStructure}) to demonstrate convergence.  Fixing
the mass ratio at $q=3$ and initial orbital angular frequency at 
$M \Omega\approx 0.033$, we then study the effects of black hole spin
by choosing the spin parameter $\tilde{a}$ between $-0.5$ and 0.75
(Cases~\D, \A\ and \B).  In all cases, the black hole spin is either aligned or
anti-aligned with the orbital angular momentum.  
Finally, we study the dependence on the mass ratio by
varying $q$ between 1 and 5 (Cases~\G, \A\ and \F).

In all simulations, AMR refinement levels are initially centered 
on the BH and NS.  After $t=0$, the levels centered on the BH track
the AH centroid, and the levels centered on the NS track the matter centroid
defined by $X^i_c$:
\beq
  X^i_c \equiv \frac{ \int_V  x^i \rho_* d^3x}{\int_V \rho_* d^3x} \ ,
\label{ns_refboxcentroid}
\eeq
where $V$ is the volume outside the innermost refinement box 
surrounding the BH. Figure~\ref{fig:refbox} shows snapshots of the 
inner refinement boxes during the Case~\A\ simulation.

In all of our BHNS simulations, we find that the integrated 
Hamiltonian constraint violation [as measured by Eq.~(40) in Paper I] is
0.1\%--0.6\% outside the BH during the inspiral phase.  During merger,
it jumps to a peak value of 1\%--2\% in Cases~\A, \D, \F, and \G\ and
to $8.8\%$ in Case~\B.  Post-merger, the violation stays roughly constant (Case~\G)
or gradually decreases (all other cases) until the simulation is
stopped. The integrated momentum constraint violation [as measured by
Eq.~(41) in Paper~I] outside the BH generally oscillates around 1\%--2\%,
except for a few peaks of 3\%--6\% during inspiral.  We measure
energy and momentum conservation by the quantities $\delta E$ and
$\delta J$ defined in Eqs.~(\ref{eq:deltae}) and (\ref{eq:deltaj}). 
We find that 
$\delta E \approx 0.02\%$ and $\delta J =$ 1--2\% (see
Table~\ref{table:results}) in all cases.  

We also monitor the rest-mass conservation by computing the quantity
\beq
  \Delta M_0 = [M_0-M_0(0)]/M_0(0) \ ,
\eeq
where $M_0(0)$ is the NS rest mass computed at $t=0$. Rest-mass 
conservation requires $\Delta M_0=0$ at all times. Most of the
violation occurs inside the BH apparent horizon when the NS 
matter falls into the BH.  
We find that $\Delta M_0 \approx$ 0.06--0.08\% during the inspiral 
phase (for all
cases except \G, which has $\Delta M_0 \approx 0.002\%$).  During and
after the merger phase, we find large ``glitches'' in $M_0$ inside
the AH due to inaccuracies in that region, which causes large $\Delta
M_0$ ($\gtrsim$3\%), but these are easily accounted for and
neglected.  We find that $\Delta M_0$ is 0.1\%--0.15\% after
the inspiral phase. This result is important
because in some cases we find a low-mass ($\sim 0.01M_0$) disk 
surrounding the BH. These results can only be trusted if the rest-mass 
violation outside the apparent horizon due to numerical error 
is much smaller than the disk mass, which is true in all of 
our simulations.

\begin{figure}
\epsfxsize=3.4in
\leavevmode
\epsffile{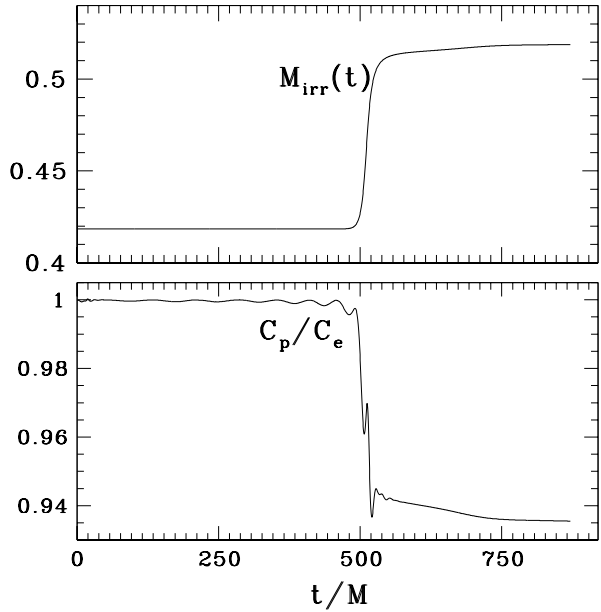}
\caption{The BH's irreducible mass $M_{\rm irr}$ and $C_p/C_e$
  versus time, for Case~\A.}
\label{fig:caseA_M_irr_CpoverCe}
\end{figure}

Figure~\ref{fig:caseA_M_irr_CpoverCe} shows the BH's irreducible
mass $M_{\rm irr}$ and $C_r=C_p/C_e$ as a function of time for
Case~\A, where $C_p$ and $C_e$ are the proper polar and equatorial
circumferences, respectively.  Notice that these two quantities remain
almost constant at early times until the NS matter falls into the
BH.  They approach constant values again at late times as the
system settles into a quasistationary state. We find the same
qualitative behavior for all of our BHNS simulations.  Given $C_r$, we
define $\tilde{a}$ using Eq.~(\ref{eq:spinbh}). When the BH is
well-approximated by the Kerr metric (e.g., the early inspiral phase,
at large separations), $\tilde{a}$ equals the BH's spin parameter 
$J_{\rm BH}/M_{\rm BH}^2$.  In its final stage, the system
consists of a BH surrounded by a disk. The spacetime near
the BH therefore deviates from Kerr, and $\tilde{a}$ does not need
to coincide with the value of the BH spin parameter defined in,
e.g.\ the isolated horizon formulation~\cite{IH_DH}.  Nevertheless,
$\tilde{a}$ is still a reasonable measure of the BH's spin when the disk
is not very massive compared to the BH's mass.
Table~\ref{table:results} lists the final value of $\tilde{a}$ in each
of our BHNS simulations.

\subsection{Convergence studies}

\begin{figure}
\epsfxsize=3.4in
\leavevmode
\epsffile{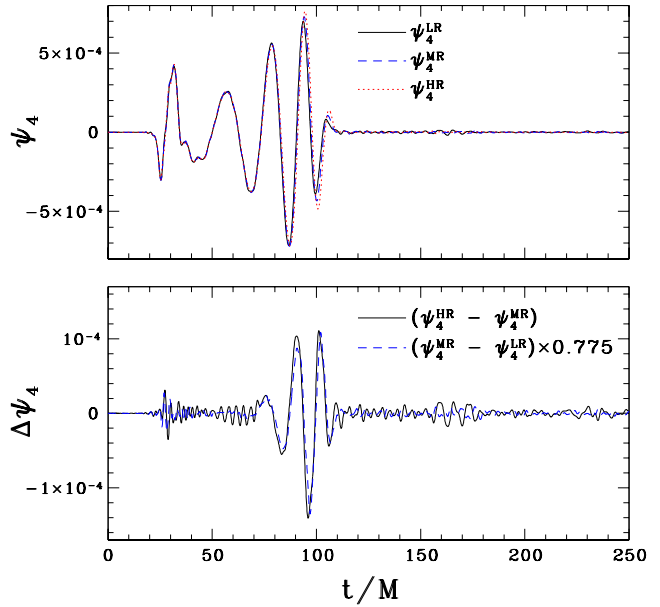}
\caption{Numerical convergence of gravitational wave signals for three 
B-SSep simulations. The highest grid resolutions are $M/41.5$, 
$M/47.9$, and $M/64.8$,
so that initially the AH diameter is covered by 38, 44, and 59
gridpoints, and the NS is covered by 58, 67, and 91 gridpoints,
respectively.  In the top panel, we show the real part of the $l=m=2$
component of $\psi_4$ for the three waveforms. In the bottom panel,
pairwise differences between the waveforms are plotted and rescaled to
demonstrate second-order convergence.}
\label{fig:caseBSSep_res_study}
\end{figure}

To demonstrate that the resolution used in our simulations is
sufficiently high, we perform a convergence test. It is known that 
as the BH spin is increased, higher resolution is required for accurate
BHBH evolutions (see e.g.~\cite{Jena_BBH}).  Hence we perform a
convergence test for the highest spinning case with
$\tilde{a}=0.75$. We use three different grid resolutions with close
initial binary separation (Cases~\B-SSep-HRes, \B-SSep-MRes and 
\B-SSep-LRes in Table~\ref{table:GridStructure}), which is 
adequate for the purpose of a convergence test. The top panel in 
Fig.~\ref{fig:caseBSSep_res_study} shows the real part of $\psi_4^{22}$, 
the $l=m=2$ mode of $\psi_4$, computed with three different resolutions, 
showing good agreement.  The bottom panel in
Fig.~\ref{fig:caseBSSep_res_study} compares the pairwise differences
between the waveforms and we see that the waveforms converge at
second-order accuracy. 

\subsection{Effects of different initial binary separations}

In addition to the convergence test, we need to verify that our
initial binary separation is sufficiently large.  That is, 
starting the simulation from any larger
initial separation should give the same outcome after a simple 
time (and, for gravitational waves, a possible phase) shift. 

\begin{figure}
\epsfxsize=3.4in
\leavevmode
\epsffile{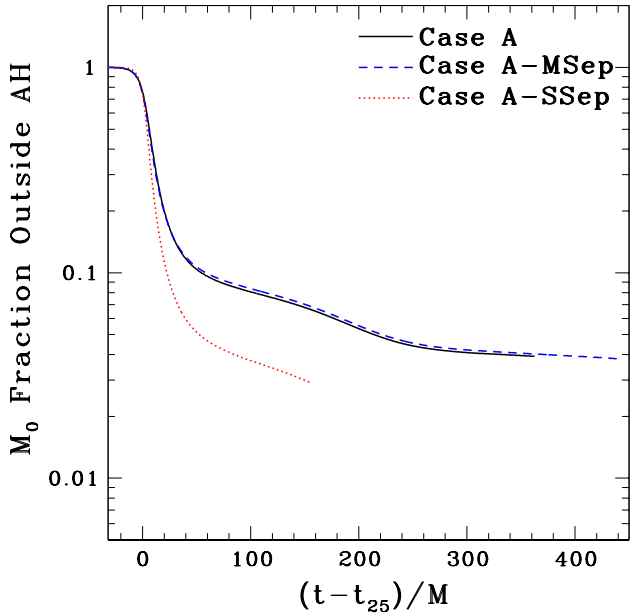}
\caption{Rest mass of NS matter outside the BH versus time for different initial
  separations for the $q=3$, nonspinning BH case (Cases~\A, \A-MSep, and
  \A-SSep).  Here, the time coordinate is shifted by $t_{25}$, where
  $t_N$ is the time at which $N\%$ of the NS rest mass has fallen into
  the apparent horizon.}
\label{fig:BH_sepstudyA_accretion}
\end{figure}

\begin{figure}
\epsfxsize=3.4in
\leavevmode
\epsffile{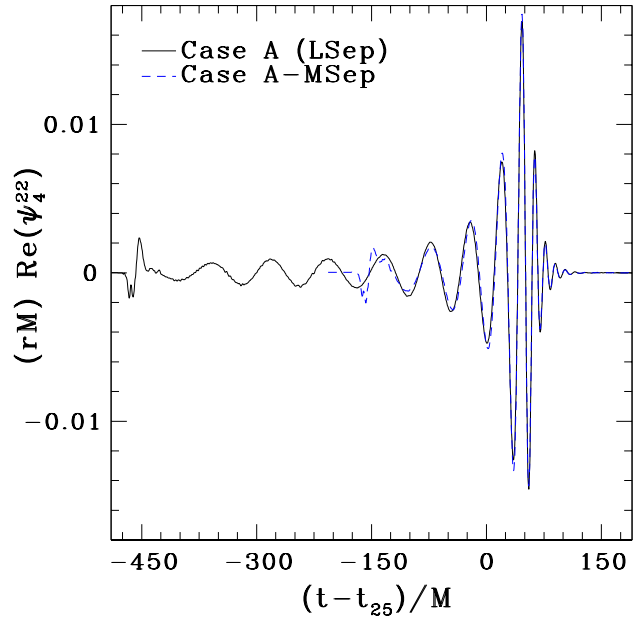}
\caption{Comparison of gravitational waveforms for Case~\A\ 
(black solid line) and \A-MSep (blue dashed line).} 
\label{fig:Psi4_comparison_caseA_sep_study}
\end{figure}

Figure~\ref{fig:BH_sepstudyA_accretion} shows the rest mass of NS
matter outside the AH as a function of time for
Cases~\A, \A-MSep, and \A-SSep. These cases correspond to 
initial binaries along the same
quasiequilibrium sequence but with different initial separations. The
time origin is shifted by $t_{25}$, where $t_N$ denotes the time at
which $N\%$ of the NS rest mass has fallen into the AH.  We see that
the results in Cases~\A\ and \A-MSep agree very well, but the result
for \A-SSep deviates from the other
two. Figure~\ref{fig:Psi4_comparison_caseA_sep_study} shows the
gravitational waveforms Re($\psi_4^{22}$) emitted in Cases~\A\ and
\A-MSep.  We again see that the waveforms agree very well after a
simple time and phase shift.  These results indicate that initial
separations corresponding to those in Cases~\A-MSep and \A\ are
sufficiently large.  For almost all other simulations, the initial
value of $[M/(q+1)] \Omega$ is similar to or smaller than that for
Case~\A\ ($[M/(q+1)] \Omega \lesssim 0.0083$), suggesting that 
the initial binary separation should be sufficiently large in 
all of these cases.
The only exception is Case~\G, for which $[M/(q+1)] \Omega\approx 0.017$ 
initially.  This value should be compared with that of
Case \A-SSep, which had $[M/(q+1)] \Omega\approx0.016$ initially.
Therefore, our results for Case~\G\ may be inaccurate, since the
initial separation may be too small.  From an astrophysical perspective, 
our Case~\G\ may  also be the least relevant, since 
the formation of a BH with a NS of equal mass seems very unlikely.

\subsection{Effects of pressure ceiling}
\label{sec:Pceiling}

Figure~\ref{fig:BH_sepstudyA_accretion} shows that a disk of rest mass
$0.04M_0$ remains outside the AH at the end of simulation, where $M_0$
is the rest mass of the initial NS. This finding disagrees with the
result reported in Paper~I. The reason for this discrepancy is due to
artificially imposed limits on pressure $P$ that were applied
everywhere in the fluid in Paper~I.  Here, we apply these limits only
in regions where the density is tiny (i.e.\ inside the artificial
``atmosphere'') or where the conformal factor $\psi$ is large (i.e.,
deep inside the AH), as discussed in
Sec.~\ref{sec:num_metric_hydro}.  Setting the pressure ceiling
everywhere has little effect on the inspiral dynamics.  However, when
the NS is tidally disrupted and forms a relatively low-density
``tail'', shock heating becomes a significant effect,  
causing $P$ to exceed $P_{\rm max}$
(see Appendix~\ref{app:shock-heating}).
In our original formulation, we reset $P$ to $P_{\rm max}$.
Artificially lowering the pressure in this way forces the low-density
material to lose a significant amount of torque and angular momentum and fall
into the BH.  Even though the density of the tail is low, it is still
many orders of magnitude larger than the artificial atmospheric
density, and hence has substantial influence on the dynamics of the
merging NS.

To more accurately model the physics of the merger phase and maintain
a stable evolution inside the BH, we now impose the limits on $P$ only
in very low-density regions (where the rest-mass density $\rho_0<
100\rho_{\rm atm}$) or when $\psi^6>\psi^6_{\rm max}$.  We set
$\psi^6_{\rm max}=10^5$ in most simulations, except Case~\G\ where we
need to set $\psi^6_{\rm max}=50$ to stabilize the evolution. We note
that even for this relatively small $\psi^6_{\rm max}$,
$\psi^6>\psi^6_{\rm max}$ holds true only deep inside the AH,
so this readjustment should not affect the dynamics outside the BH.  By relaxing
this artificial pressure ceiling, we now observe a small remnant disk
in many cases and achieve significantly better angular momentum
conservation. For example, for Case~\A\ we previously had 
$\delta J \approx 13\%$ but now 
$\delta J \approx 2\%$ after relaxing the pressure ceiling.

\subsection{Effects of black-hole spin}
\label{sec:bhspin}

\begin{figure*}
\vspace{-4mm}
\begin{center}
\epsfxsize=2.15in
\leavevmode
\epsffile{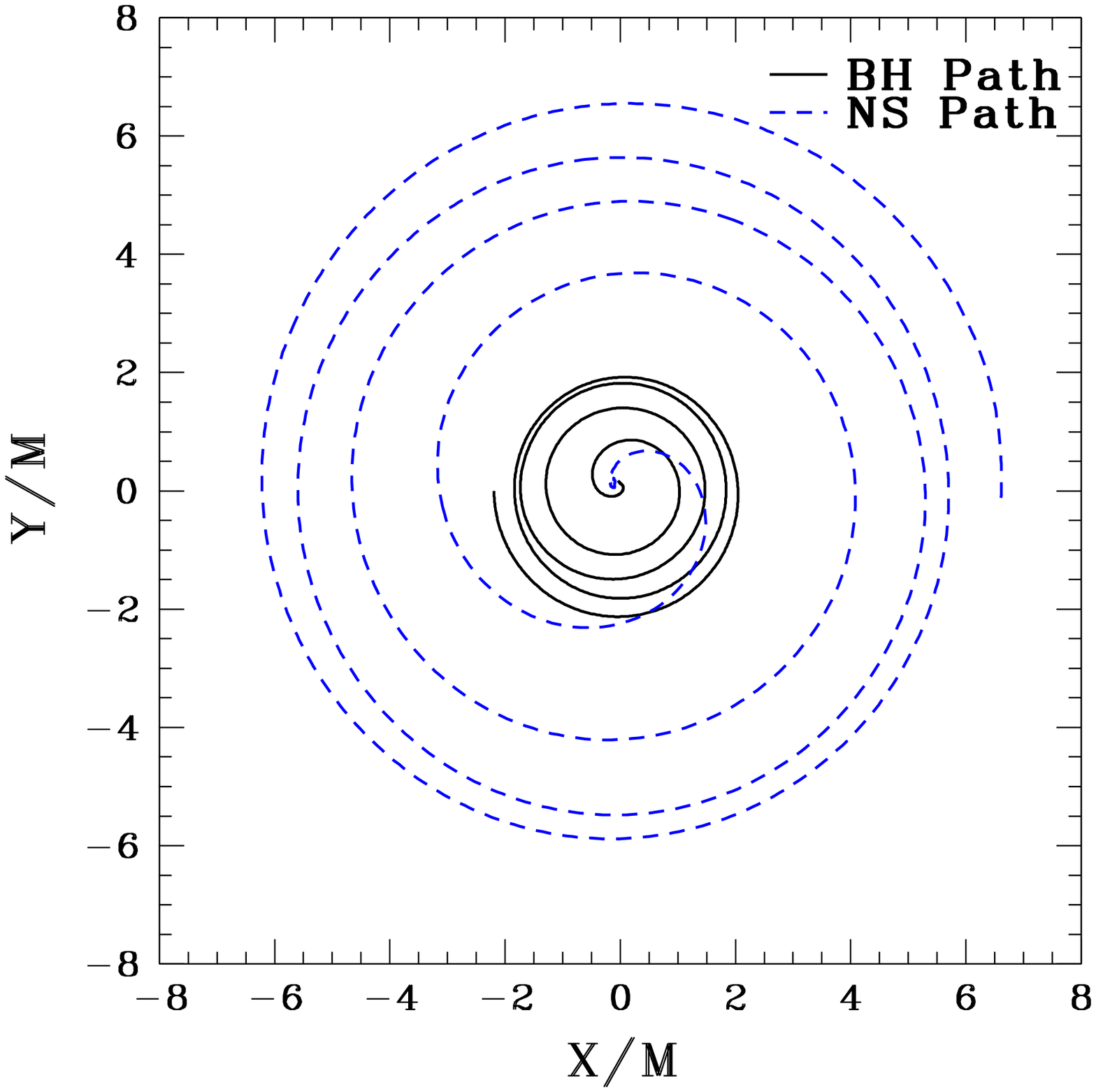}
\epsfxsize=2.15in
\leavevmode
\epsffile{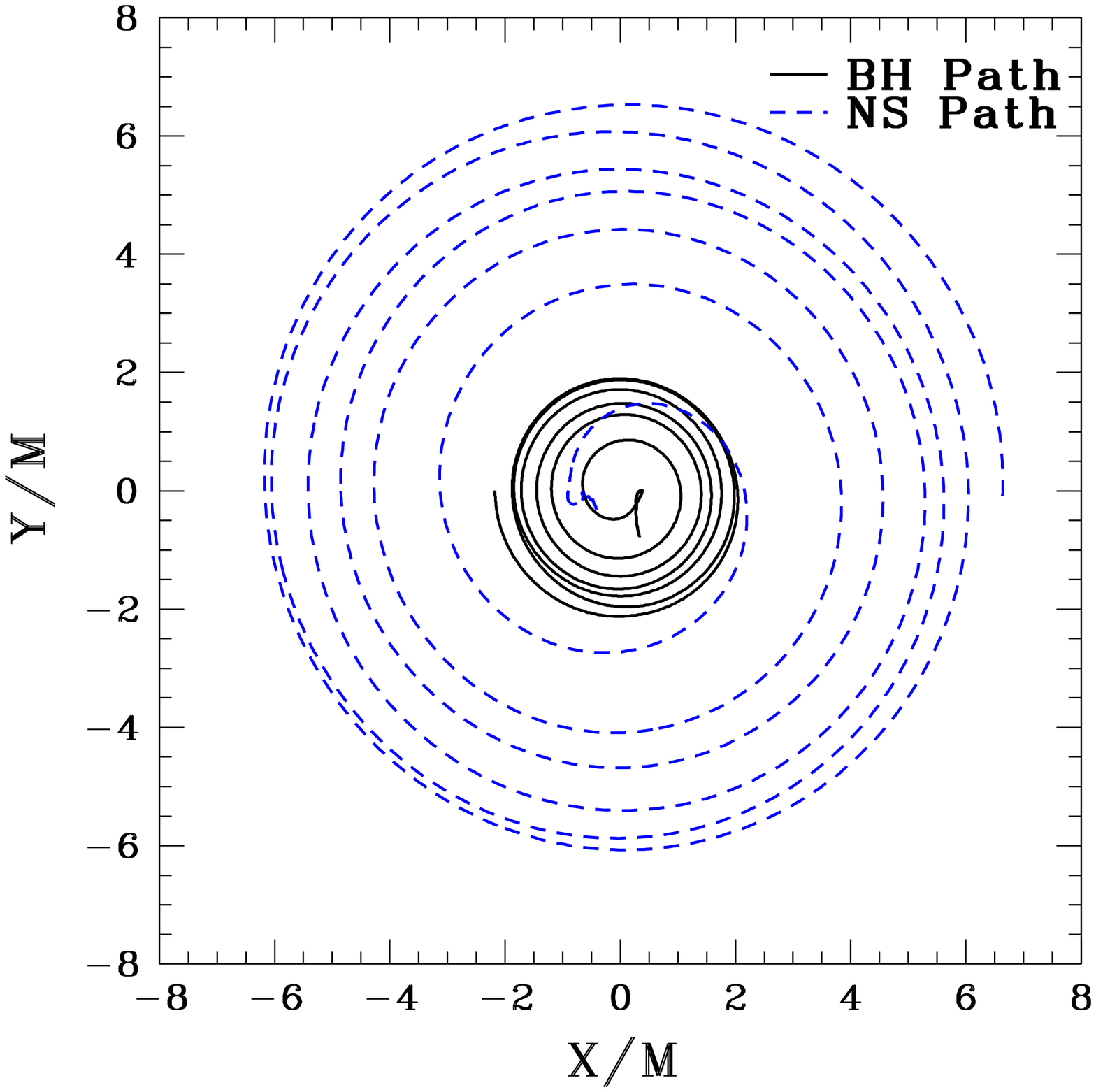}
\caption{Trajectories of the BH and NS coordinate centroids for
  Cases~\A\ (left) and \B\ (right).  The BH coordinate centroid
  corresponds to the centroid of the AH, and the NS coordinate
  centroid is computed via: $\tilde{X}^i_c \equiv (\int_V x^i \rho_*
  d^3x)/(\int_V \rho_* d^3x)$, where $V$ is the {\it total} simulation
  volume.  Note that this equation contains a different $V$ than
  Eq.~(\ref{ns_refboxcentroid}).}
\label{fig:casesAandB_traj}
\end{center}
\end{figure*}

\begin{figure*}
\vspace{-4mm}
\begin{center}
\epsfxsize=2.15in
\leavevmode
\epsffile{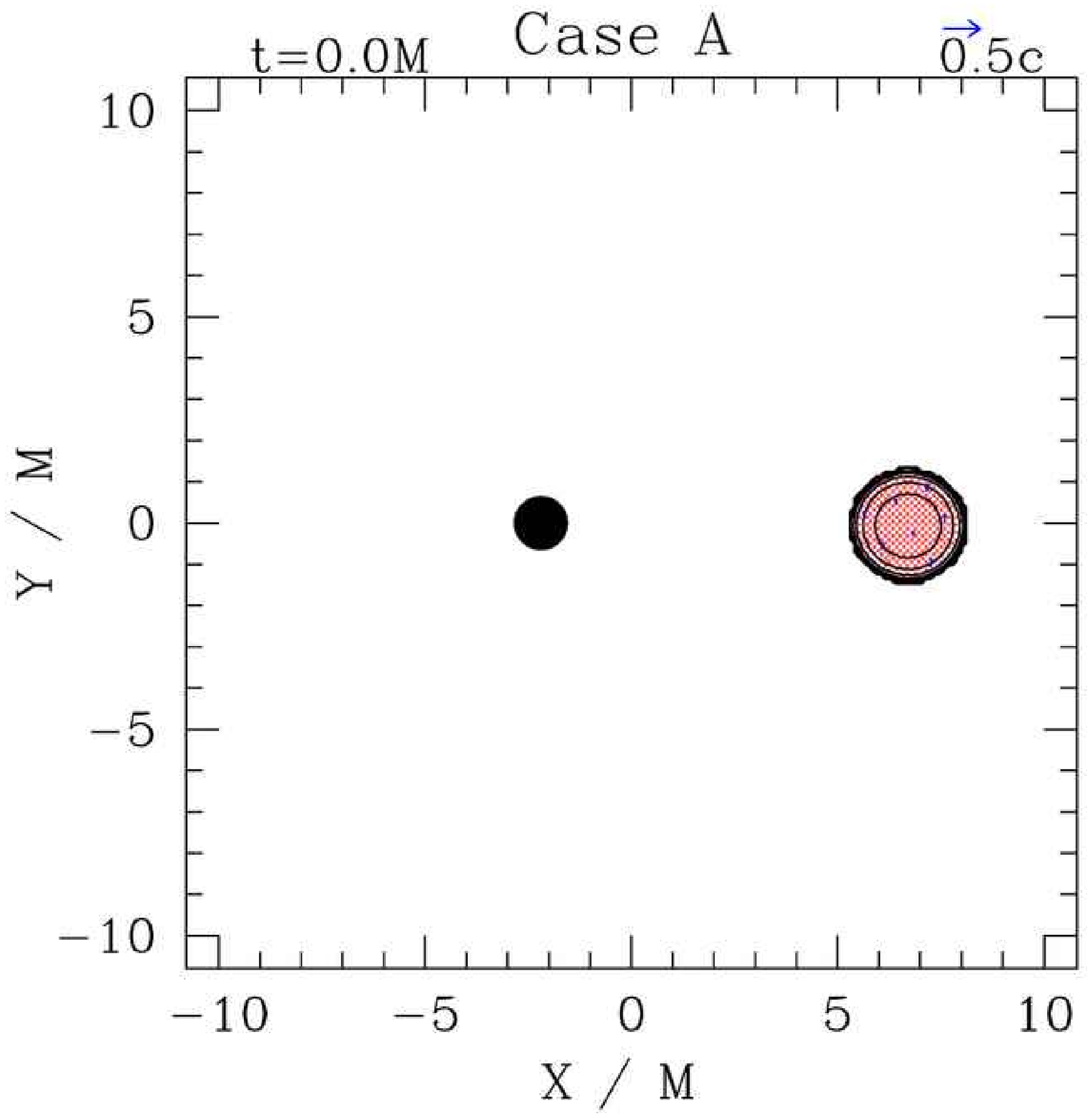}
\epsfxsize=2.15in
\leavevmode
\epsffile{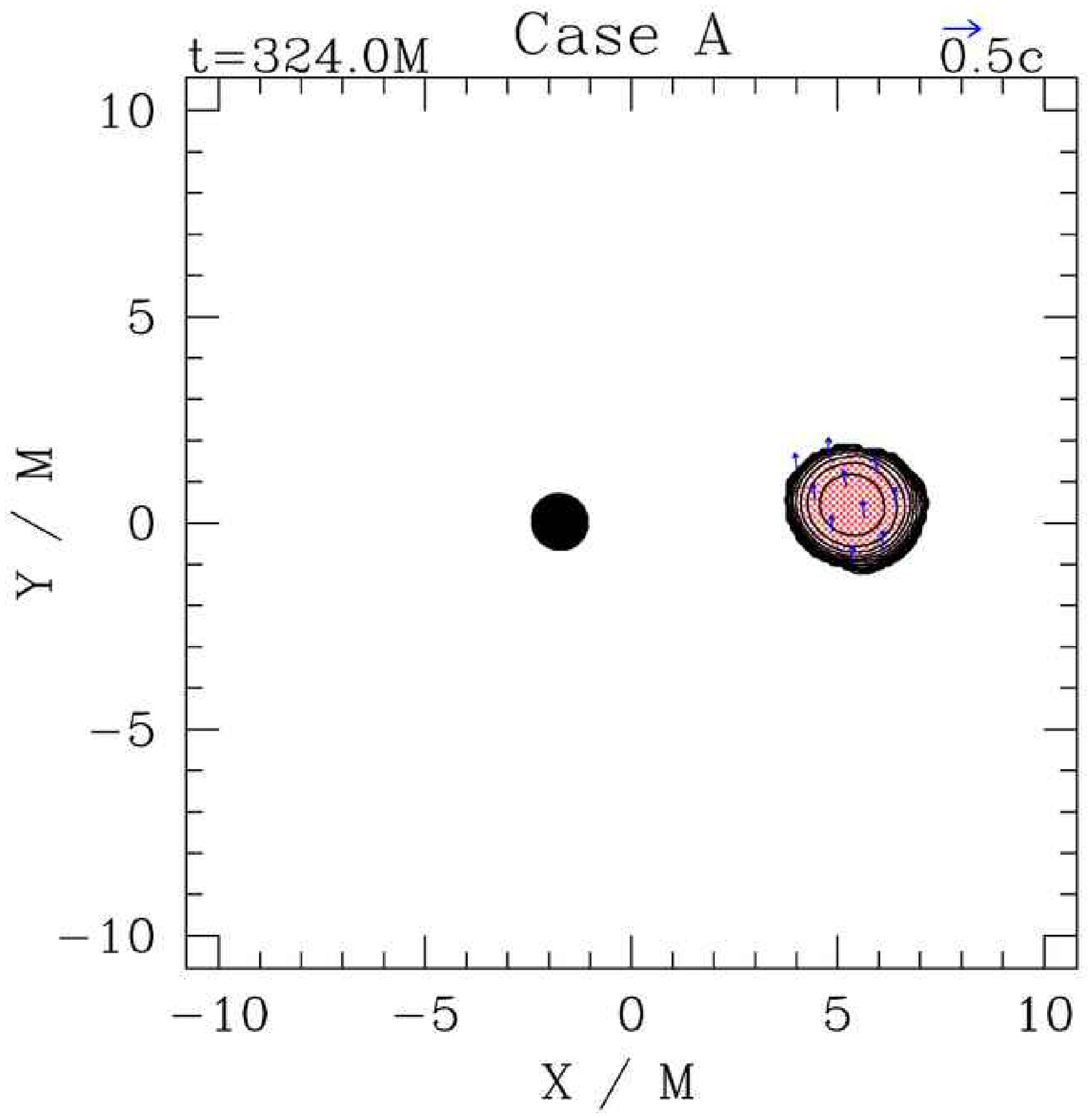}
\epsfxsize=2.15in
\leavevmode
\epsffile{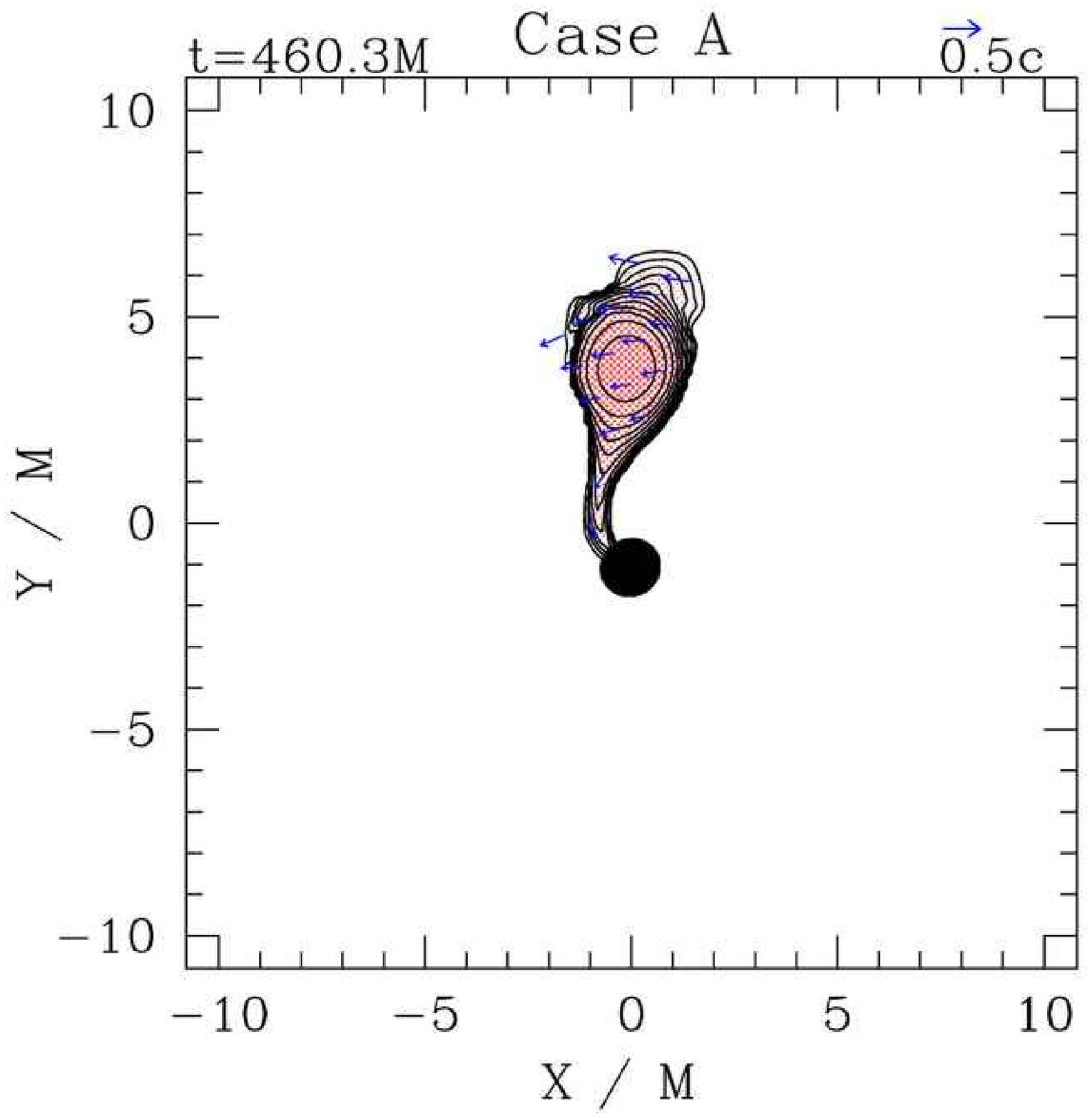}\\
\epsfxsize=2.15in
\leavevmode
\epsffile{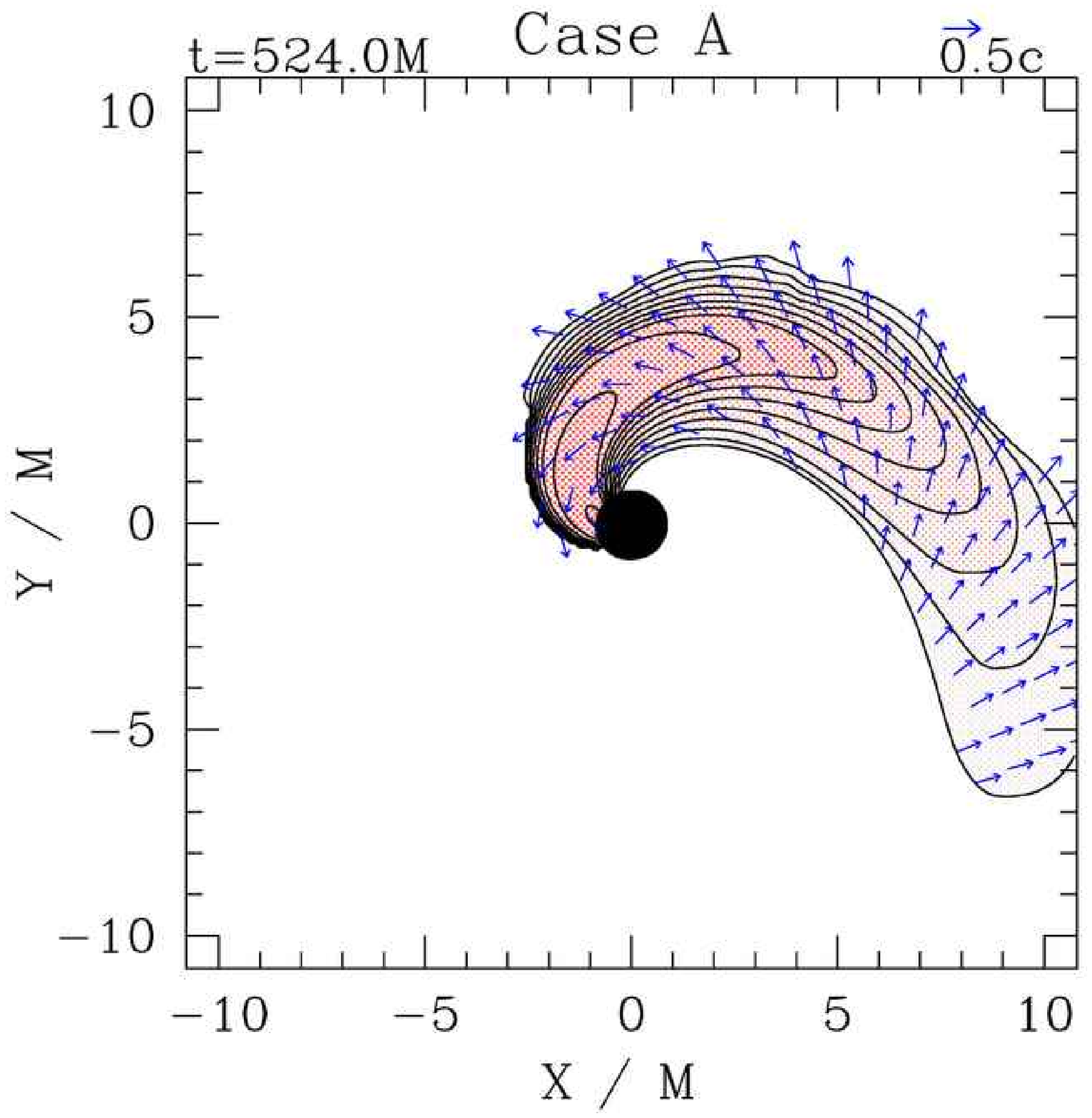}
\epsfxsize=2.15in
\leavevmode
\epsffile{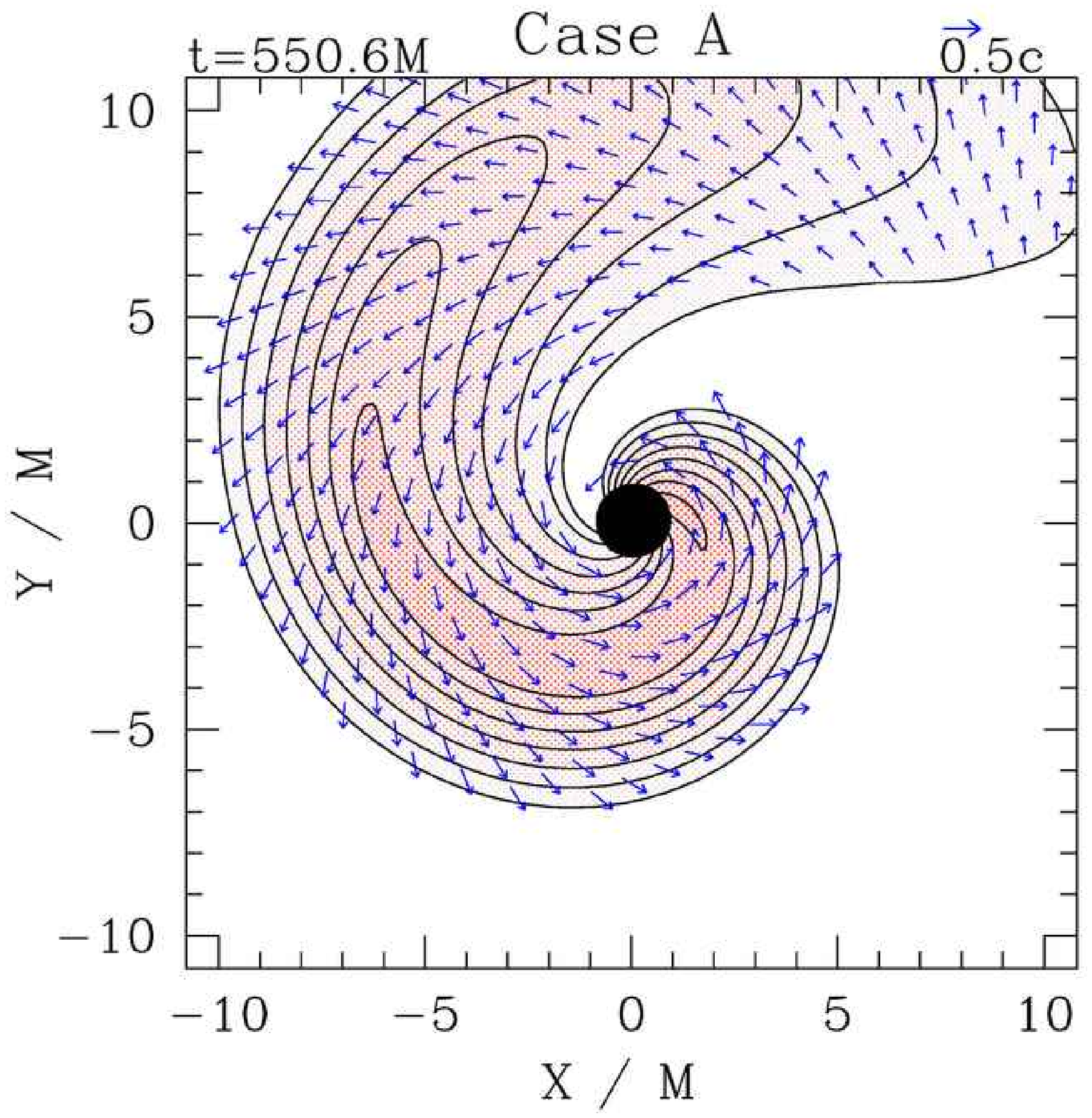}
\epsfxsize=2.15in
\leavevmode
\epsffile{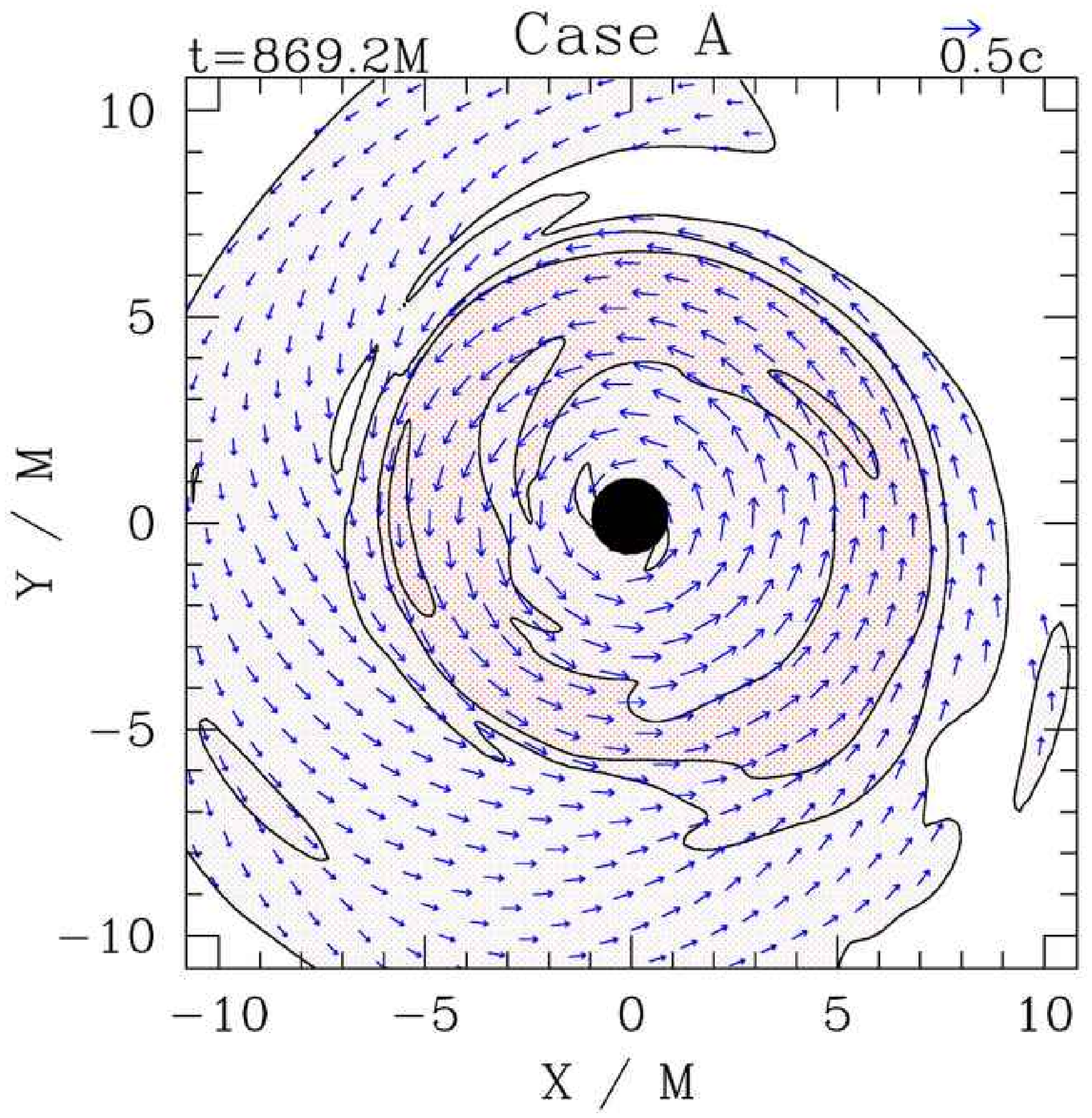}
\caption{Snapshots of density and velocity profiles at selected times
  for Case~\A. The contours represent the density in the orbital plane,
  plotted according to $\rho_0 = \rho_{0,{\rm max}}10^{-0.38j-0.04}$, 
  ($j$=0, 1, ... 12), with darker greyscaling for higher density.  
  The maximum initial NS density is $\kappa \rho_{0,{\rm max}} = 0.126$, 
  or $\rho_{0,{\rm max}}=9\times 10^{14}\mbox{g cm}^{-3}(1.4M_\odot/M_0)^2$.  
  Arrows represent the velocity field
  in the orbital plane.  The black hole AH interior is marked 
  by a filled black circle.  In cgs units, the initial 
  ADM mass for this case is $M=2.5\times 10^{-5}(M_0/1.4M_\odot)$
  s$=7.6(M_0/1.4M_\odot)$km.}
\label{fig:A_snapshots}
\end{center}
\end{figure*}

\begin{figure*}
\vspace{-4mm}
\begin{center}
\epsfxsize=2.15in
\leavevmode
\epsffile{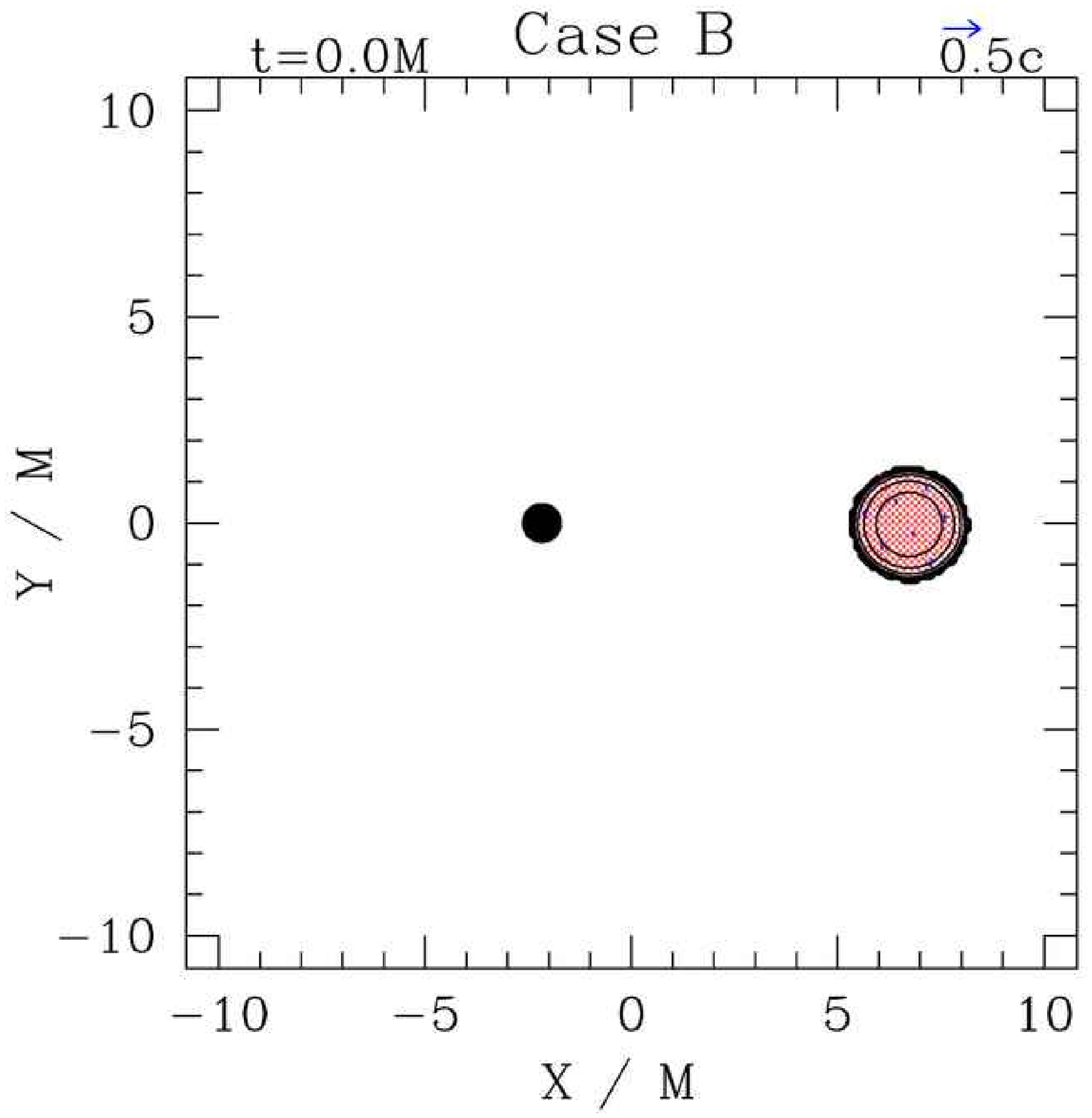}
\epsfxsize=2.15in
\leavevmode
\epsffile{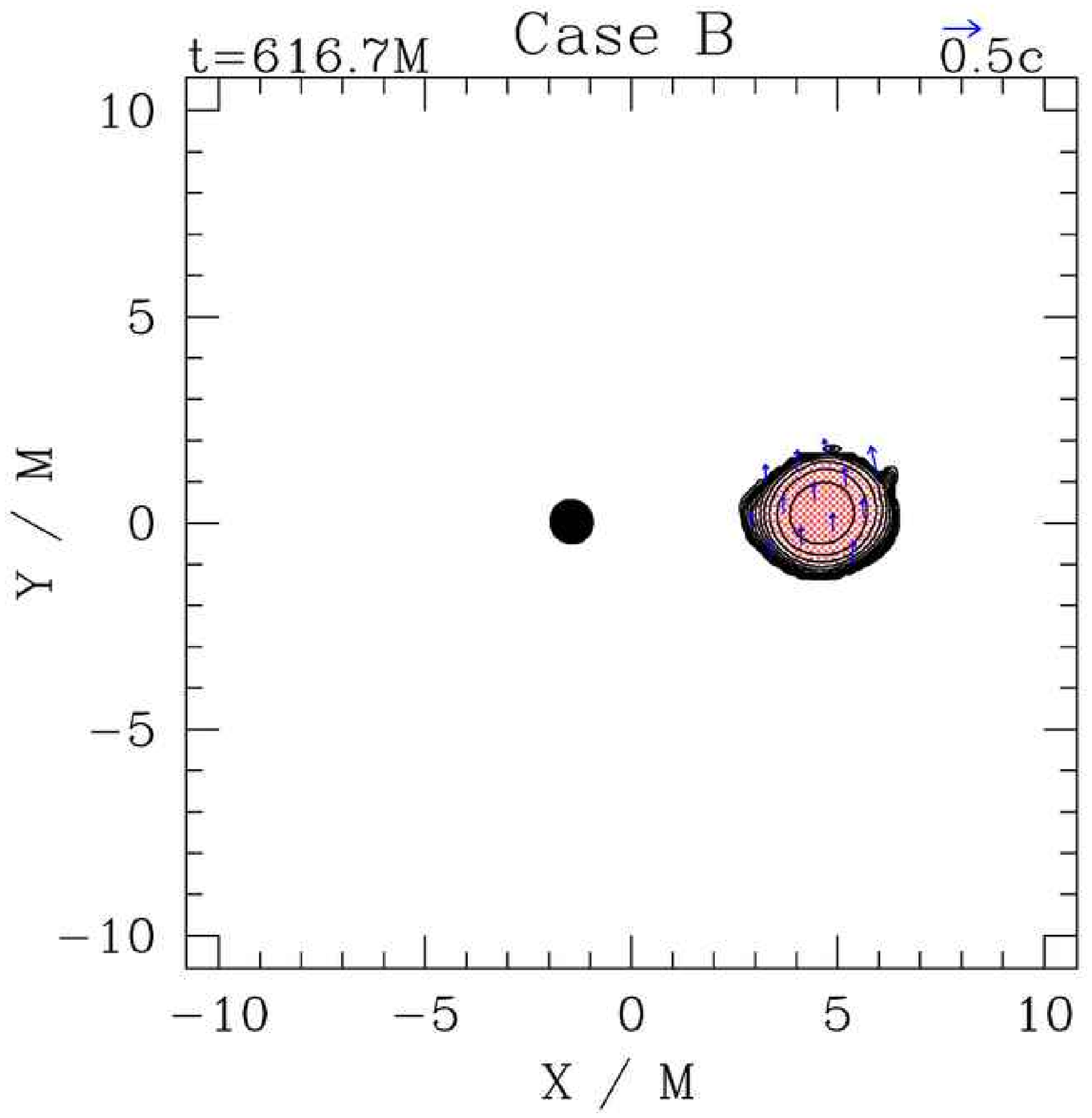}
\epsfxsize=2.15in
\leavevmode
\epsffile{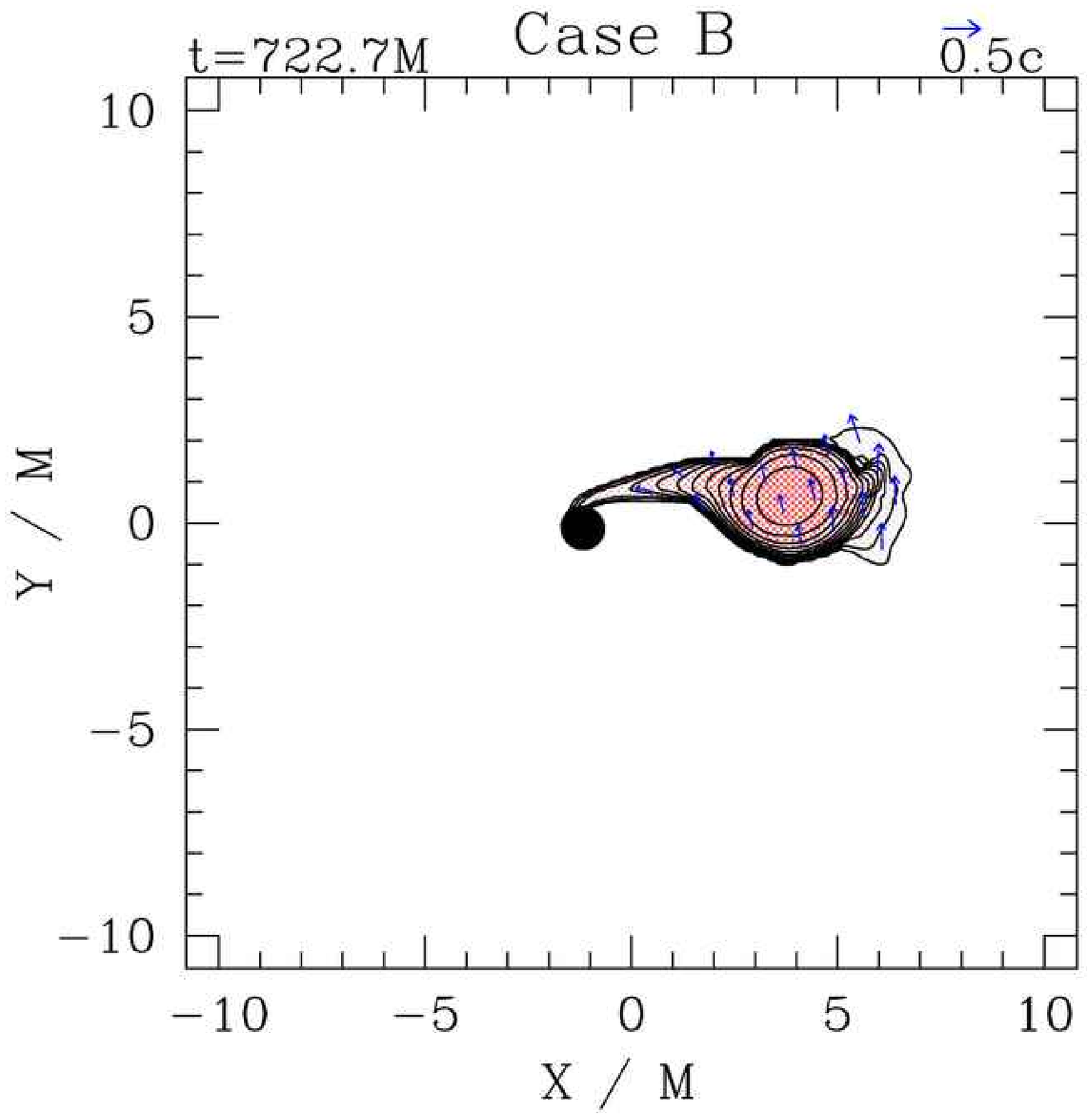}\\
\epsfxsize=2.15in
\leavevmode
\epsffile{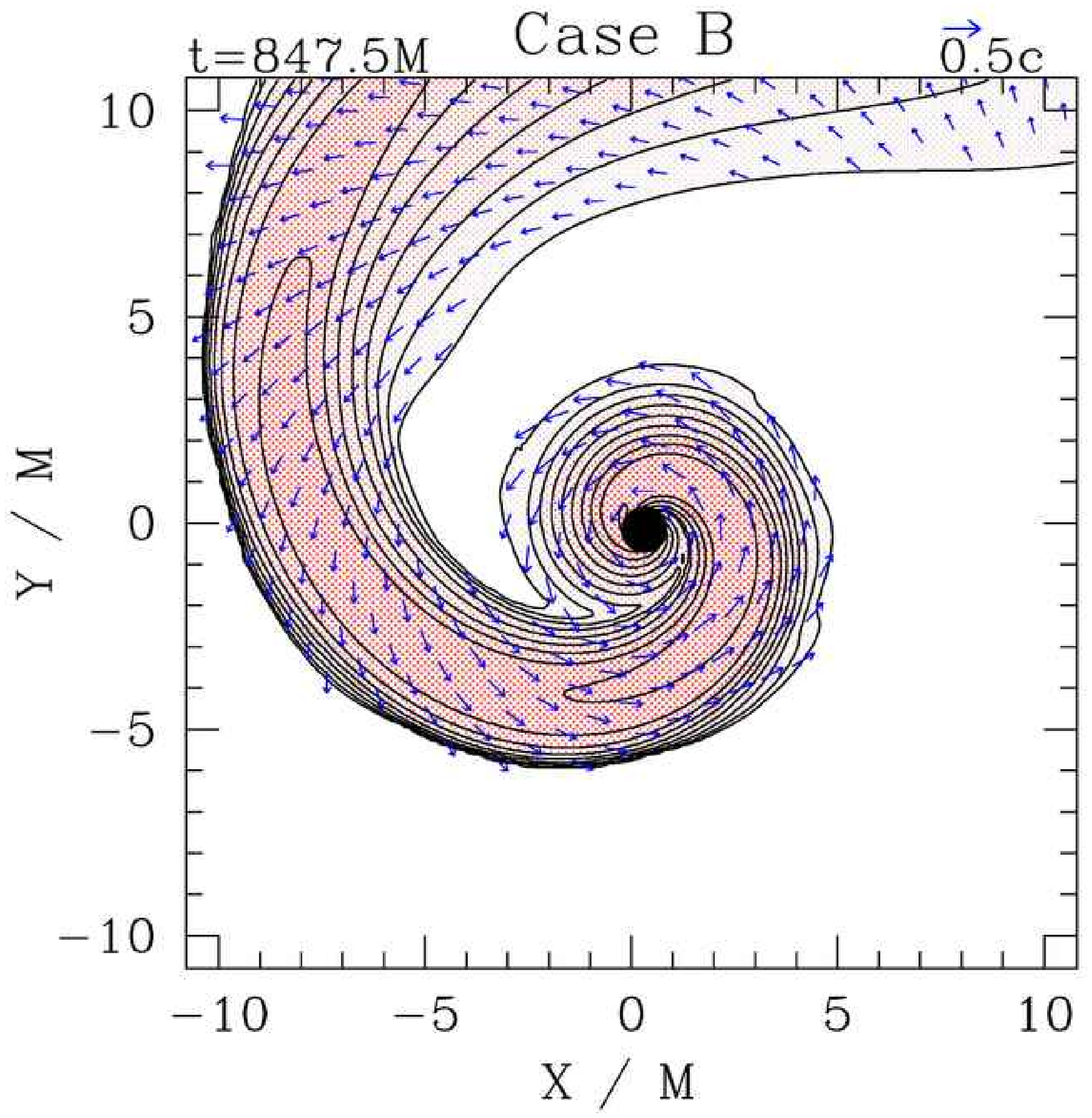}
\epsfxsize=2.15in
\leavevmode
\epsffile{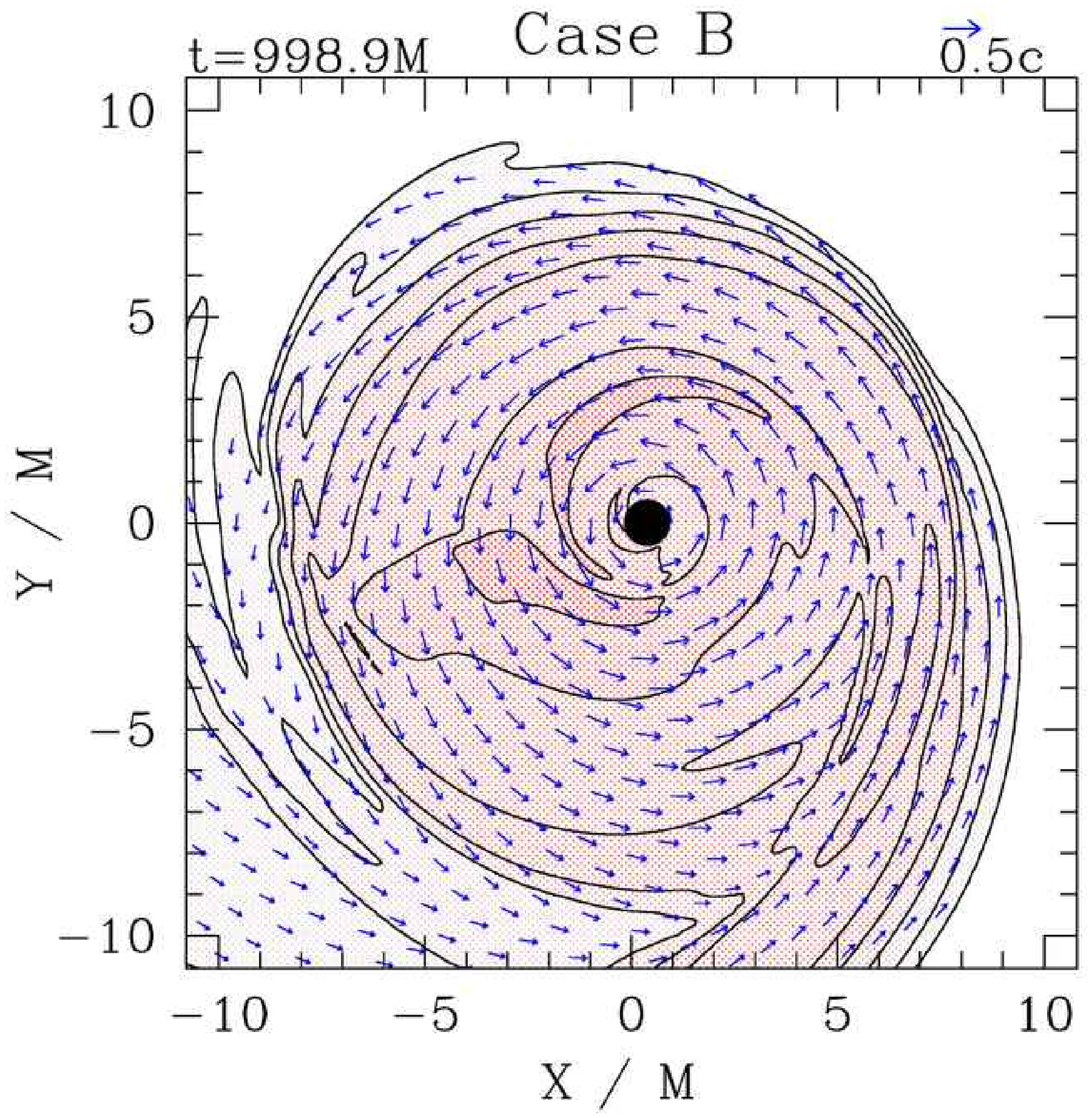}
\epsfxsize=2.15in
\leavevmode
\epsffile{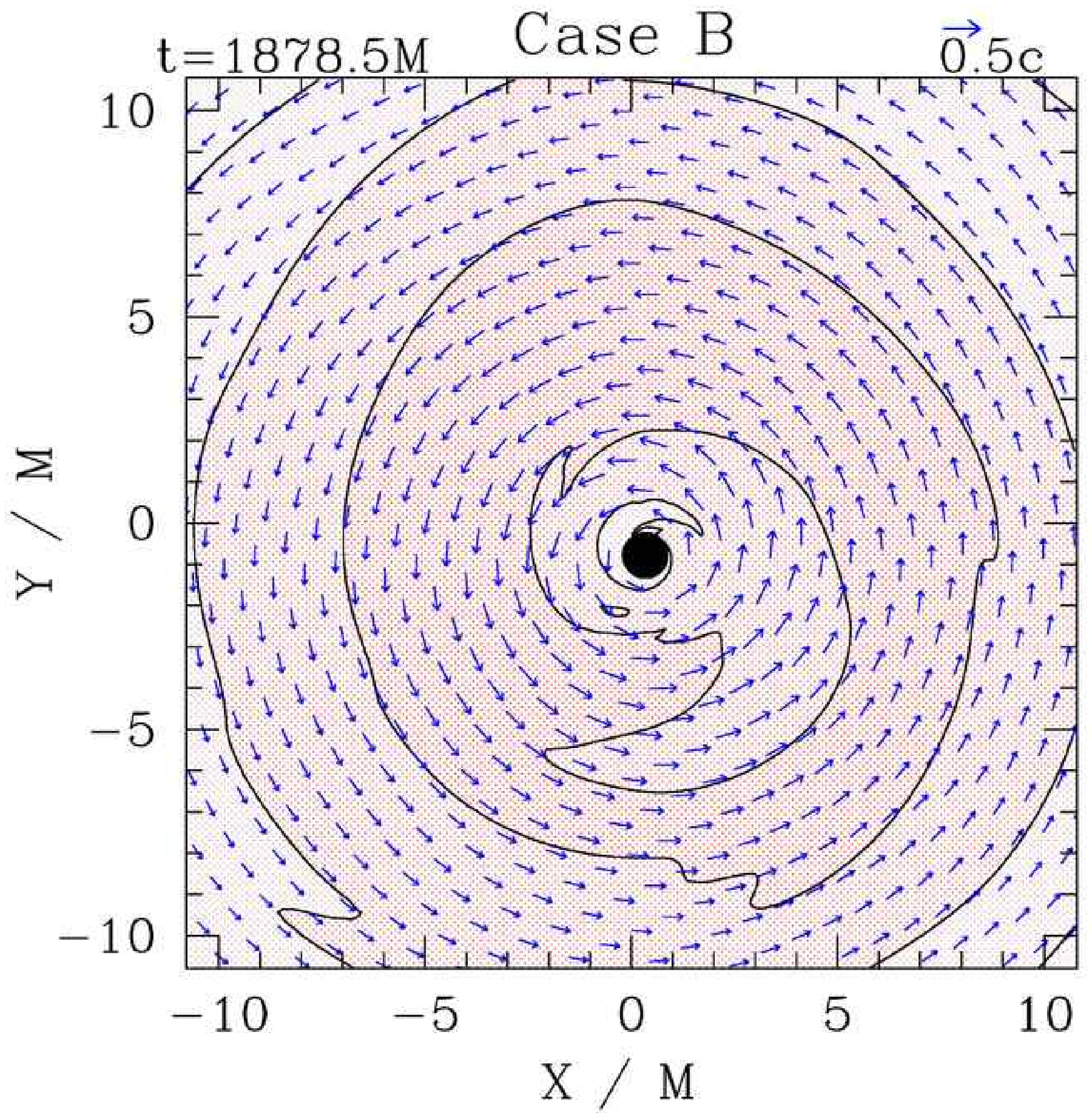}
\caption{Snapshots of density and velocity profiles at selected times
  for Case~\B. The contours represent the density in the orbital plane,
  plotted according to $\rho_0 = \rho_{0,{\rm max}} 10^{-0.38j-0.04}$ 
  ($j$=0, 1, ... 12), with darker
  greyscaling for higher density. 
  The maximum initial NS density is $\kappa \rho_{0,{\rm max}} = 0.126$, or 
  $\rho_{0,{\rm max}}=9\times 10^{14}\mbox{g cm}^{-3}(1.4M_\odot/M_0)^2$.
  Arrows represent the velocity field
  in the orbital plane.  We specify the black hole AH interior in
  each snapshot with a filled black circle.  In cgs units, the initial
  ADM mass for this case is $M= 2.5\times 10^{-5}(M_0/1.4M_\odot)$s$
  =7.6(M_0/1.4M_\odot)$km.}
\label{fig:B_snapshots}
\end{center}
\end{figure*}

\begin{figure*}
\vspace{-4mm}
\begin{center}
\epsfxsize=2.15in
\leavevmode
\epsffile{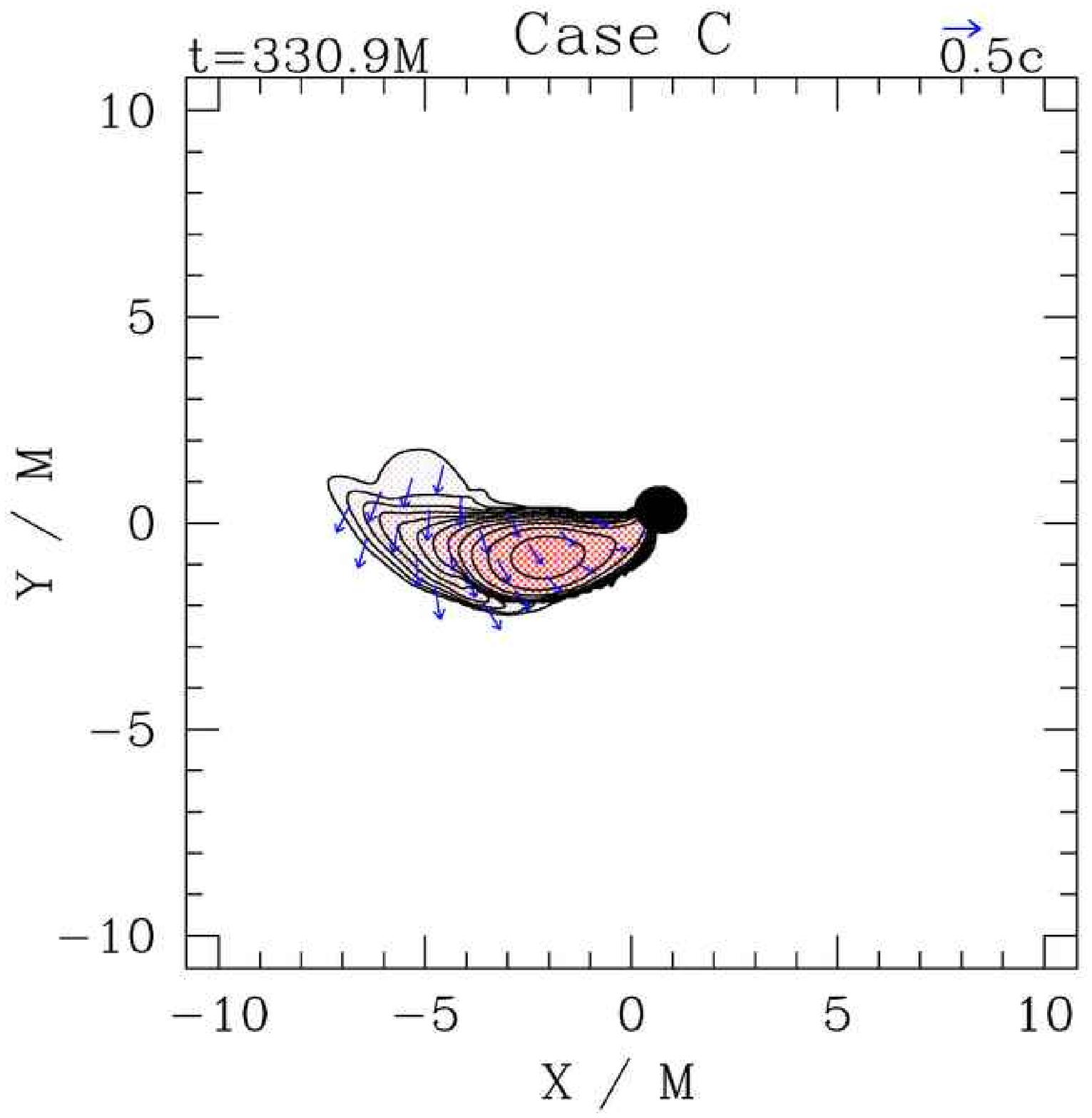}
\epsfxsize=2.15in
\leavevmode
\epsffile{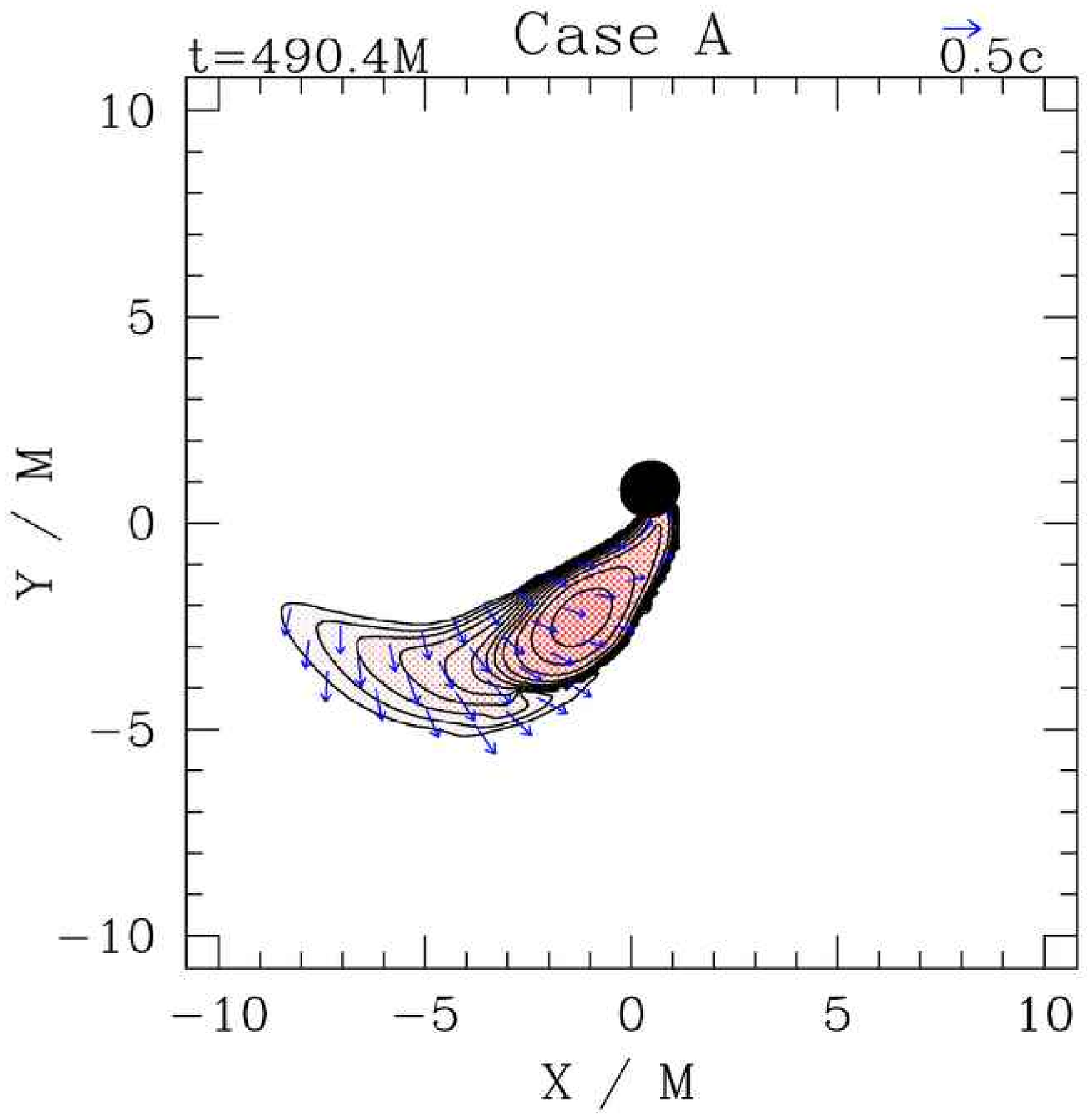}
\epsfxsize=2.15in
\leavevmode
\epsffile{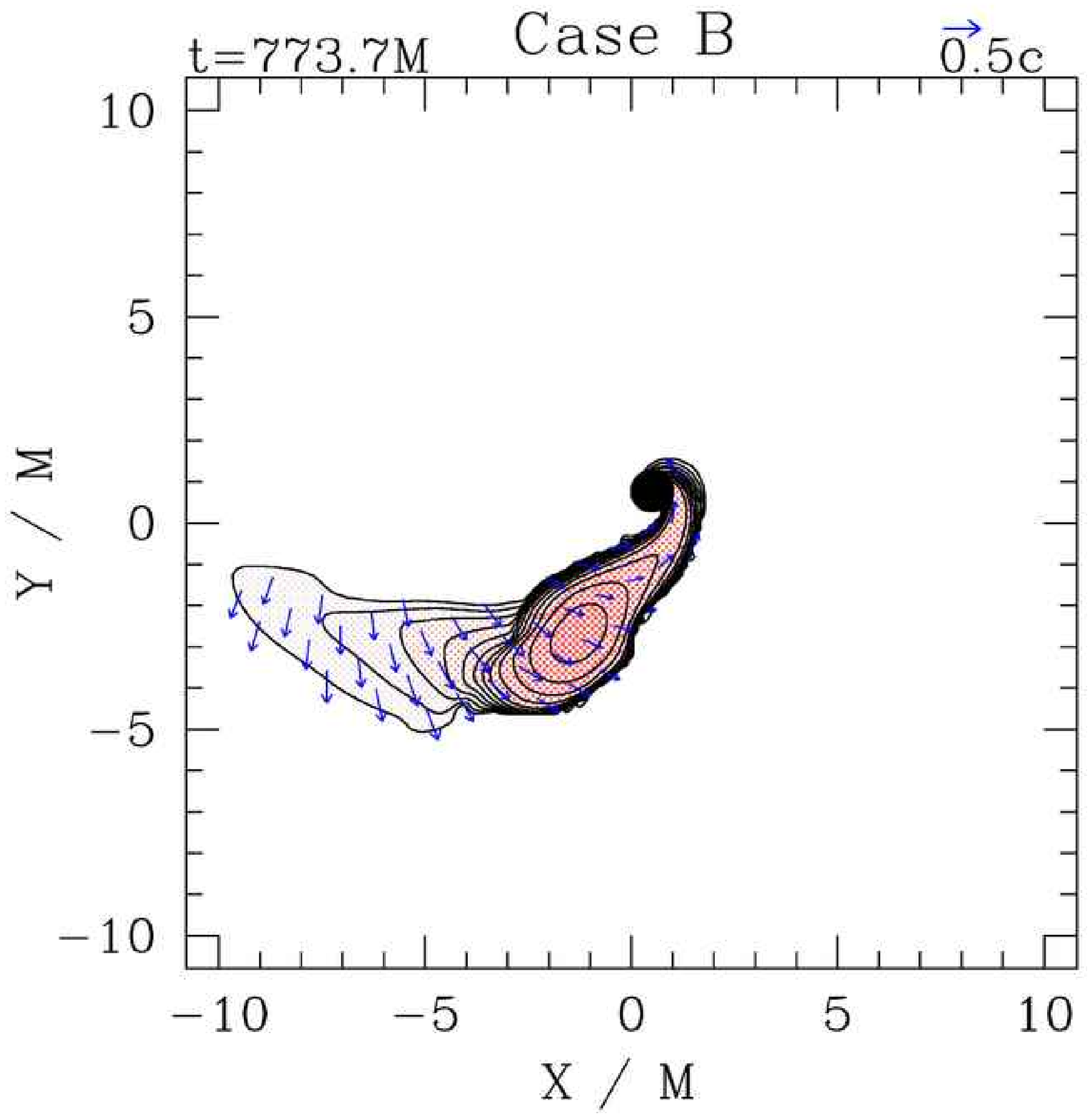} \\
\epsfxsize=2.15in
\leavevmode
\epsffile{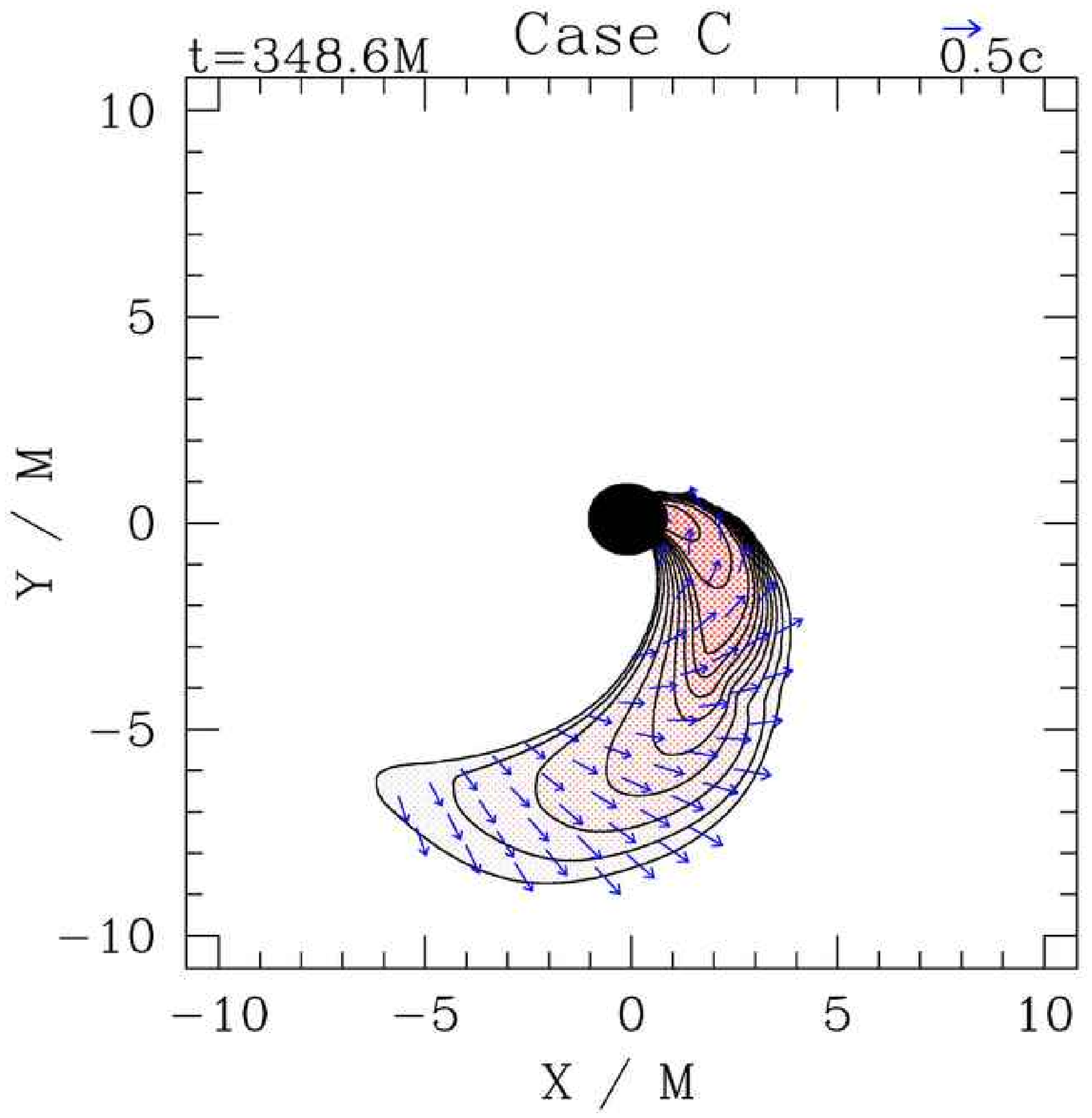}
\epsfxsize=2.15in
\leavevmode
\epsffile{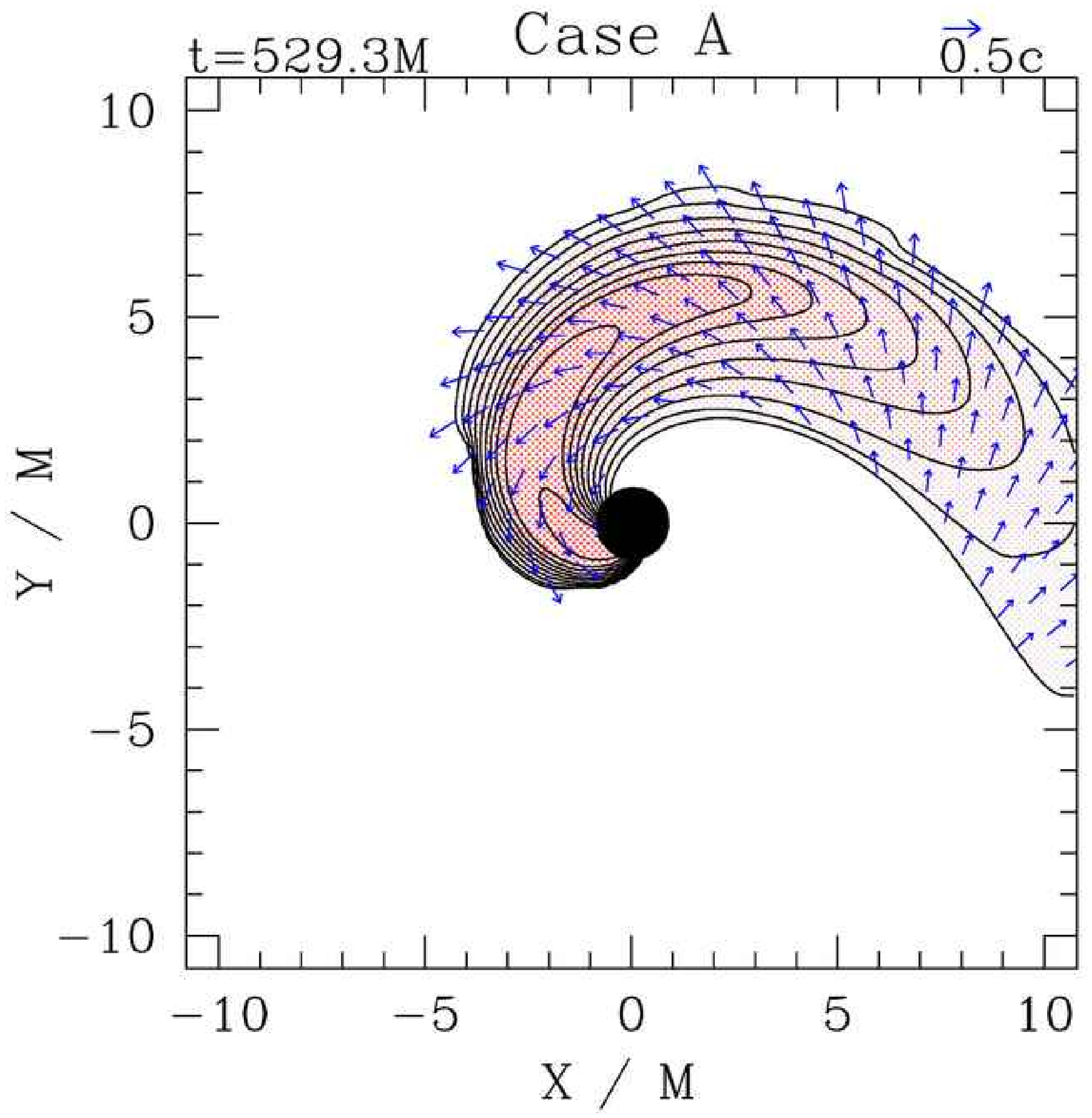}
\epsfxsize=2.15in
\leavevmode
\epsffile{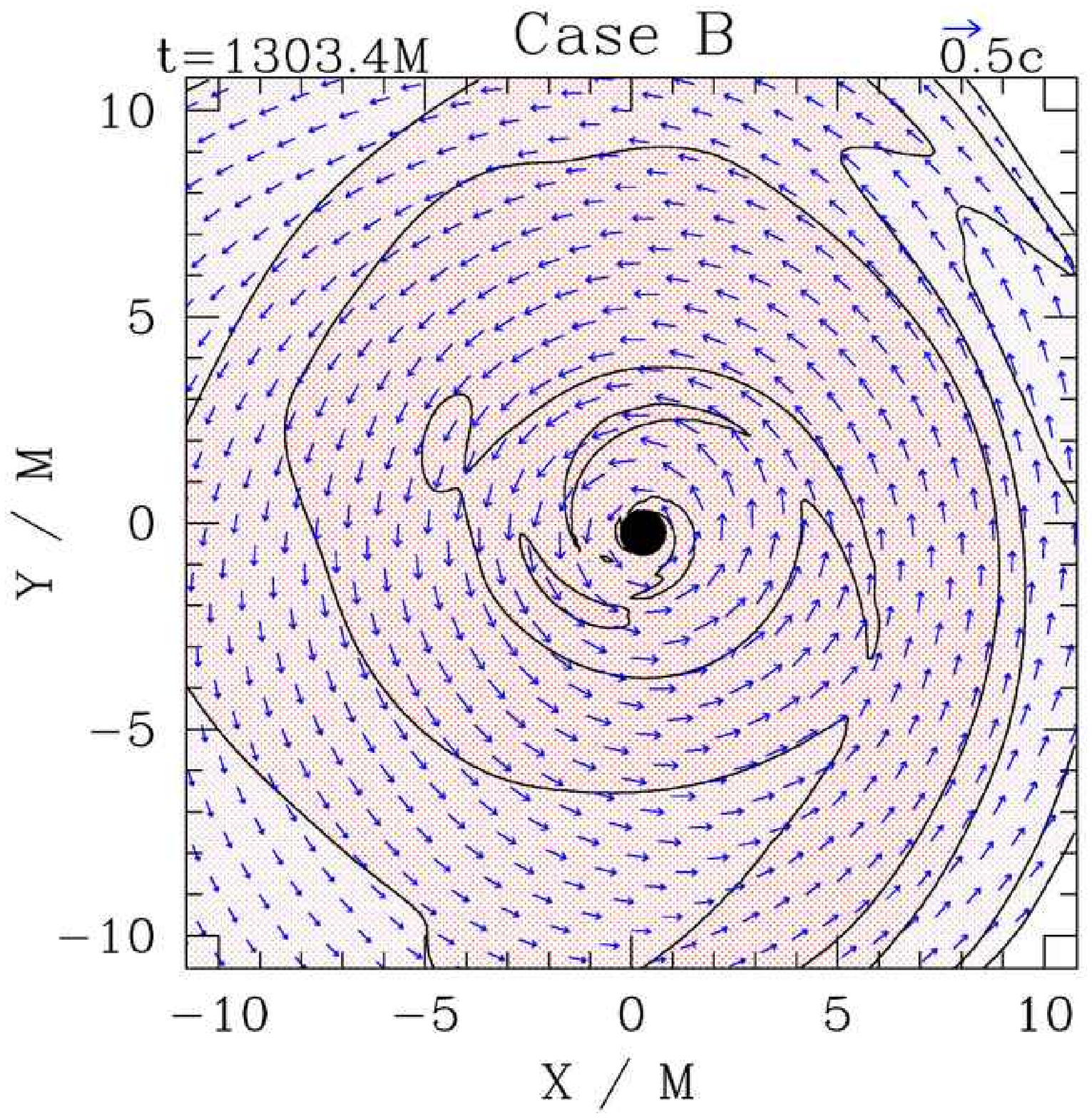}
\caption{Snapshots of density and velocity profiles at
  $t_1$ (top row) and $t_{80}$ (bottom row) for 
  Cases~\D, \A, and \B, which have initial BH spins of $\tilde{a}=-0.5$,
  $0.0$, and $0.75$, respectively.  The
  contours represent the density in the orbital plane, plotted
  according to $\rho_0 = \rho_{0,{\rm max}} 10^{-0.38j-0.04}$ 
  ($j$=0, 1, ... 12), with darker
  greyscaling for higher density.  
  The maximum initial NS density is $\kappa \rho_{0,{\rm max}} = 0.126$ for all
  cases, or 
  $\rho_{0,{\rm max}}=9\times 10^{14}\mbox{g cm}^{-3}(1.4M_\odot/M_0)^2$.
  Arrows represent the velocity field in the orbital plane.
  The black hole AH interior is marked by a filled
  black circle.  In cgs units, the initial ADM mass for these cases is
  $M=2.5\times 10^{-5}(M_0/1.4M_\odot)$s$=7.6(M_0/1.4M_\odot)$km.}
\label{fig:BH_spin_snapshots}
\end{center}
\end{figure*}

\begin{figure}
\epsfxsize=3.4in
\leavevmode
\epsffile{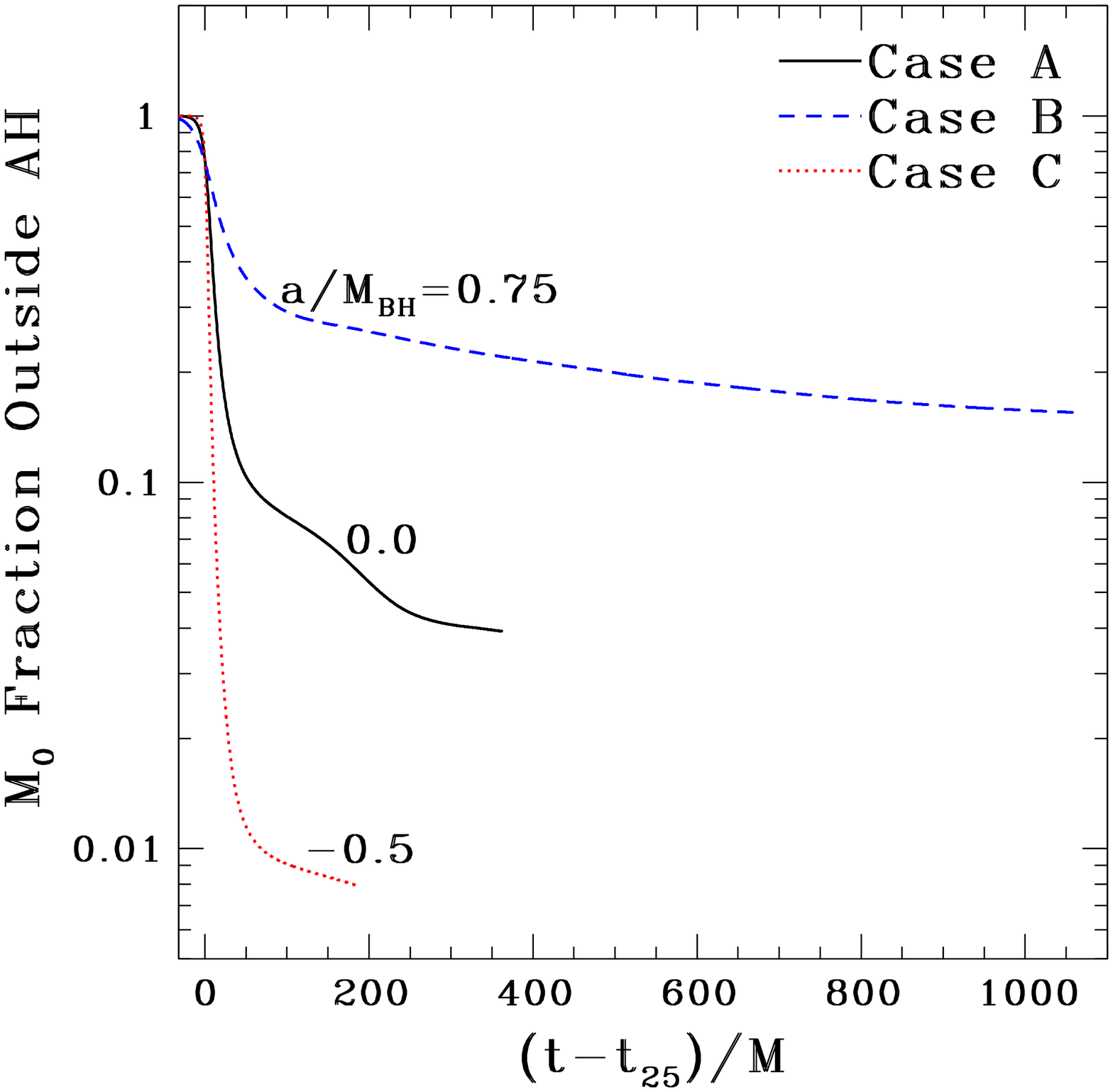}
\caption{Rest-mass fraction outside the BH for different initial BH spins
  (Cases~\D, \A, and \B).  Here, the time coordinate is shifted by
  $t_{25}$, the time at which $25\%$ of the NS rest mass has fallen
  into the apparent horizon.}
\label{fig:BH_spins_accretion}
\end{figure}

In Cases~\A--\D, we vary the initial spin parameter of the black hole
(parallel to the orbital angular momentum) from
$J_{\rm BH}/M_{\rm BH}^2$=$-0.5$ to 0.75, fixing $q=3$ and the 
initial orbital angular velocity $M\Omega \approx 0.033$.  Notice in
Table~\ref{table:initconf} that as the initial BH spin parameter
$\tilde{a}$ increases, the total initial angular momentum increases,
requiring more gravitational-wave cycles to emit angular momentum and
bring the BH and NS close enough to merge.  Thus we expect the binary
to undergo more orbits before merger as $J_{\rm BH}/M_{\rm BH}^2$
increases.  This is precisely what we find: for spins $-0.5$, 0.0, and
0.75, the binary inspiral phase lasts for 3.25, 4.5, and 6.5
orbits, respectively.  Figure~\ref{fig:casesAandB_traj} shows 
the coordinate trajectories of the AH and NS centroids for Cases~\A\ and
\B. 

Figure~\ref{fig:A_snapshots} shows snapshots of rest-mass density
contours for Case~\A.  The upper panels show the binary at the start
of the simulation, two orbits into the simulation, and 3.5
orbits into the simulation, corresponding to the time at which the
first few density contours have crossed into the AH.  Notice that the
equilibrium shape of the NS is for the most part undisturbed after
2 orbits.  The lower panels in Fig.~\ref{fig:A_snapshots} are 
snapshots taken at $t_{44}$, $t_{74}$, and $t_{96}$. 
They demonstrate how the NS tail deforms into a quasistationary disk,
as the bulk of the matter is accreted onto the hole.

Similarly, the upper plots of Fig.~\ref{fig:B_snapshots} demonstrate
that for Case~\B, the initial NS (upper left) retains its shape after 4
orbits (upper middle), and begins shedding its outer layers due to
tidal disruption at about 5 orbits (upper right).  Notice however, in
this case, the tail at $t_{60}$ (lower left) is quite massive,
and at $t_{74}$ (lower center) it is much larger than the tail at
$t_{74}$ in Case~\A\ (lower left plot in Fig.~\ref{fig:A_snapshots}).
The lower right plot is a snapshot taken near the end of the
simulation, when a quasistationary disk resides outside the BH, at
$t=t_{85}=1878.5M$.

To compare the effects of BH spin for all cases in this study, we
plot the density contours at $t_{1}$ and $t_{80}$ for Cases~\A--\D\ in
Fig.~\ref{fig:BH_spin_snapshots}.  In the $\tilde{a}=-0.5$ case
(Case~\D), the NS is basically ``swallowed whole'' by the BH during
merger, leaving $<1\%$ of the NS rest mass as a disk.  As the spin
increases, NS tidal disruption becomes more pronounced, resulting in
long tidal tails that eventually form disks
with rest mass $\approx 4\%$ and $\approx$ 15\% the 
rest mass of the NS in Cases~\A\ and \B, respectively 
(see Fig.~\ref{fig:BH_spins_accretion}). This result is not 
surprising. The ISCO decreases as the BH spin increases, and 
hence the tidal disruption of the NS occurs farther from the 
ISCO as the BH spin increases. Thus BHs with higher spin 
would likely lead to a more massive disk. 

Figure~\ref{fig:BH_spins_accretion} shows the rest mass outside 
the AH as a function of time. We see that there 
are two phases of matter falling into the BH: a plunge phase 
and an accretion phase. The plunge phase occurs early in the 
merger as part of the NS matter streams onto the BH and the rest
deforms into a tidal tail. The plunge phase ends when 70\%--90\% of the
NS matter falls into the BH, resulting in a sudden drop in the slope of 
the exterior $M_0$ vs.~time plot in Fig.~\ref{fig:BH_spins_accretion}. 
The matter in the tail, having larger specific angular 
momentum, spreads out and forms a disk. Material with lower specific 
angular momentum in the disk accretes onto the BH. Since there is 
neither viscosity nor magnetic fields in our simulation, the accretion 
should eventually cease as the evolution continues.  However, in
realistic astrophysical environments, magnetic fields do exist and
could have substantial influence on the dynamics of the disk on
secular timescales. In addition, some material in the disk is shock
heated to high temperature (see Sec.~\ref{sec:disk}).  Copious amounts
of neutrinos and antineutrinos may be generated as a result.  Both
magnetic fields and neutrino-antineutrino pairs could drive energetic
jets, resulting in an SGRB (see Sec.~\ref{sec:disk} for further
discussion). 

\subsection{Effects of varying the mass ratio}
\label{sec:q}

\begin{figure}
\epsfxsize=3.4in
\leavevmode
\epsffile{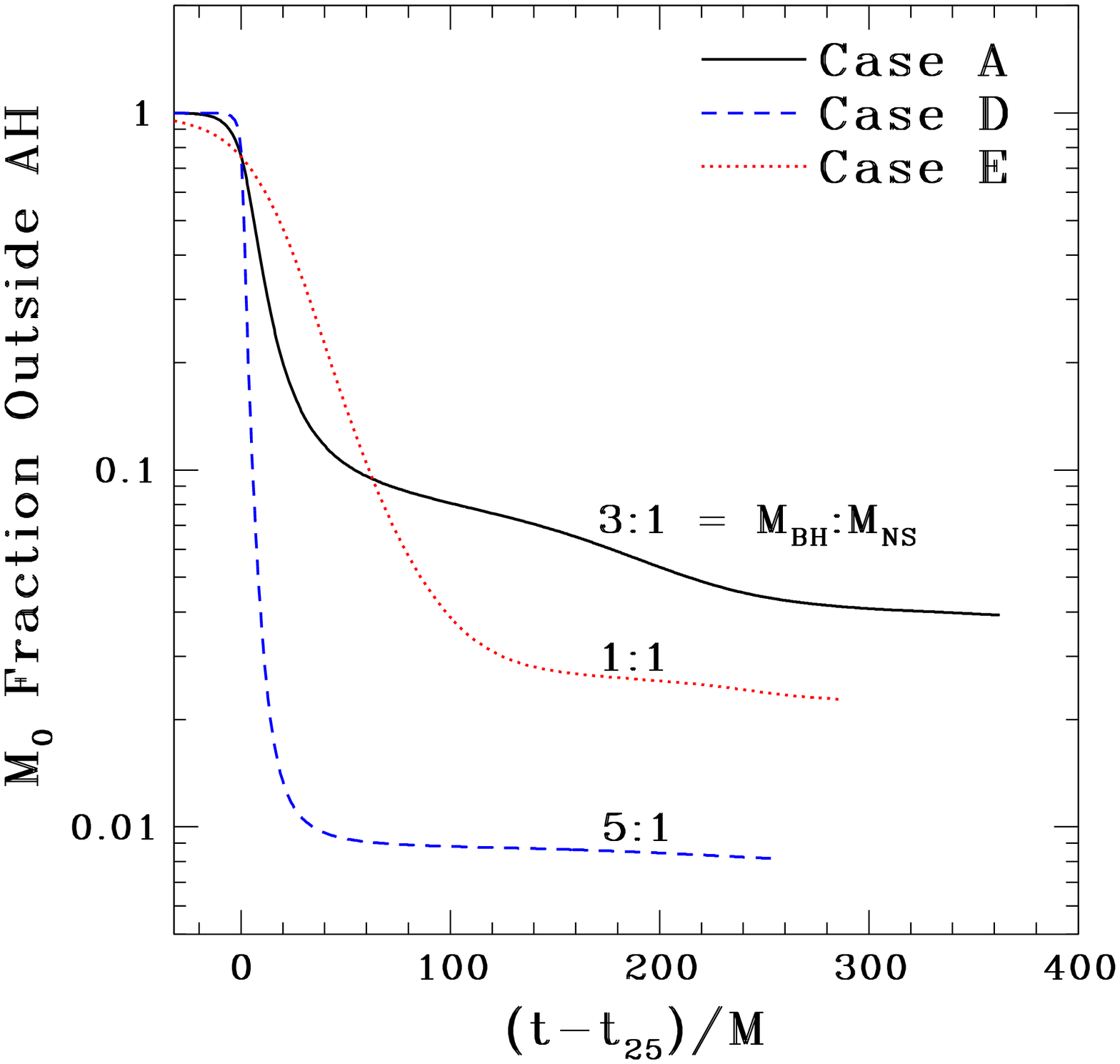}
\caption{Rest-mass fraction outside the BH for different BHNS mass
  ratios.  Here, the time coordinate is shifted by $t_{25}$,
  the time at which $25\%$ of the NS rest mass has fallen into the
  apparent horizon.}
\label{fig:BH_massratios_accretion}
\end{figure}

Figure~\ref{fig:BH_massratios_accretion} demonstrates how different 
mass ratios alter the NS material falling into the BH.  
For the $q=1$ and $q=3$ cases (Cases~\G\ and \A), we clearly identify
the plunge and accretion phases mentioned in the previous subsection.
However, for the $q=5$ case (Case~\F), the NS basically plunges into
the BH, leaving $<1\%$ of its rest mass outside the
BH at the end of the simulation. This result is not surprising since,
at the ISCO,  the tidal effect of the BH is smaller for larger $q$,
resulting in tidal disruption occurring closer to the ISCO. Moreover,
for a fixed NS compaction, the horizon size of the BH is larger for
larger $q$.  Hence more NS matter is expected to fall into the BH.

\begin{figure*}
\vspace{-4mm}
\begin{center}
\epsfxsize=2.15in
\leavevmode
\epsffile{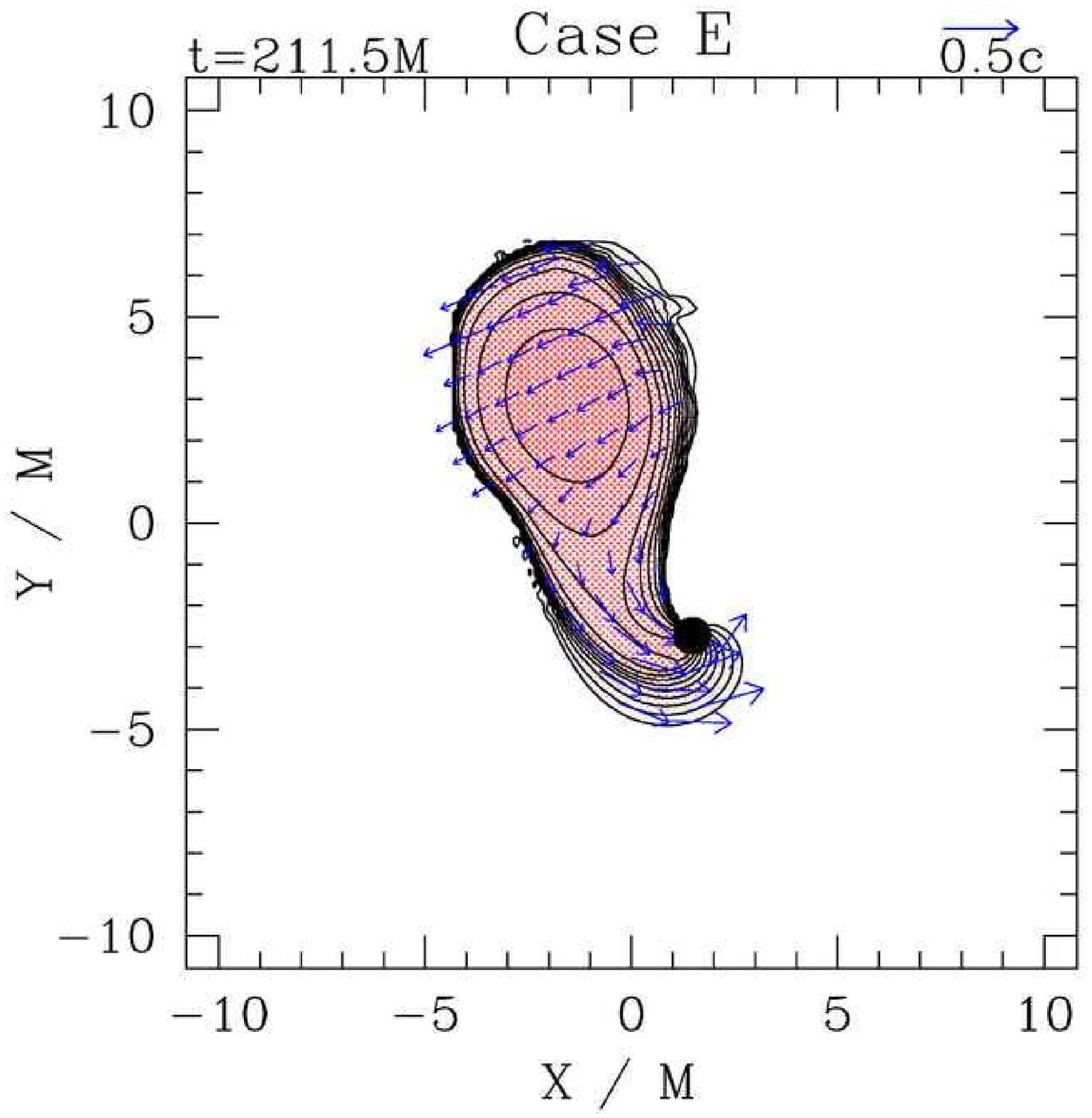}
\epsfxsize=2.15in
\leavevmode
\epsffile{A_t1.ps}
\epsfxsize=2.15in
\leavevmode
\epsffile{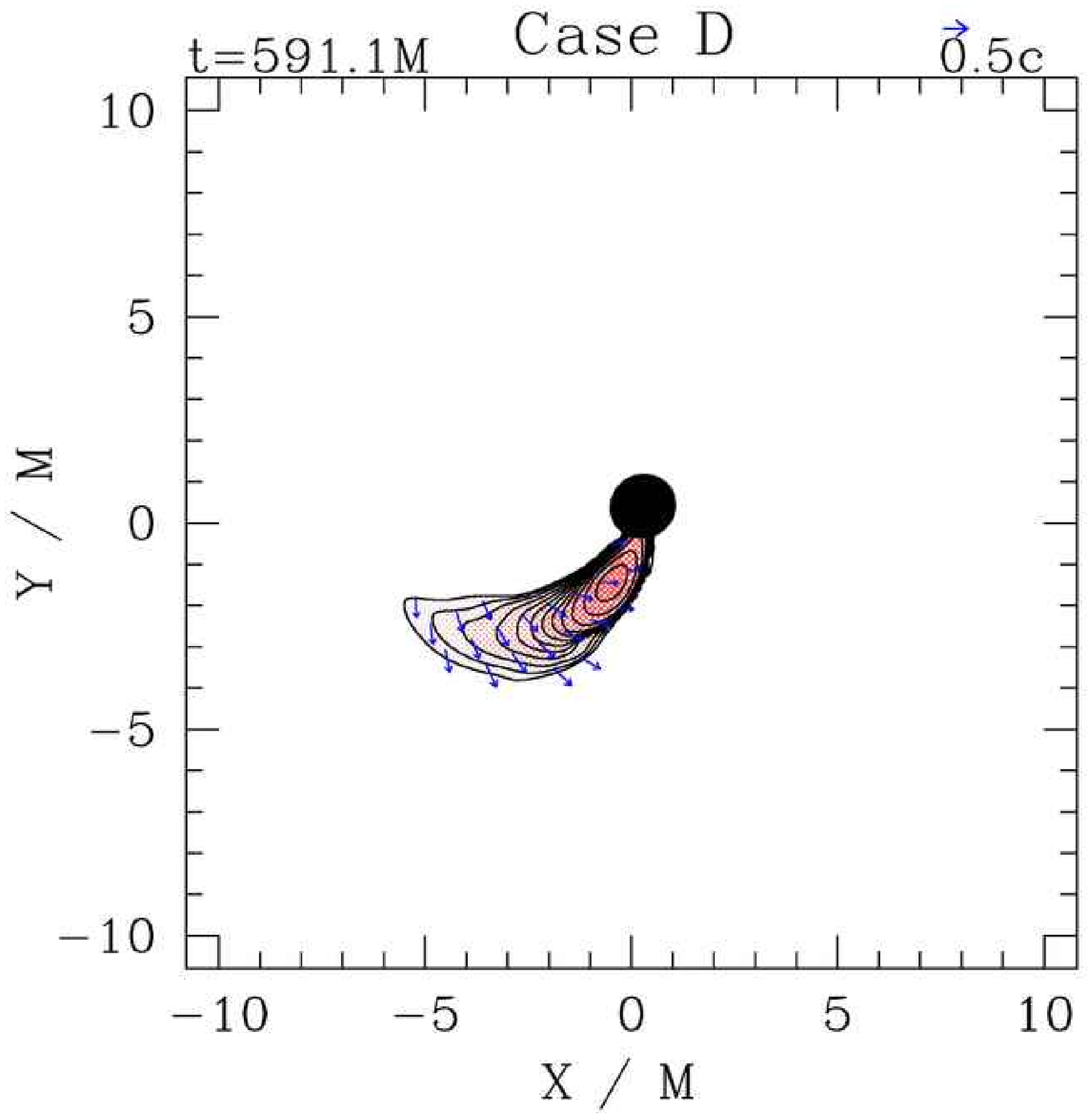} \\
\epsfxsize=2.15in
\leavevmode
\epsffile{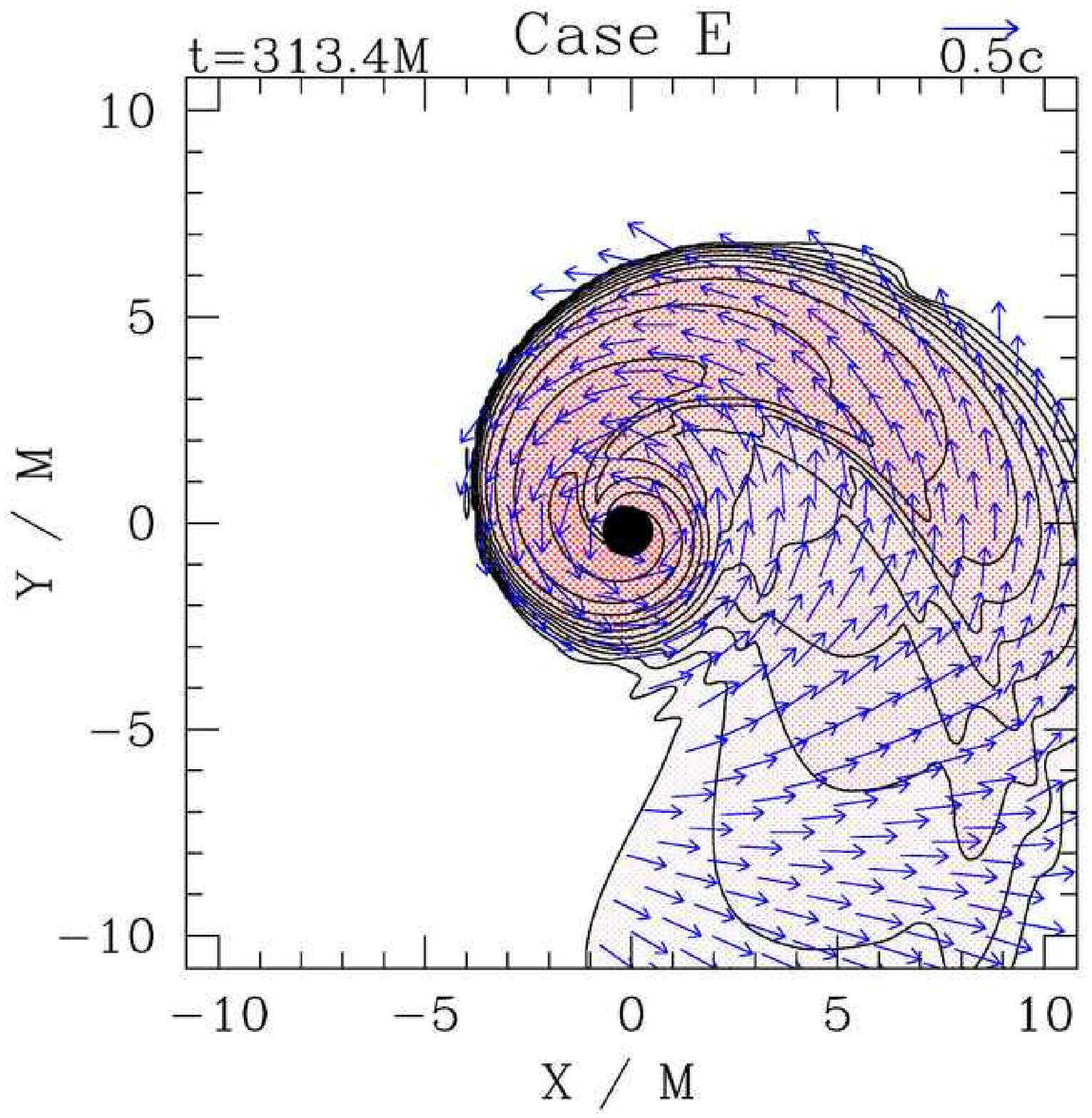}
\epsfxsize=2.15in
\leavevmode
\epsffile{A_t80.ps}
\epsfxsize=2.15in
\leavevmode
\epsffile{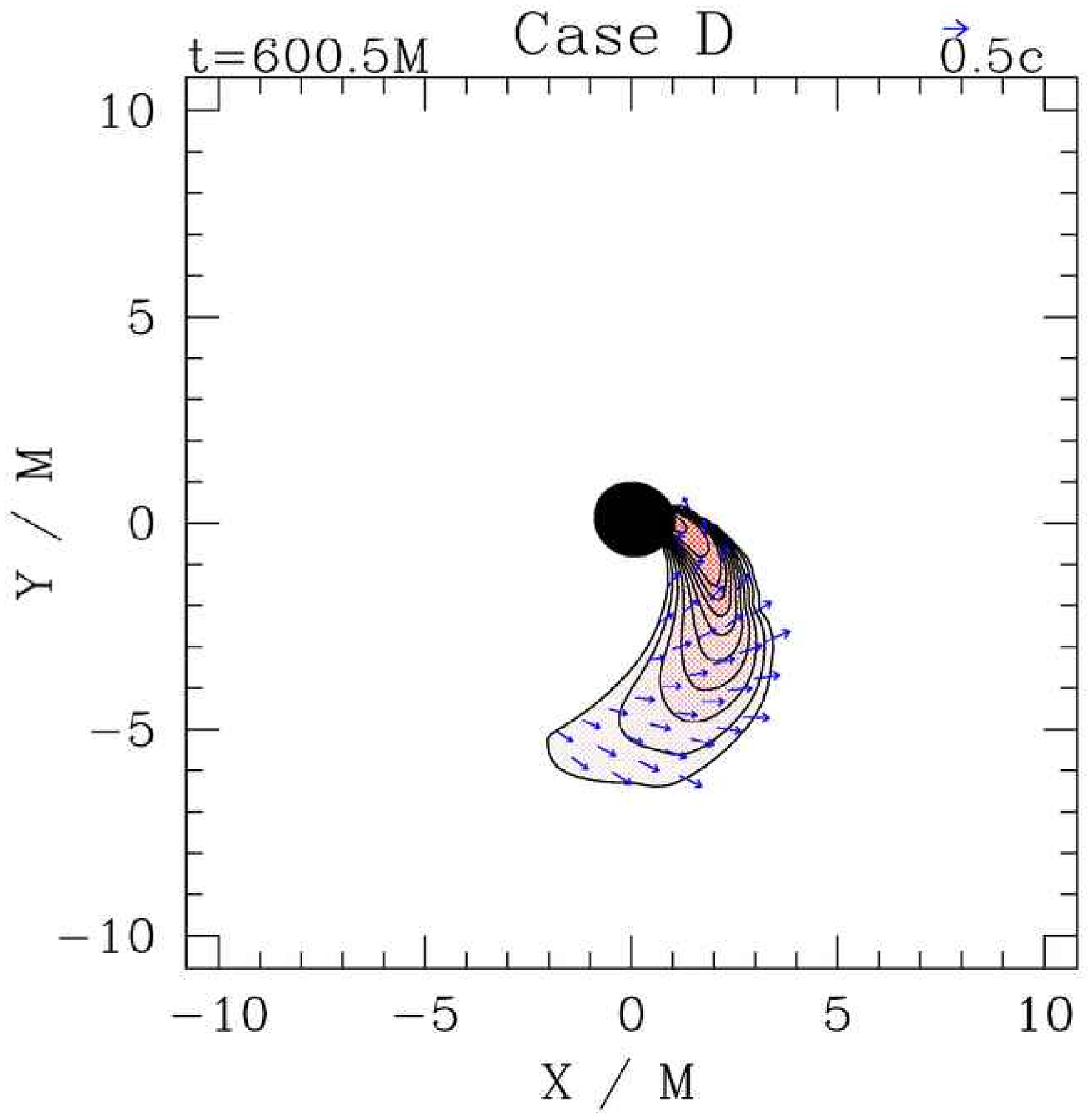}
\caption{Snapshots of density and velocity profiles at
  $t_1$ (top row) and $t_{80}$ (bottom row) for Cases~\G, \A, and \F,
  which have initial mass ratios of $q=1$, 3, and 5, respectively. The
  contours represent the density in the orbital plane, plotted
  according to $\rho_0 = \rho_{0,{\rm max}} 10^{-0.38j-0.04}$
  ($j$=0, 1, ... 12), with darker greyscaling for higher density.  
  The maximum initial NS density is $\kappa \rho_{0,{\rm max}}= 0.126$ 
  for all cases, or 
  $\rho_{0,{\rm max}}=9\times 10^{14}\mbox{g cm}^{-3}(1.4M_\odot/M_0)^2$. 
  Arrows represent the
  velocity field in the orbital plane.  The black hole AH
  is marked by a filled black circle.  In cgs units,
  the total ADM mass for these cases is $M= 6.25\times
  10^{-6}(q+1)(M_0/1.4M_\odot)$s$=1.9(q+1)(M_0/1.4M_\odot)$km.}
\label{fig:massratiostudysnapshots}
\end{center}
\end{figure*}

\begin{figure*}
\vspace{-4mm}
\begin{center}
\epsfxsize=2.15in
\leavevmode
\epsffile{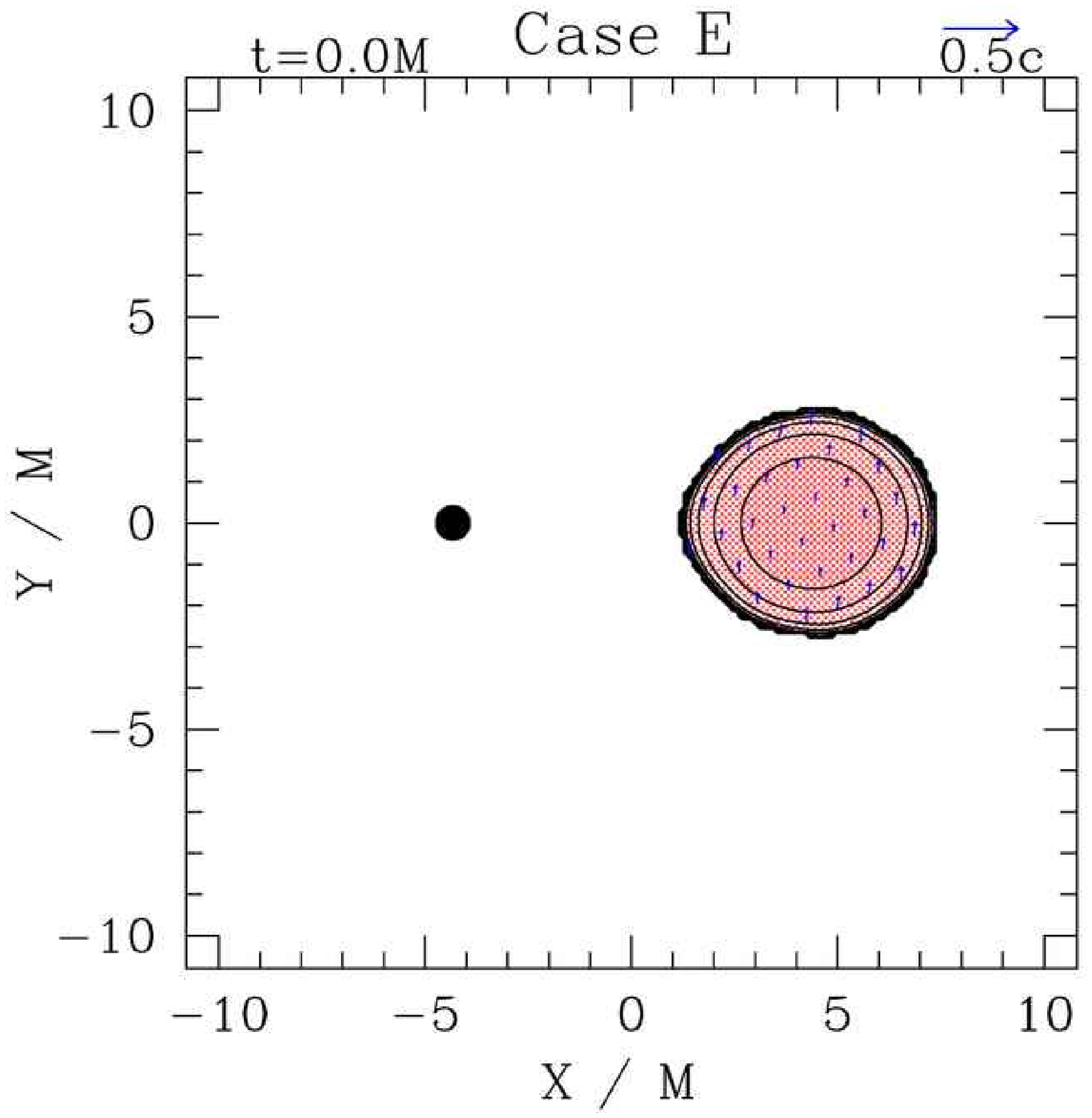}
\epsfxsize=2.15in
\leavevmode
\epsffile{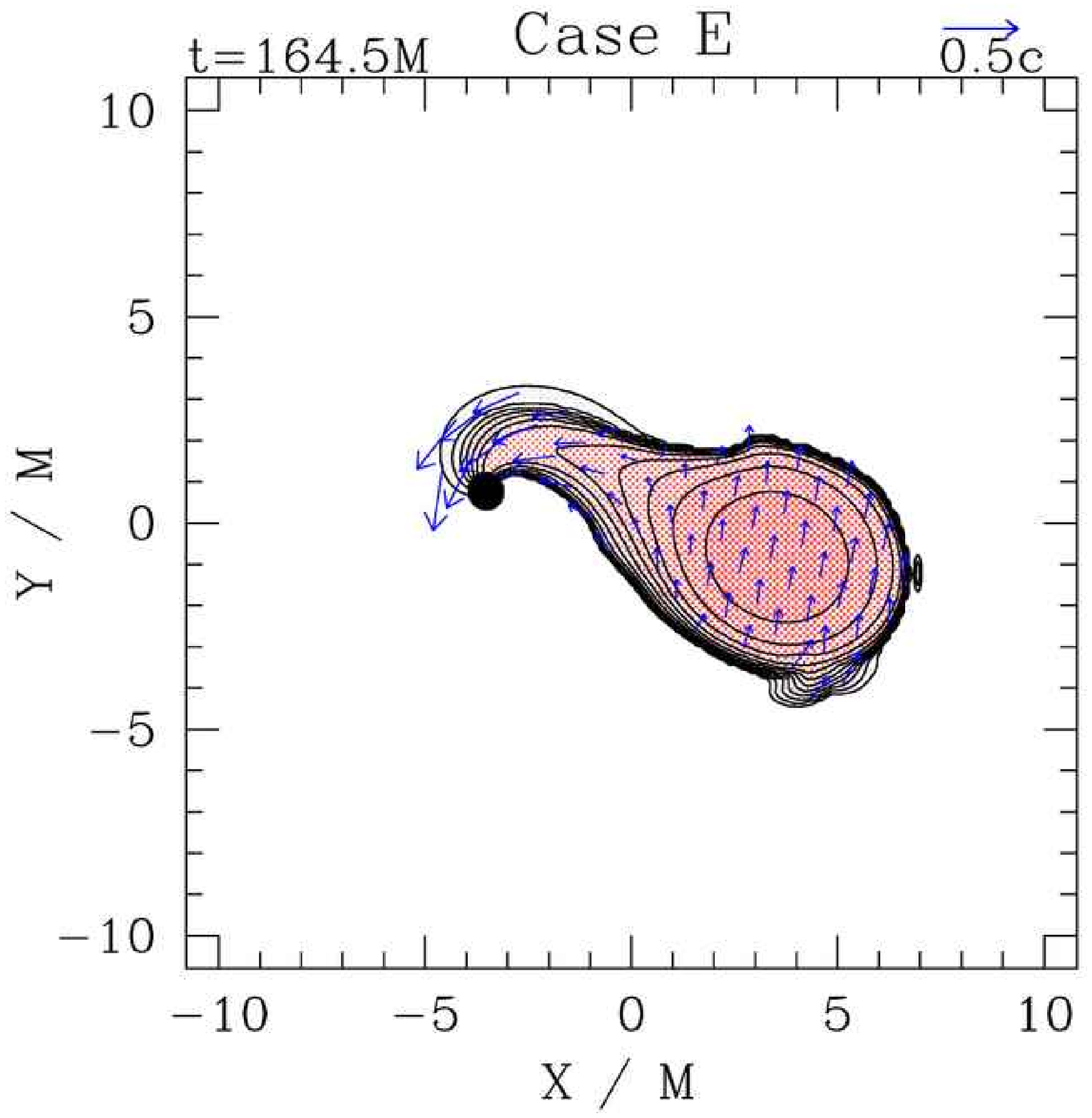} 
\epsfxsize=2.15in
\leavevmode
\epsffile{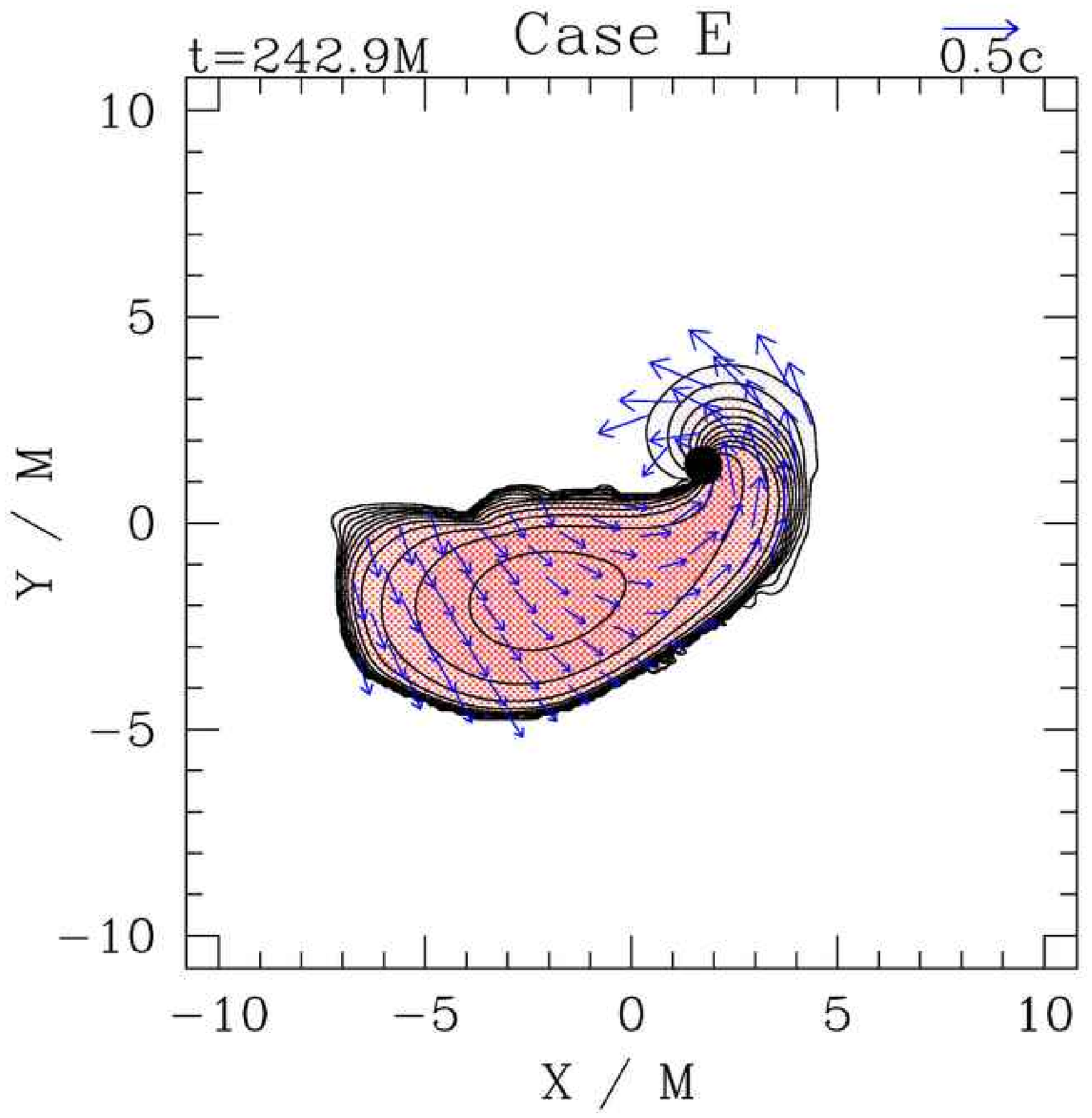} \\
\epsfxsize=2.15in
\leavevmode
\epsffile{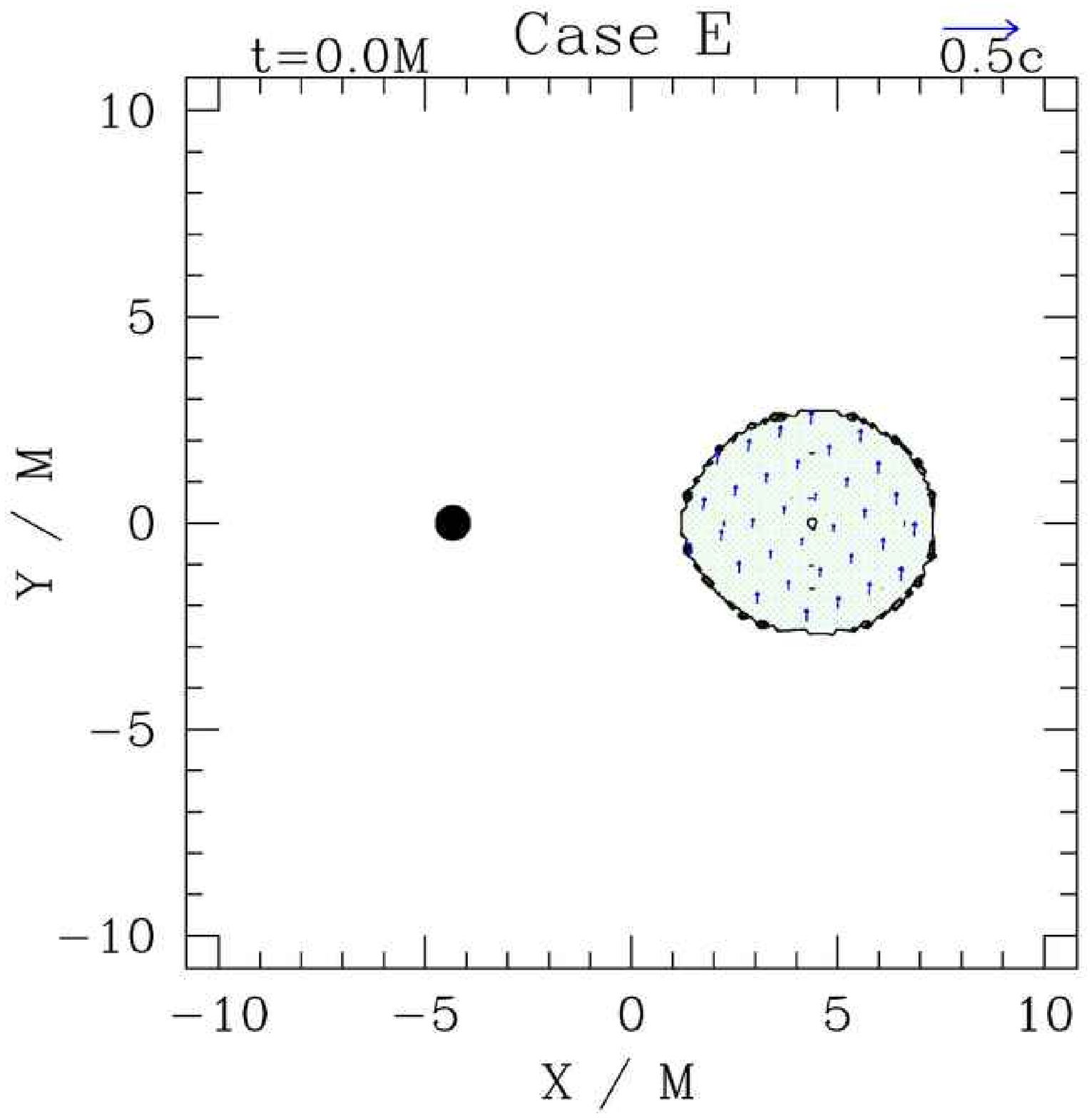}
\epsfxsize=2.15in
\leavevmode
\epsffile{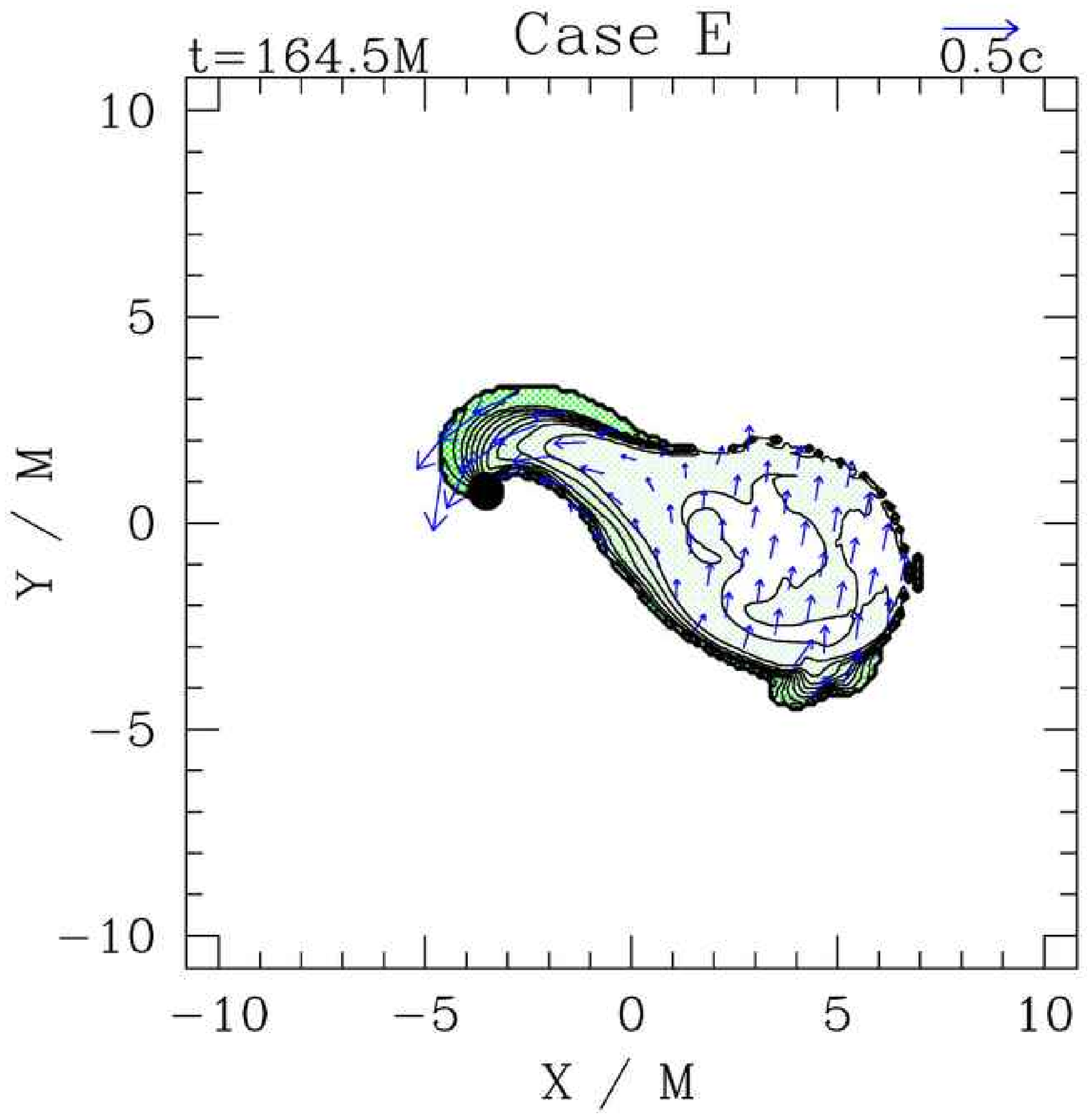}
\epsfxsize=2.15in
\leavevmode
\epsffile{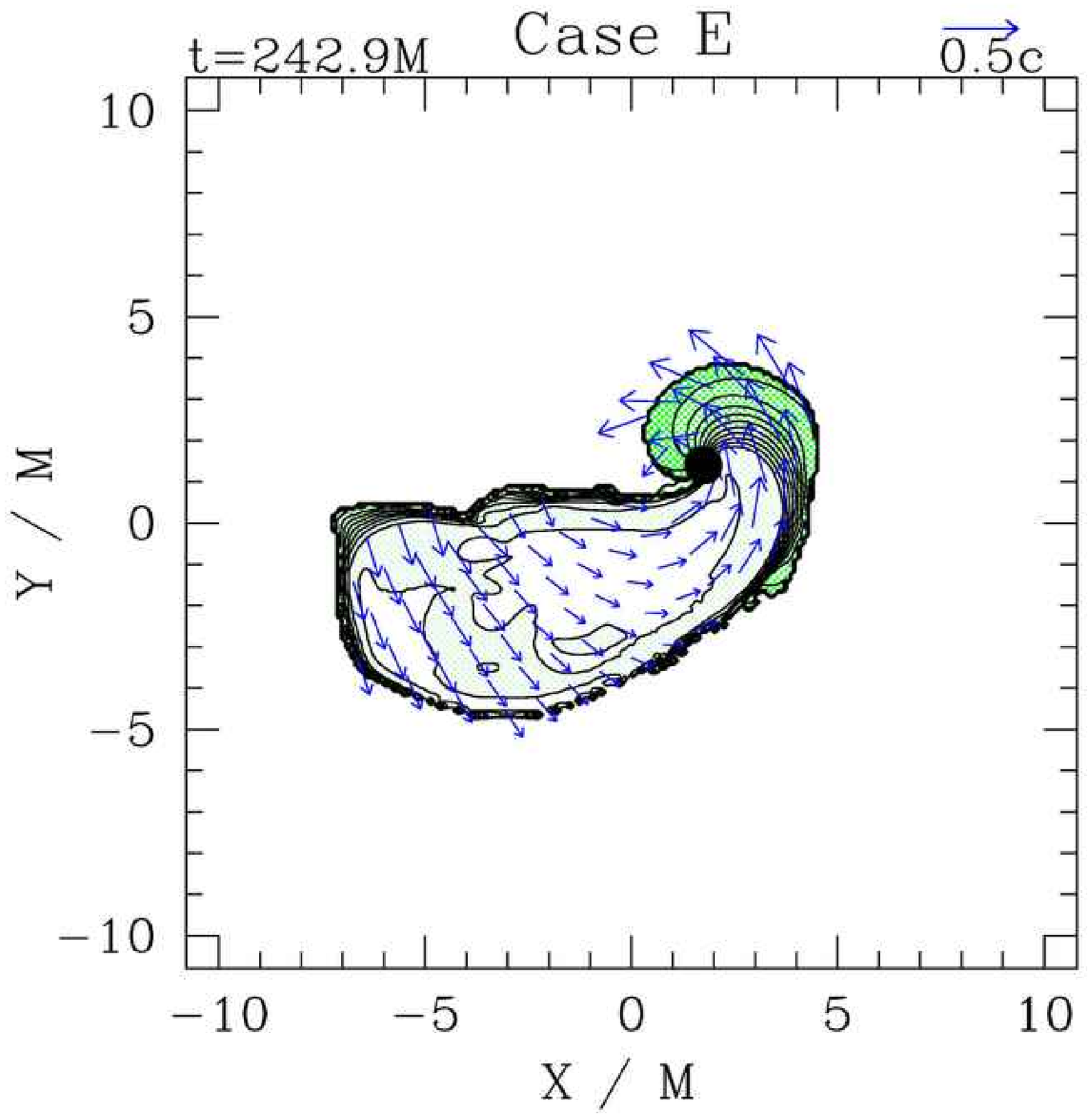} \\
\vspace{2mm}
\line(2,0){500}
\vspace{2mm}
\epsfxsize=2.15in
\leavevmode
\epsffile{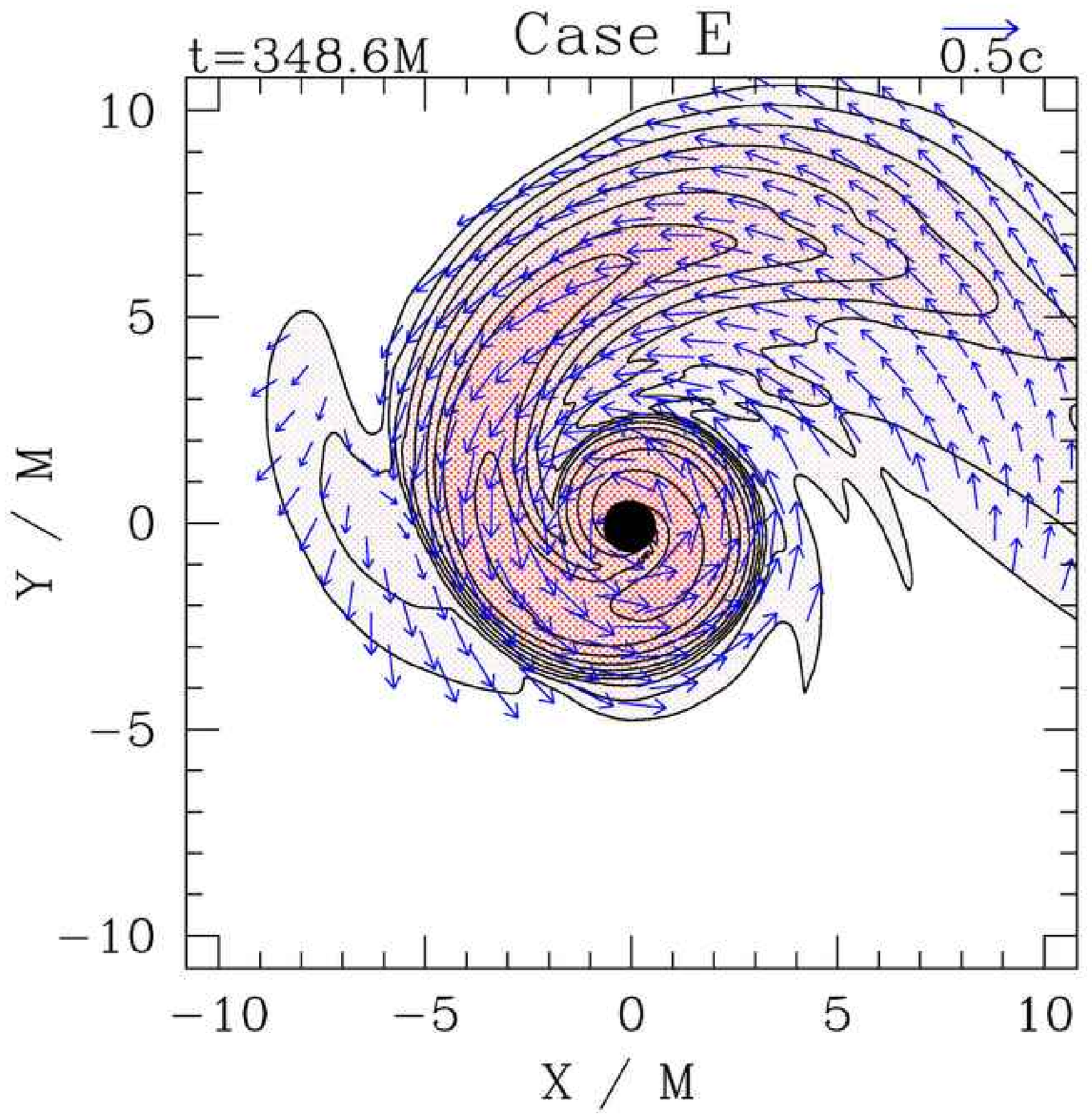}
\epsfxsize=2.15in
\leavevmode
\epsffile{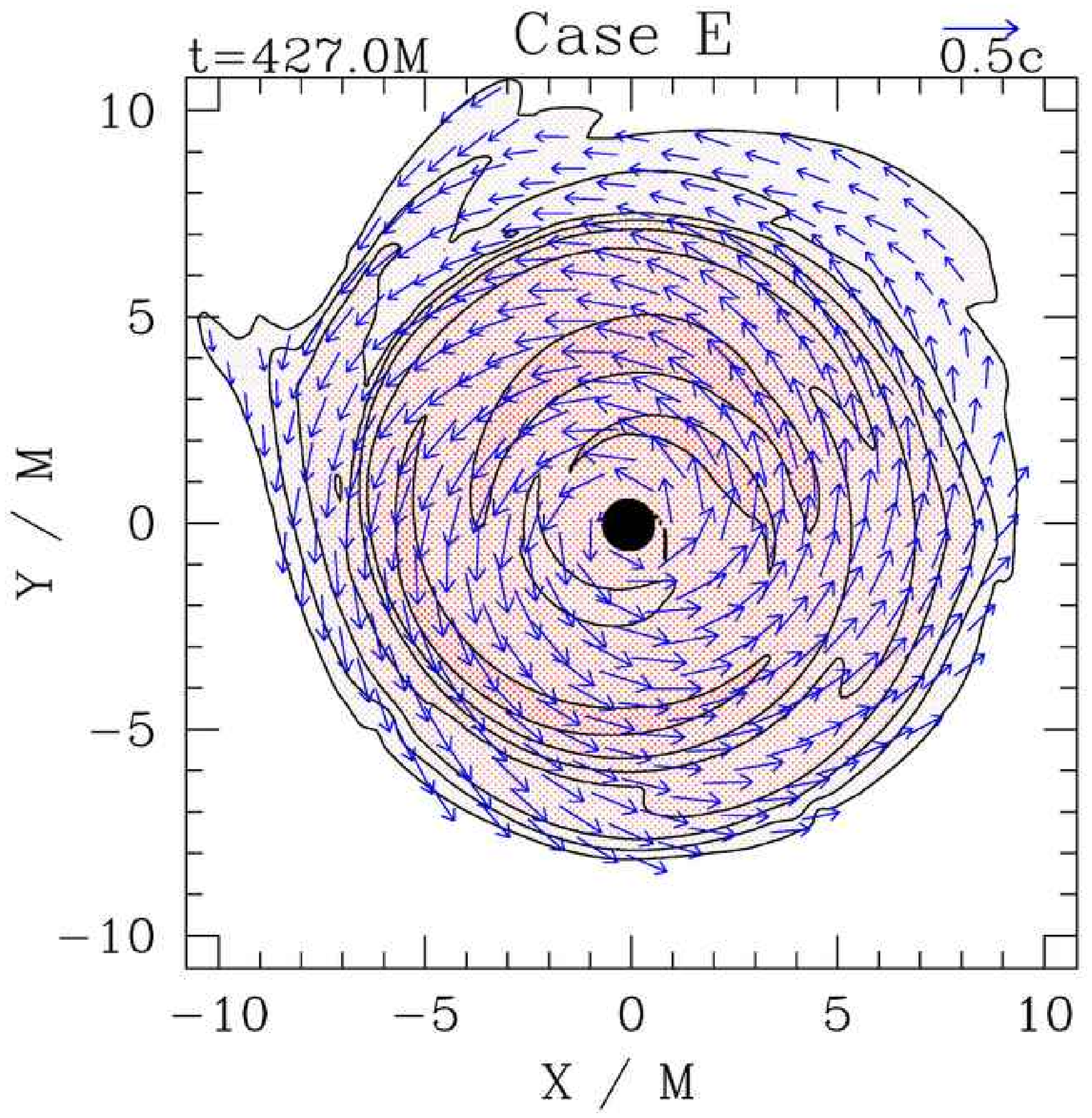}
\epsfxsize=2.15in
\leavevmode
\epsffile{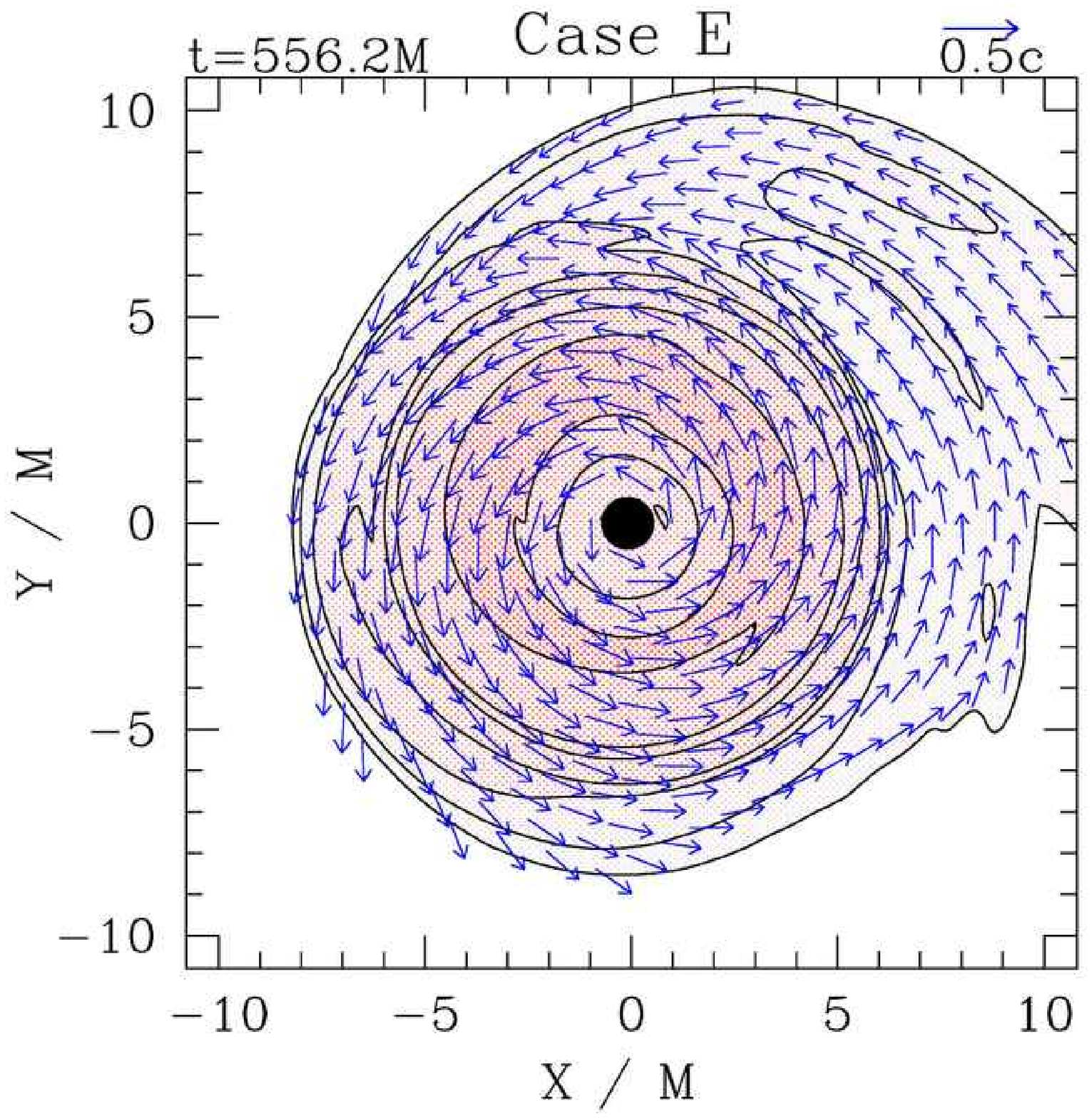} \\
\epsfxsize=2.15in
\leavevmode
\epsffile{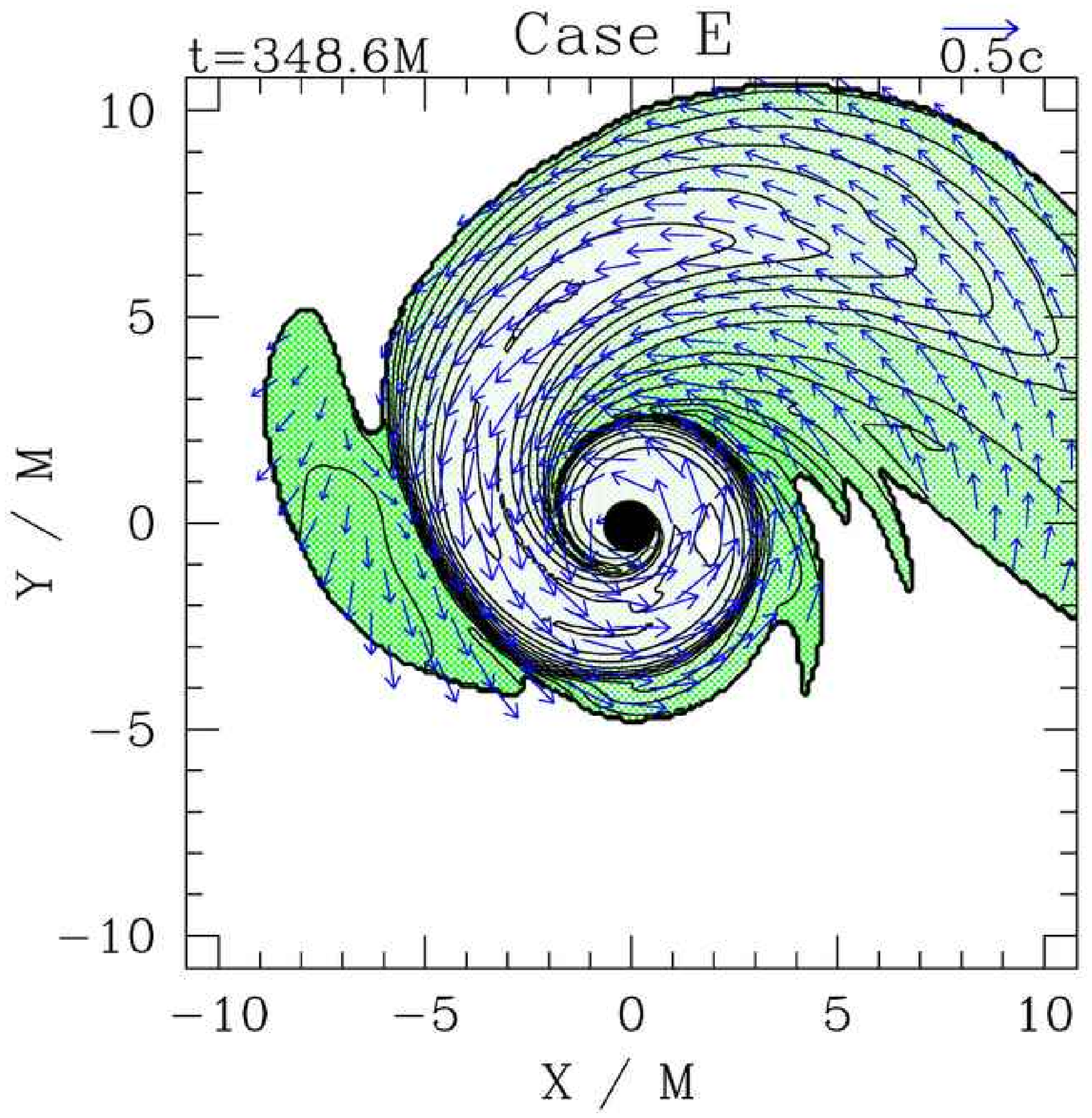} 
\epsfxsize=2.15in
\leavevmode
\epsffile{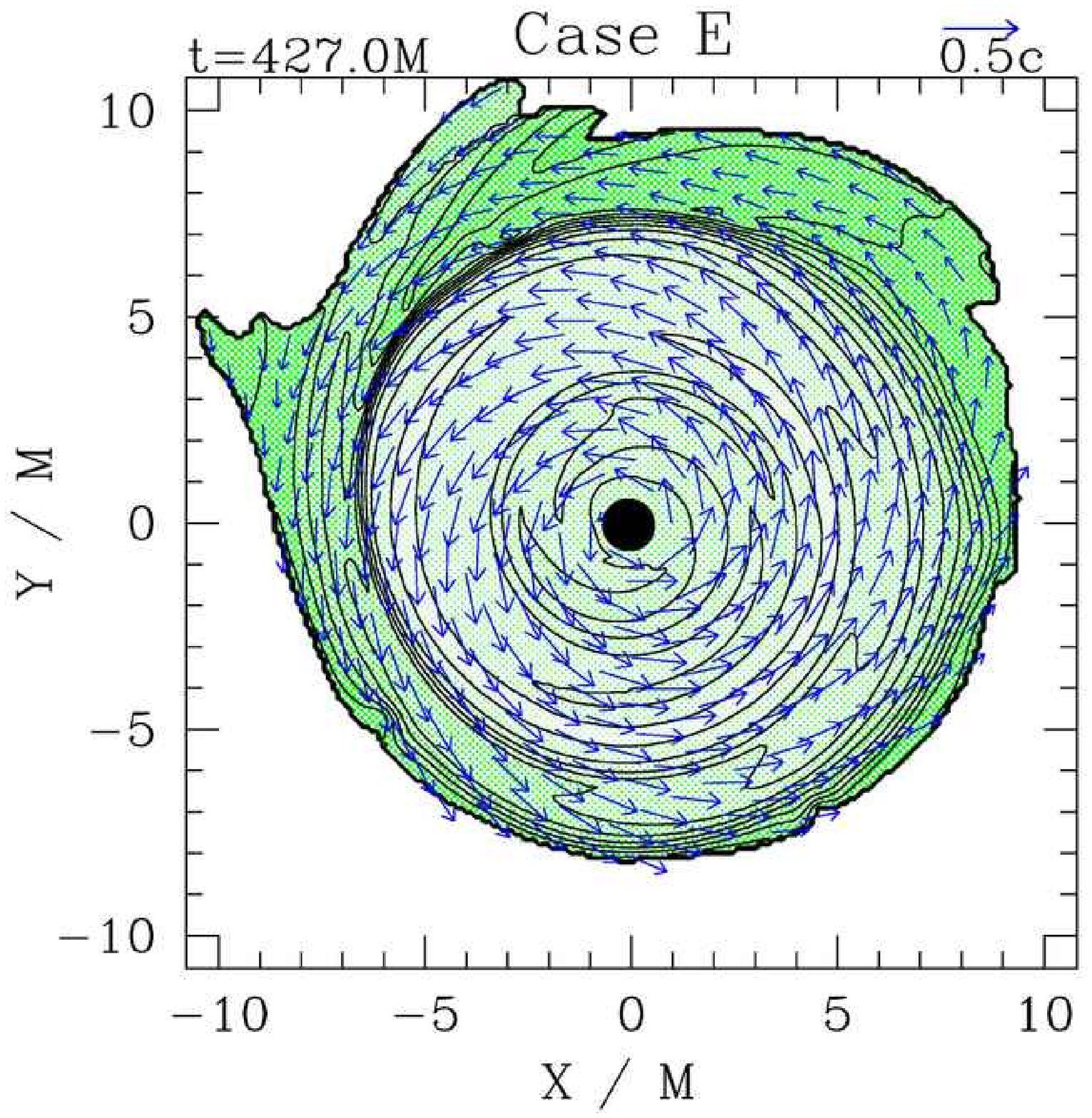}
\epsfxsize=2.15in
\leavevmode
\epsffile{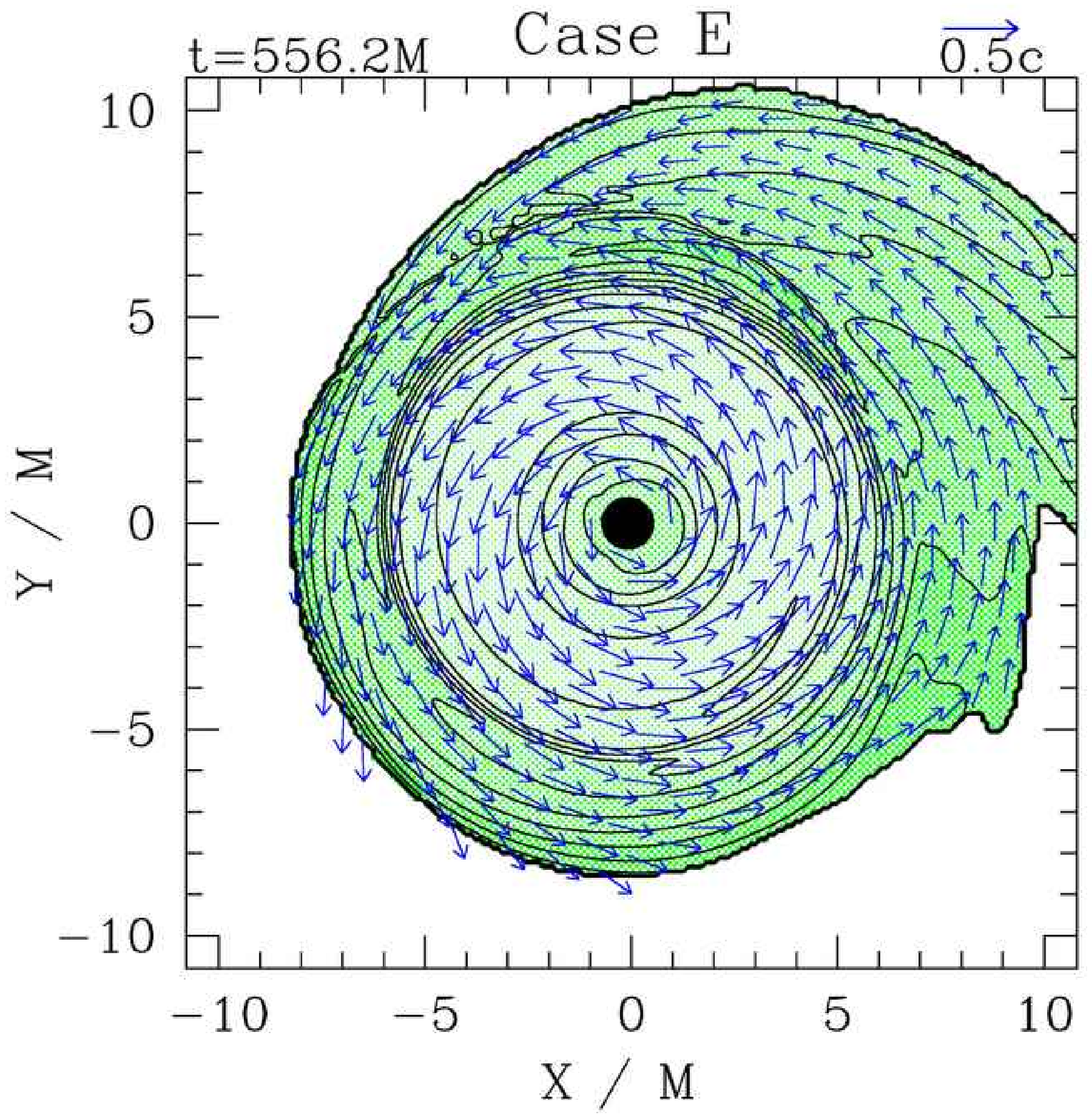}
\caption{Snapshots of rest-mass density $\rho_0$ and $K$ contours for
  Case~\G\ ($q=1$ mass ratio).  The first and third rows show
  snapshots of density contours and velocity profiles in the orbital
  plane.  The second and fourth rows show snapshots of $K=P/\kappa
  \rho_0^\Gamma$. The density is plotted according to $\rho_0 =
  \rho_{0,{\rm max}}10^{-0.38j-0.04}$ ($j$=0, 1, ... 12), with darker
  greyscaling for higher density. The maximum density of the initial 
  NS is $\kappa \rho_{0,{\rm max}}=0.126$, or 
  $\rho_{0,{\rm max}}=9\times 10^{14}\mbox{g cm}^{-3}(1.4M_\odot/M_0)^2$.
  Arrows represent the
  velocity field in the orbital plane. $K$ is plotted according to
  $K=10^{0.32j}$ ($j$=0, 1, ... 12), with darker greyscaling for
  higher $K$.  These figures neglect $K$ in regions where the density
  is less than $\rho_{0,\rm{min}}$.  The black hole AH interior is
  marked by a filled black circle. In cgs units, $M=1.3\times
  10^{-5}(M_0/1.4M_\odot)$s=$3.8(M_0/1.4M_\odot)$km.}
\label{fig:G_snapshot}
\end{center}
\end{figure*}

Figure~\ref{fig:massratiostudysnapshots} shows snapshots of 
density and velocity profiles of the three cases at $t_1$ and 
$t_{80}$. We again see different structures of the NS tail
among the three cases. 
The merger of Case~\G\ ($q=1$ case) is particularly interesting. 
Figure~\ref{fig:G_snapshot} shows snapshots of rest-mass density 
and $K$ contours. 
Since the BH is less massive in Case~\G, tidal
disruption occurs at a farther binary separation (upper middle plots).
The disrupted NS matter curls around the BH, forming a hot, 
low-density spiral (upper right plots) that winds 
around the AH and smashes into the tidal tail, generating a large amount
of shock heating.  Some of the heated NS matter
transfers angular momentum to the other part of the tail and falls
into the BH.  The remaining matter in the tail deforms into an
inhomogeneous disk (lower center plots) before settling into a
quasistationary state, in which a high density, relatively low
temperature torus of matter surrounds the BH (lower right plots).
Analysis of the accretion versus time
(Fig.~\ref{fig:BH_massratios_accretion}) demonstrates that the plunge
phase is relatively long compared to Case~\A.  Because of the complicated
interaction of the disrupted NS matter in the merging process, the
disk mass is less than that of Case~\A.  

\subsection{Disk profile}
\label{sec:disk}

\begin{figure*}
\vspace{-4mm}
\begin{center}
\epsfxsize=2.1in
\leavevmode
\epsffile{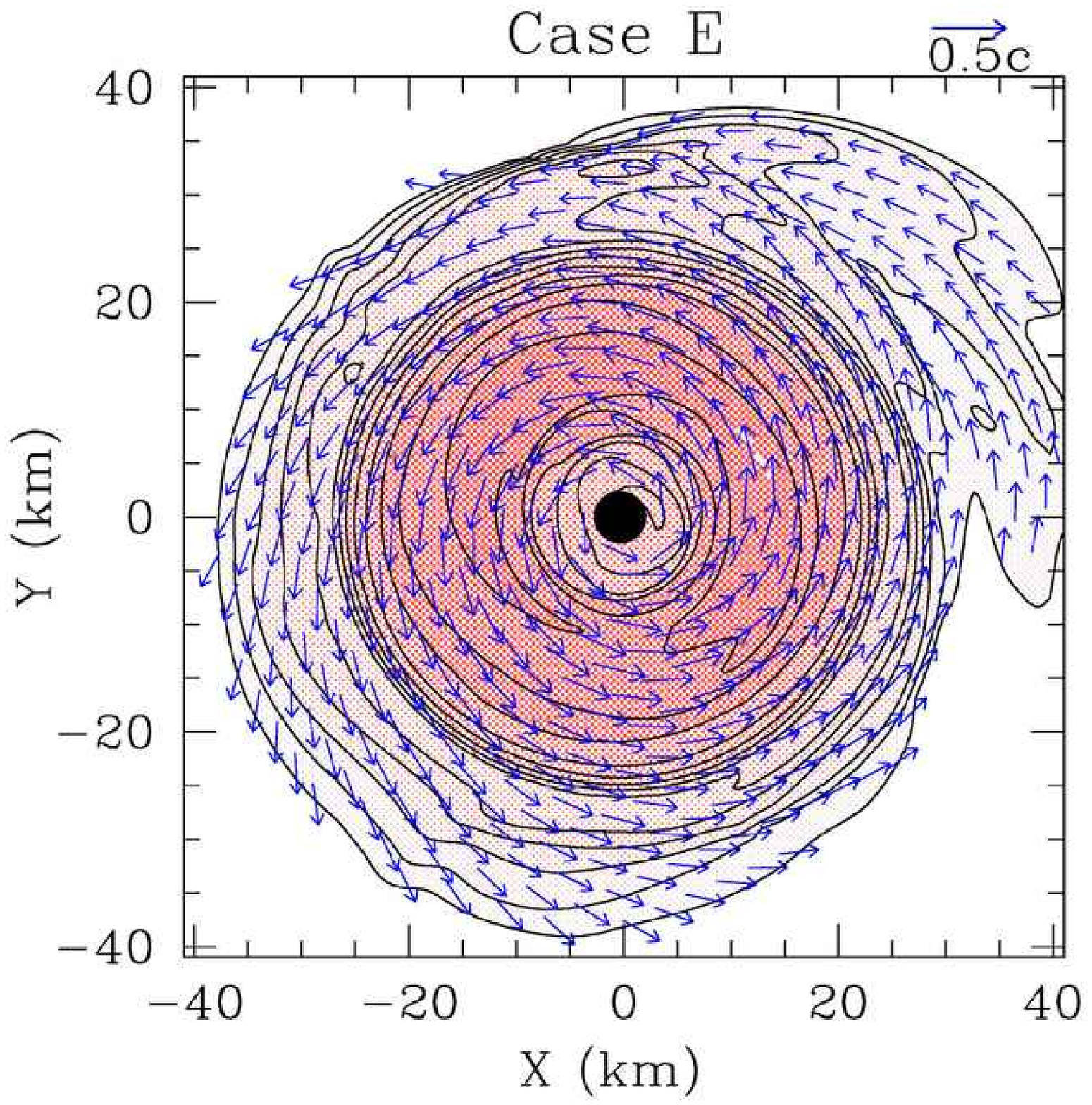}
\epsfxsize=2.1in
\leavevmode
\epsffile{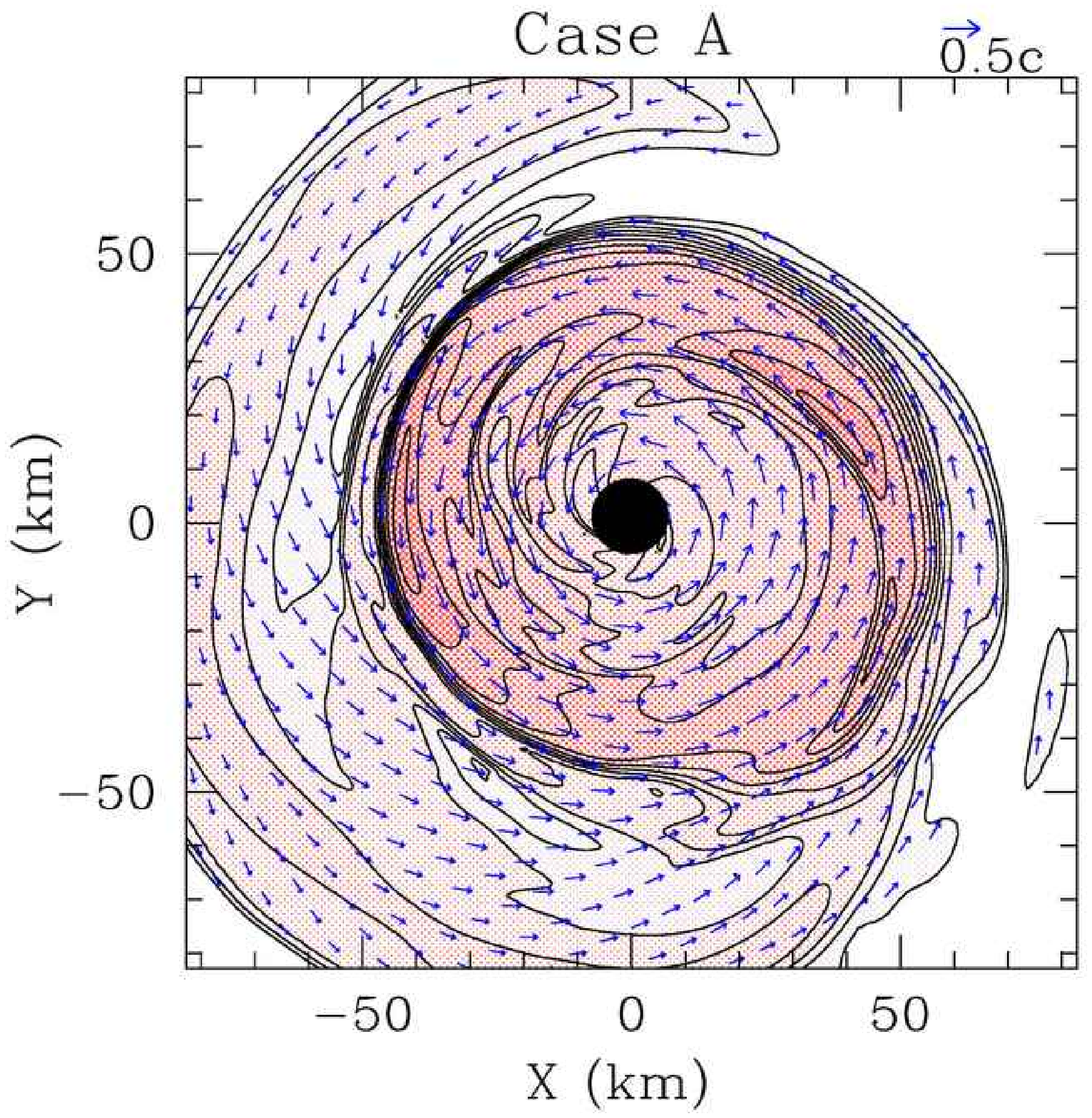}
\epsfxsize=2.1in
\leavevmode
\epsffile{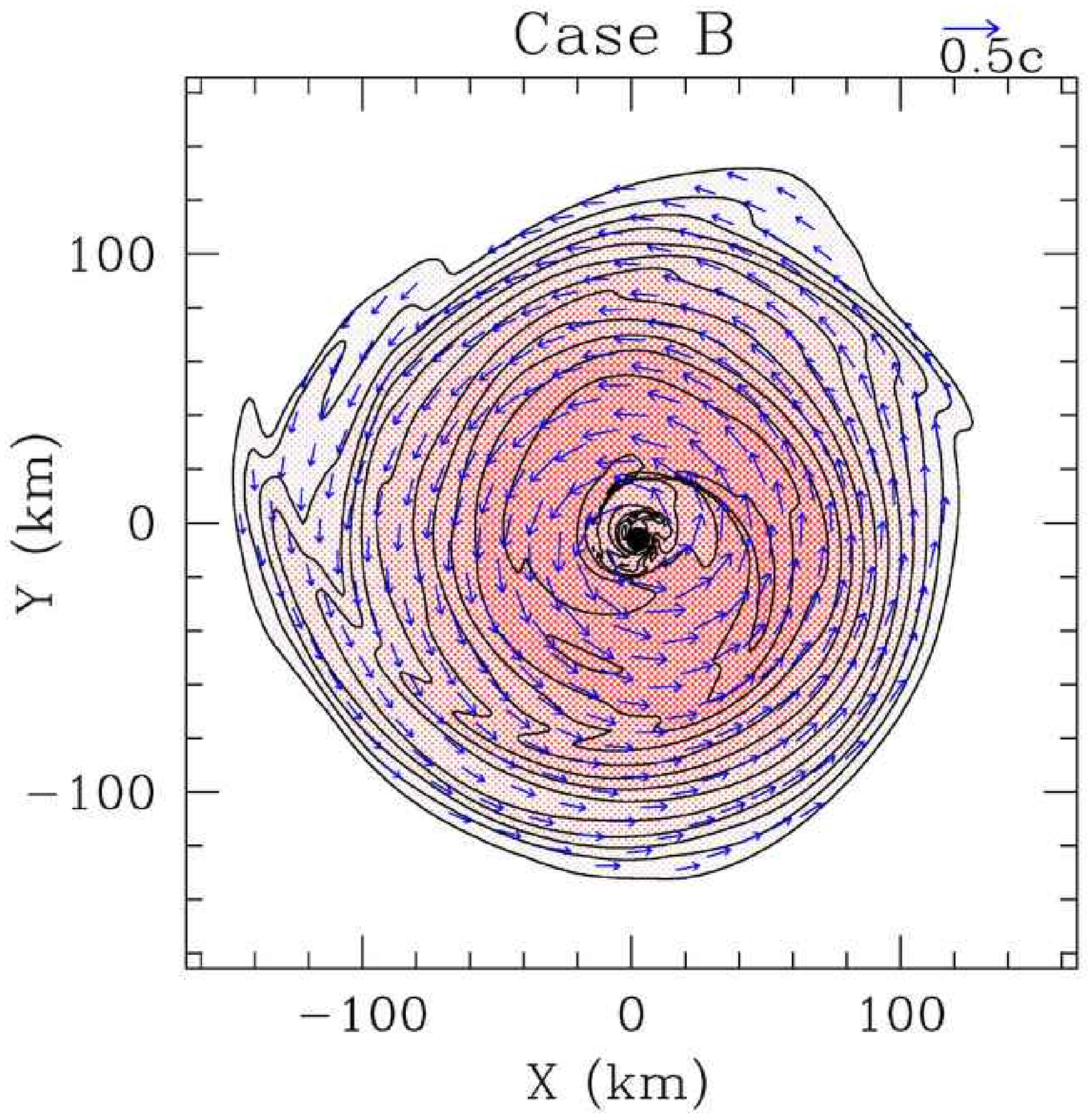} \\
\epsfxsize=2.1in
\leavevmode
\epsffile{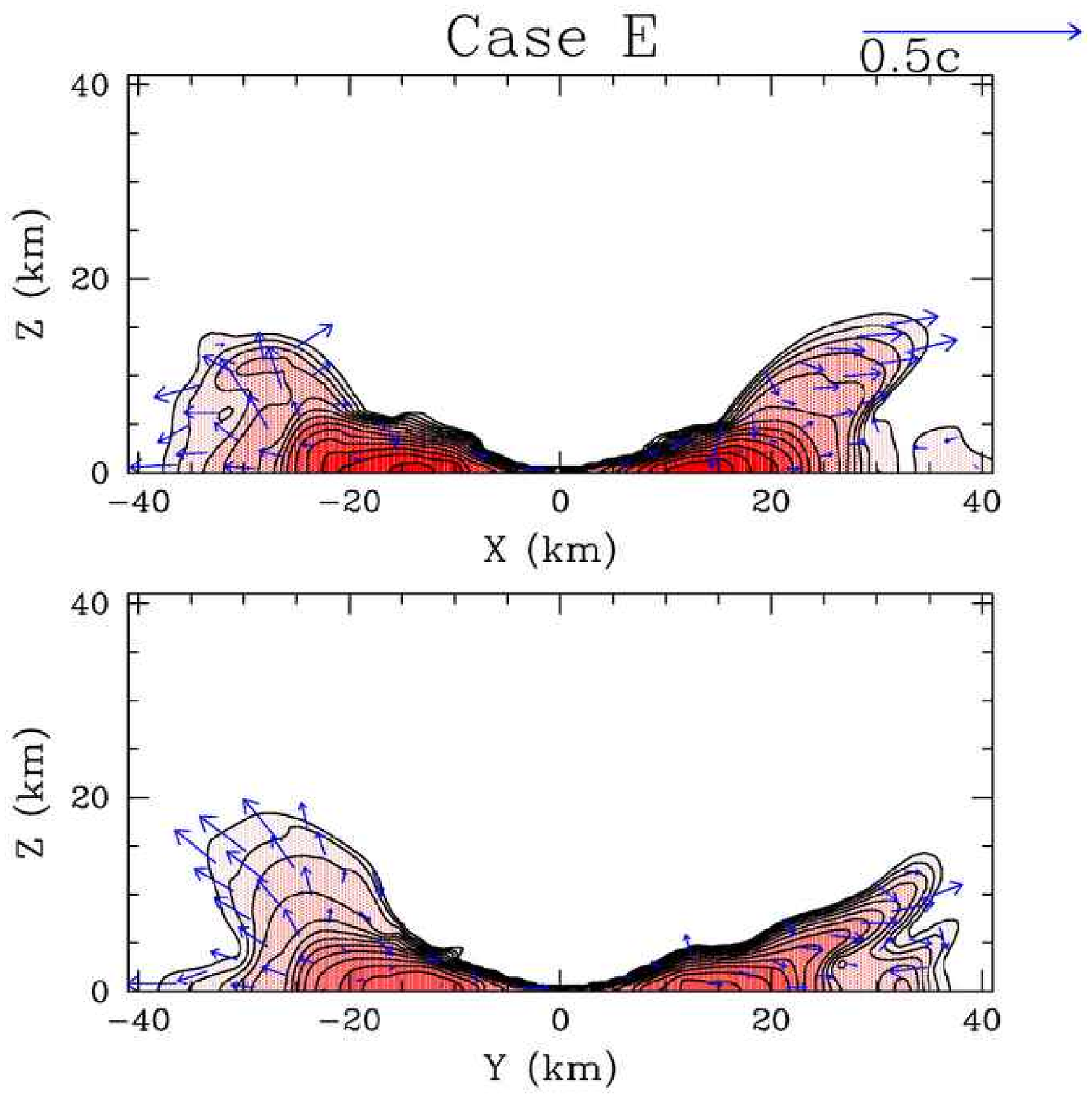}
\epsfxsize=2.1in
\leavevmode
\epsffile{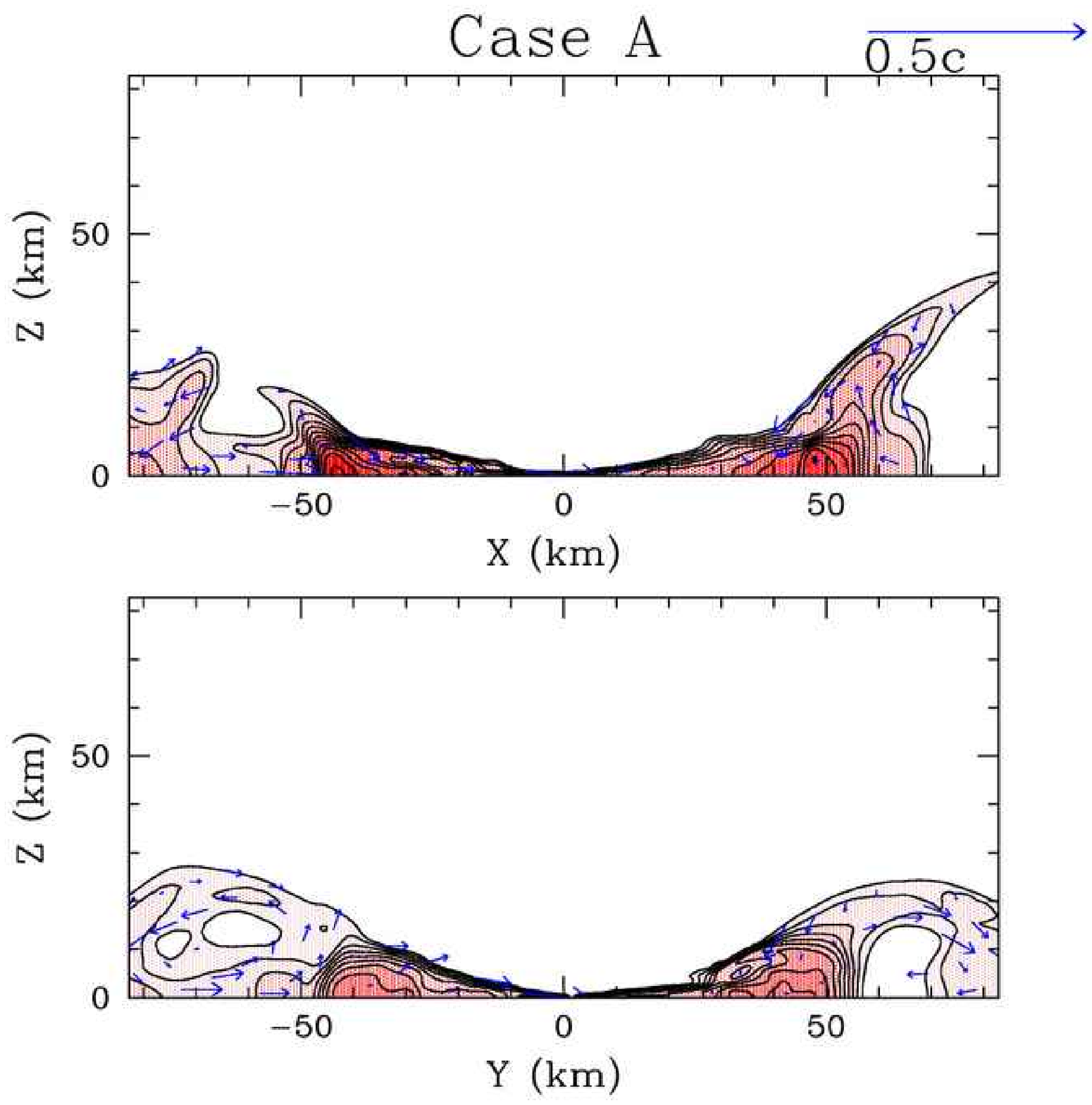}
\epsfxsize=2.1in
\leavevmode
\epsffile{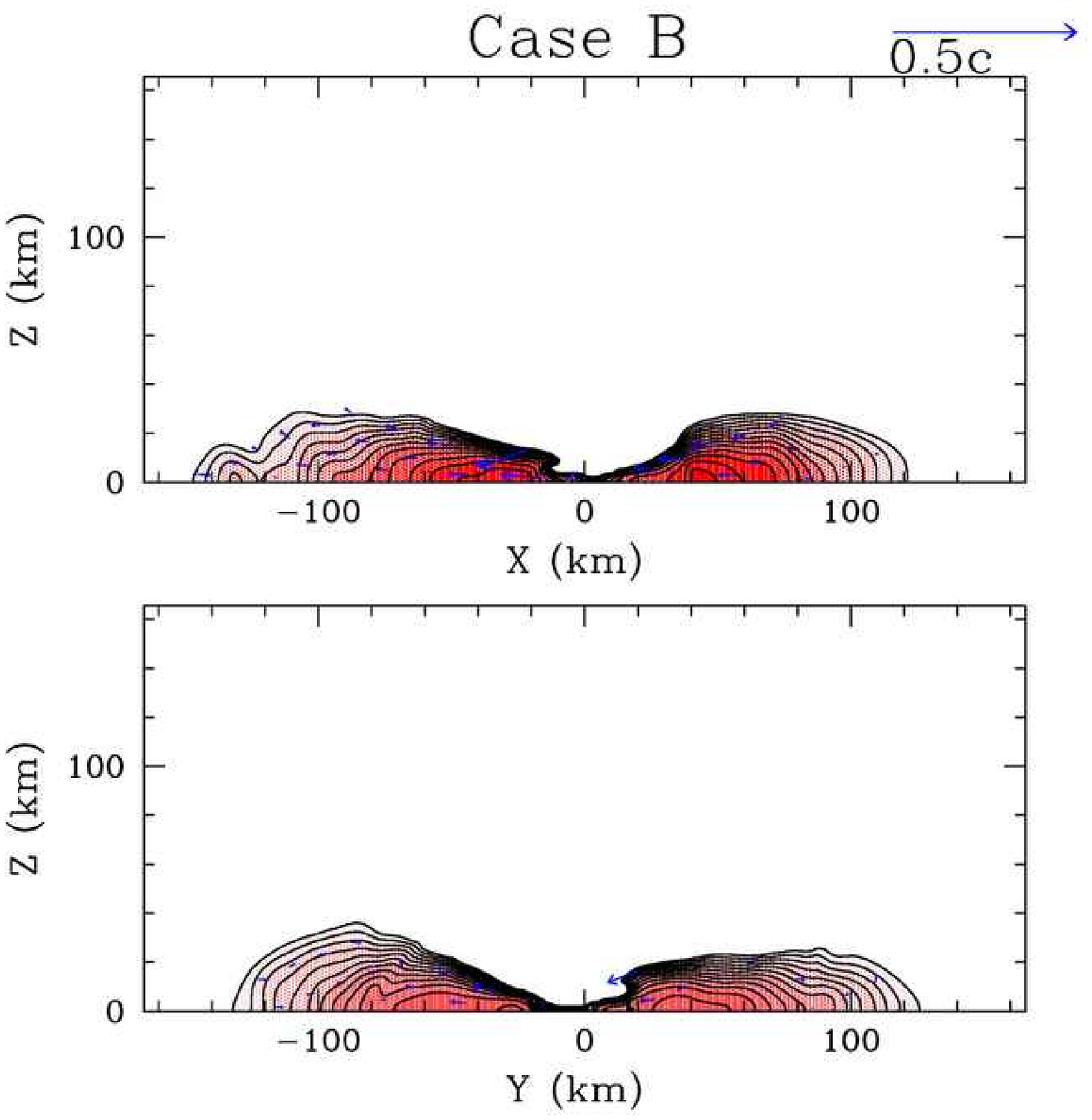} \\
\vspace{2mm}
\line(2,0){500}
\vspace{2mm}
\epsfxsize=2.1in
\leavevmode
\epsffile{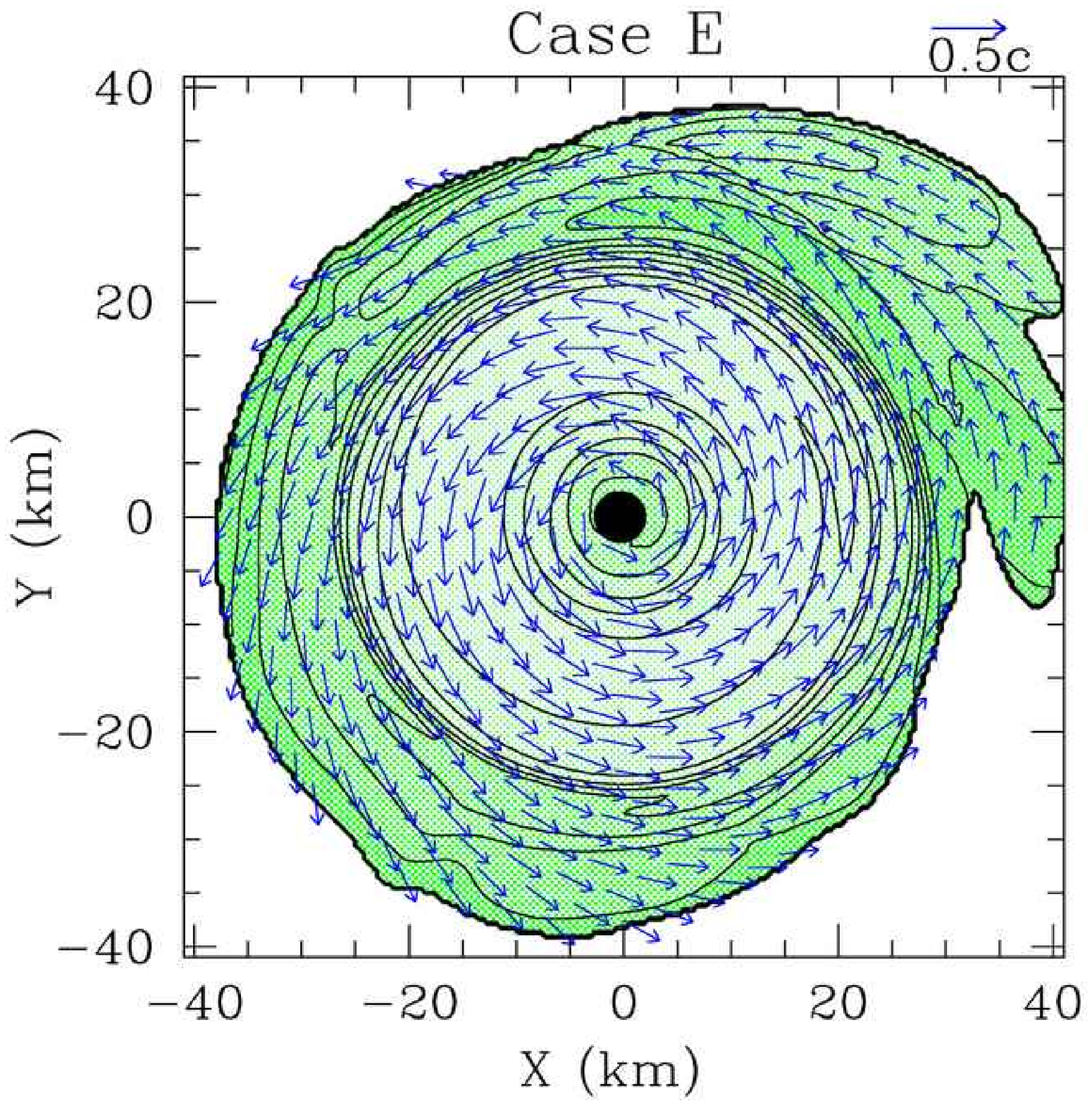}
\epsfxsize=2.1in
\leavevmode
\epsffile{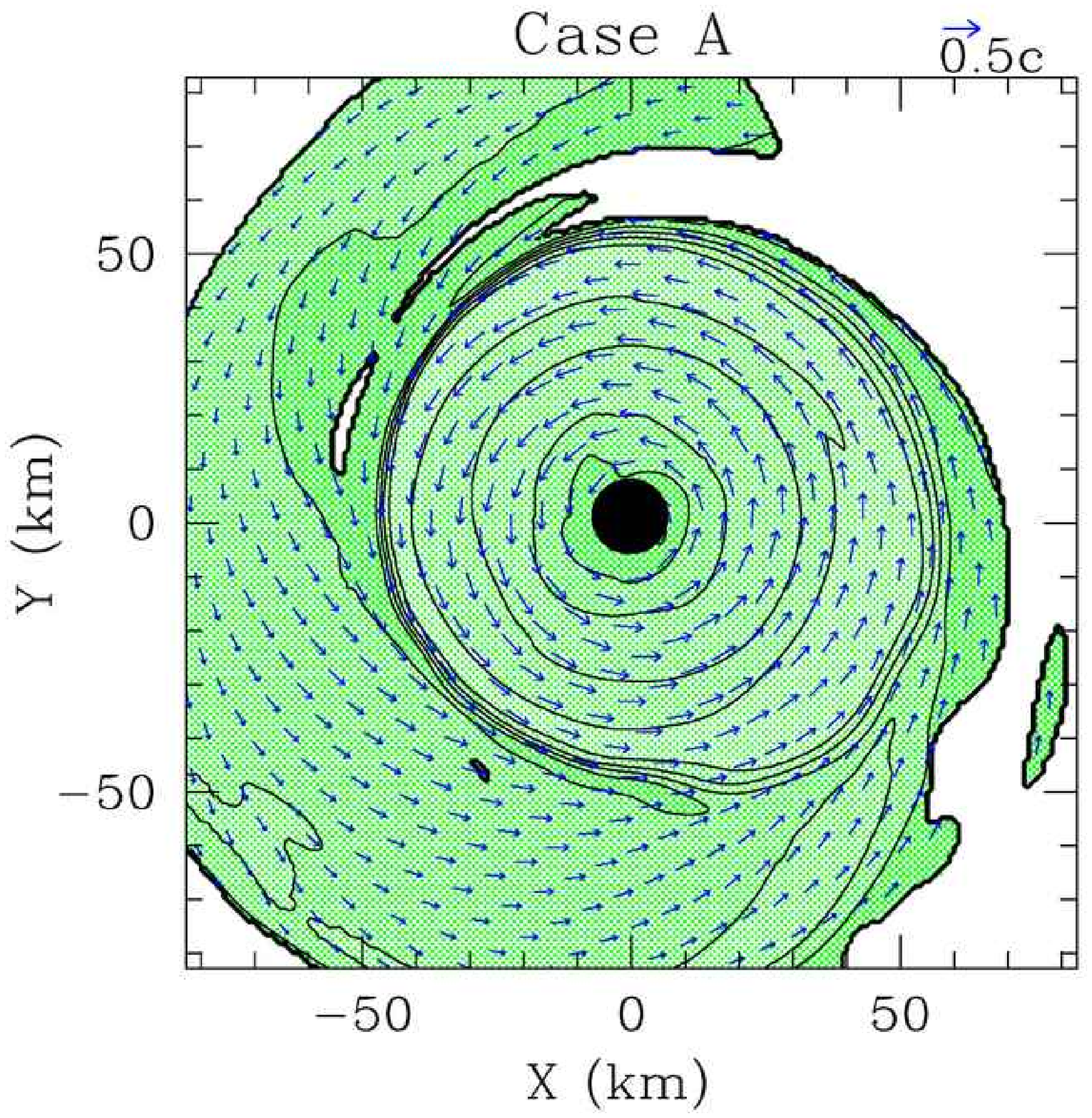}
\epsfxsize=2.1in
\leavevmode
\epsffile{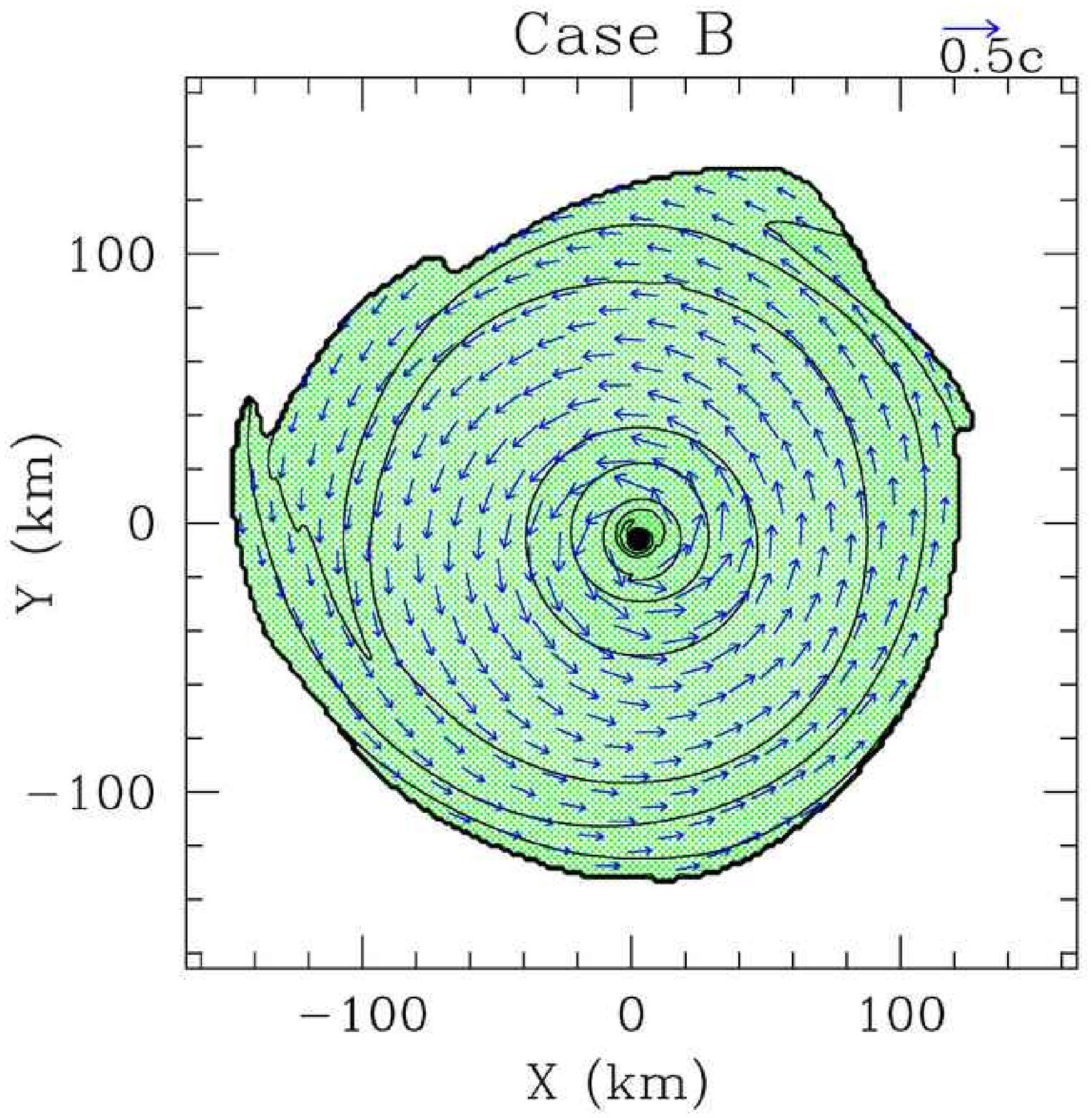} \\
\epsfxsize=2.1in
\leavevmode
\epsffile{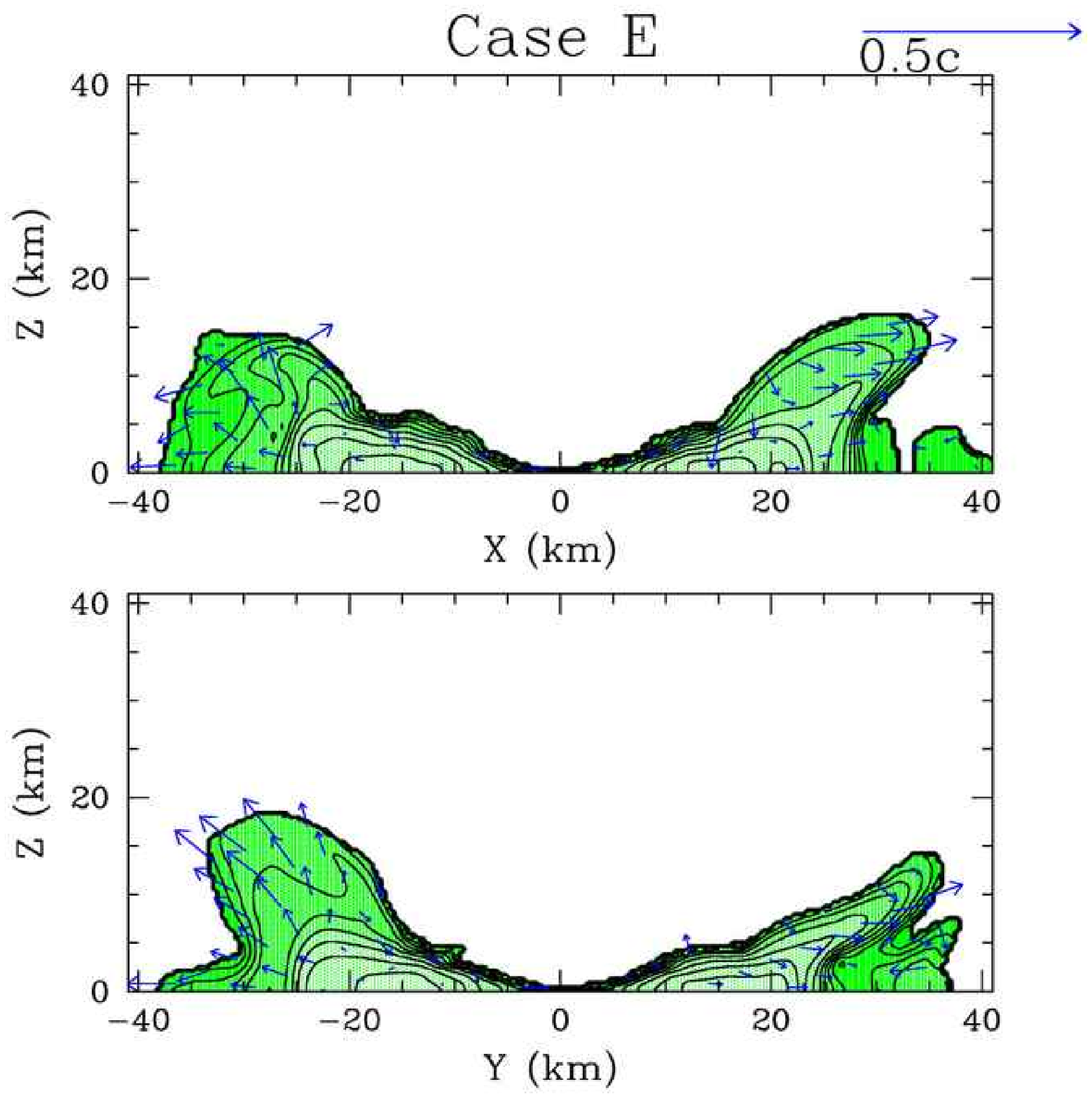}
\epsfxsize=2.1in
\leavevmode
\epsffile{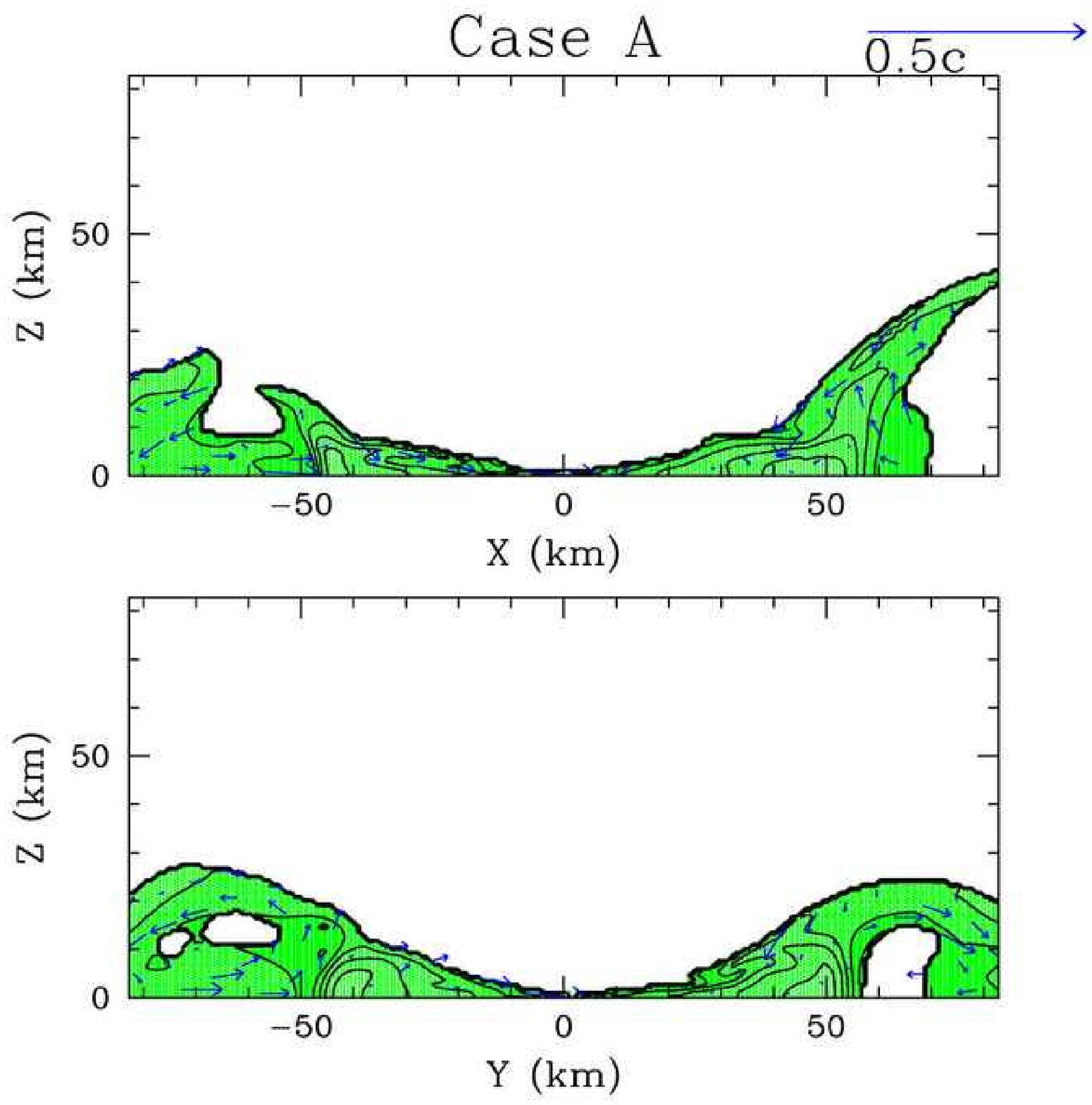}
\epsfxsize=2.1in
\leavevmode
\epsffile{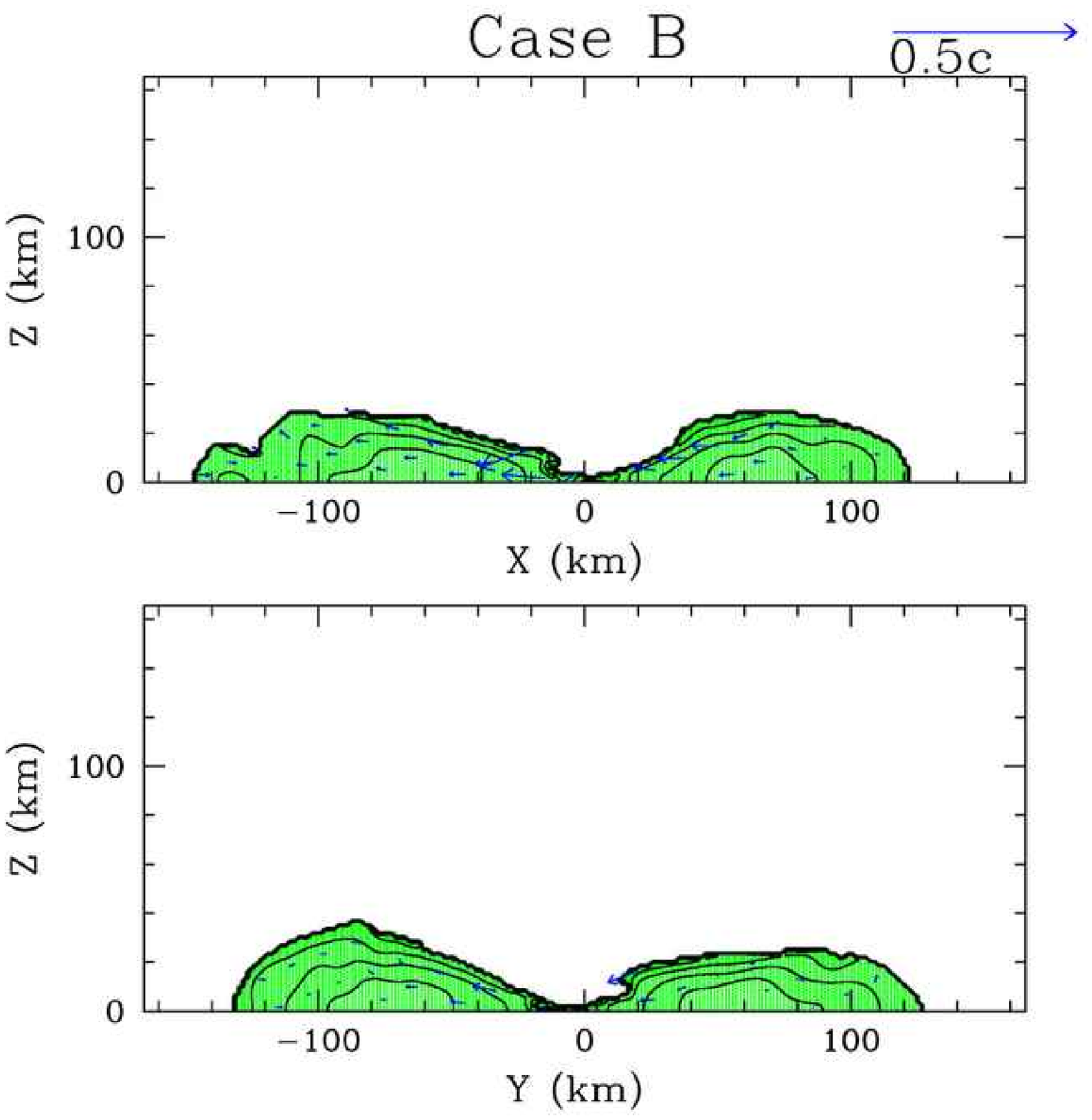}
\caption{Snapshots of rest-mass density $\rho_0$, $K=P/\kappa
   \rho_0^\Gamma$, and velocity profiles at the end of the simulation, 
   for Cases \G\ (left column), \A\ (middle column), and
   \B\ (right column) on the orbital ($xy$, first and third
   rows) and meridional ($xz$ and $yz$, second and fourth rows) planes.
   In the top six plots, density contours are shown according to 
   $\rho_0 = \rho_{0,{\rm max}} 10^{-0.1j-3.3}$ ($j$=0, 1, ... 13) 
   for Cases~\A\ and \B, $\rho_0 = \rho_{0,{\rm max}} 10^{-0.19j-2.32}$ 
   ($j$=0, 1, ... 12) for Case~\G. The maximum density of the initial 
   NS is $\kappa \rho_{0,{\rm max}}=0.126$ for all cases, or 
   $\rho_{0,{\rm max}}=9\times 10^{14}\mbox{g cm}^{-3}(1.4M_\odot/M_0)^2$. 
   The bottom six plots show contours of $K$, plotted according to 
   $K=10^{0.32j}$ ($j$=0, 1, ... 12). The 
   $K$ figures neglect regions where the density is less than
   $\rho_{0,\rm{min}}$.  All contour plots show darker greyscaling for 
   higher $\rho_0$ and $K$.  Arrows represent the velocity
   field in the given plane.  The AH interior in the orbital
   plane is marked by a filled black circle.  Length scales are specified in
   km, assuming the NS has a rest mass of 1.4$M_{\odot}$; it can be converted 
   to units of $M$ via the formula $M=1.9(q+1)$km.}
\label{fig:diskstudysnapshots}
\end{center}
\end{figure*}

In Cases~\A, \B, and \G, a substantial disk forms after the merger. 
Figure~\ref{fig:diskstudysnapshots} shows the density and thermal
energy profiles of the disk in both the equatorial and meridional
plane at the end of the simulation for Cases~\A, \B, and \G. For ease 
of comparison, we specify the length scale in km, assuming the NS has 
a rest mass of 1.4$M_{\odot}$; it can be converted to units of $M$ 
via the formula $M=1.9(q+1)$km. The most distinguishable
features of the disks are their characteristic radius
$r_{\rm disk}$ and final mass, $M_{\rm disk}$.
The characteristic disk radii lie between 20km and 100km, 
as seen in the top row of plots.
The $q=1$ mass ratio (Case~\G) produces the smallest
($r_{\rm disk}=20$km), densest (maximum density
$\rho_{0,{\rm max}}\approx 6\times 10^{12}\mbox{g cm}^{-3}$),
and least massive disk ($M_{\rm disk}/M_0\approx 2\%$).   
Meanwhile, the resulting disk in the canonical 
$q=3$ case (Case~\A) is about 2.5 times
larger than Case~\G\ ($r_{\rm disk}=50$km), but possesses a
lower maximum density by an order of magnitude 
($\approx 4\times 10^{11}\mbox{g cm}^{-3}$).  
Due in part to its larger volume,
the mass of the disk in Case~\A\ is about 70\% larger than in Case~\G\
($M_{\rm disk}/M_0\approx 4\%$). The maximum density of Case~\B's 
disk is $\rho_{0,{\rm max}}\approx 5\times 10^{11}\mbox{g cm}^{-3}$, 
which is similar to Case~\A. However, the disk is 
about twice as large in size, 
with a characteristic radius of $r_{\rm disk}=100$km, yielding a
total disk mass of $M_{\rm disk}/M_0\approx 15\%$.

Despite their differences, the disks in
Fig.~\ref{fig:diskstudysnapshots} share many common features. 
For example, the top row of plots show that the disk forms a 
torus whose density plummets at the BH
ISCO (roughly at coordinate radius 7km, 25km, and 12km for 
Cases~\G, \A, and \B, respectively, in the equatorial
plane~\cite{ft2}). 
Also, the characteristic height of the disks is about 15\%--20\% of the
characteristic radius in each case (top two rows of plots).  Finally,
the higher density regions tend to be colder than lower density
regions, but $K\gtrsim 85$ everywhere in the disks in
Cases~\A\ and \B, and $K\gtrsim 6$ everywhere in Case~\G's disk.  
The characteristic $K$ in the disks in Cases~\A, \B, and \G\ 
are roughly 200, 200, and 10, respectively.

As in Paper~I, the temperature $T$ can be estimated roughly by modeling 
the temperature dependence of the specific thermal 
internal energy density $\epsilon_{\rm th}$ as
\begin{equation}
\epsilon_{\rm th} = \frac{3kT}{2m_n} +
f\frac{aT^4}{\rho_0}
\end{equation}
(compare \cite{PWF}), where $m_n$ is the mass of a nucleon, $k$ is the
Boltzmann constant, and $a$ is the radiation constant.  The first term
represents the approximate thermal energy of the nucleons, and the
second term accounts for the thermal energy due to radiation and 
(thermal) relativistic particles.  The
factor $f$ reflects the number of species of ultrarelativistic
particles that contribute to thermal radiation. When $T \ll 2m_e/k
\sim 10^{10}$K, where $m_e$ is the mass of electron, thermal radiation
is dominated by photons and $f=1$. When $T \gg 2m_e/k$, electrons and
positrons become ultrarelativistic and also contribute to radiation,
and $f=1+2\times (7/8) = 11/4$.  At sufficiently high temperatures and
densities ($T \gtrsim 10^{11}$K, $\rho_0 \gtrsim 10^{12}~{\rm g}~{\rm
cm}^{-3}$), thermal neutrinos are copiously generated and become trapped, so,
taking into account three flavors of neutrinos and anti-neutrinos,
$f=11/4 + 6\times (7/8) = 8$.

The characteristic density and $K$ in the disk for Cases~\A\ and \B\  
are $\kappa \rho_0 \sim 4\times 10^{-5}$ (or 
$\rho_0\sim 3\times 10^{11} \mbox{g cm}^{-3}$) 
and $K\sim 200$. We find $T\sim 5\times 10^{10}K$ (or $\sim$4MeV). 
For Case~\G, we have $\kappa \rho_0 \sim 2\times 10^{-4}$ 
(or $\rho_0\sim 10^{12} \mbox{g cm}^{-3}$). 
We find $T\sim 10^{10}$K (or $\sim$1MeV) for Case~\G's disk.
According to numerical models of rotating BHs with ambient disks
in~\cite{SRJ}, the remnant disks could produce a total gamma-ray
energy $E\sim 10^{47}$--$10^{50}$erg from neutrino-antineutrino 
annihilation. This result is promising for generating a SGRB. 

\subsection{Gravitational wave emission}

\begin{figure}
\epsfxsize=3.4in
\leavevmode
\epsffile{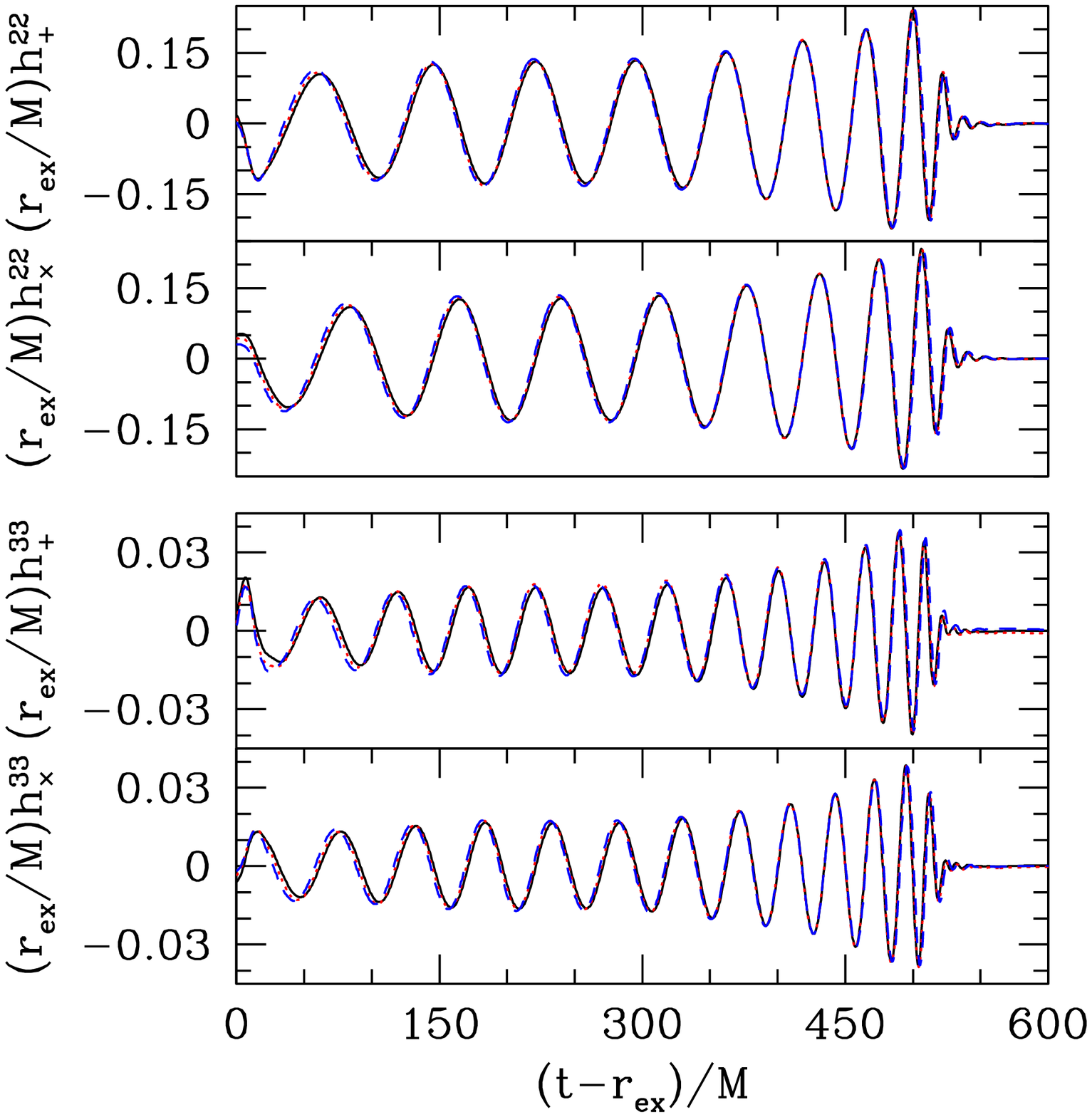}
\caption{Gravitation wave signal from Case~\A. Shown here are the (2,2) 
and (3,3) modes of $rh_+(t-r)$ and $rh_\times(t-r)$ extracted at radii 
$r_{\rm ex}=43.4M$ (black solid lines), $54.2M$ (red dotted lines) 
and $83.2M$ (blue dashed lines). Note that the three lines almost 
overlap in each case.}
\label{fig:hphc_A}
\end{figure}

\begin{figure}
\epsfxsize=3.4in
\leavevmode
\epsffile{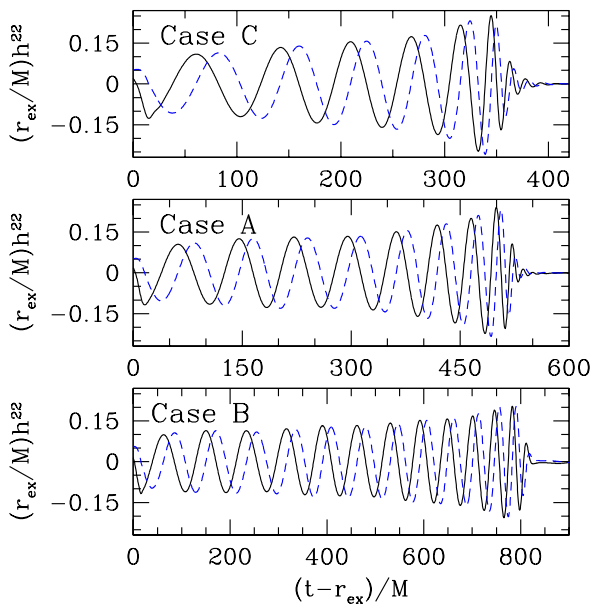}
\caption{Gravitational wave signal from (top to bottom) Cases~\D, \A\ and 
\B. Black solid (blue dash) line denotes the (2,2) mode of 
$h_+$ ($h_\times$) extracted at $r_{\rm ex}=43.4M$.}
\label{fig:hphc_DAB}
\end{figure}

\begin{figure}
\epsfxsize=3.4in
\leavevmode
\epsffile{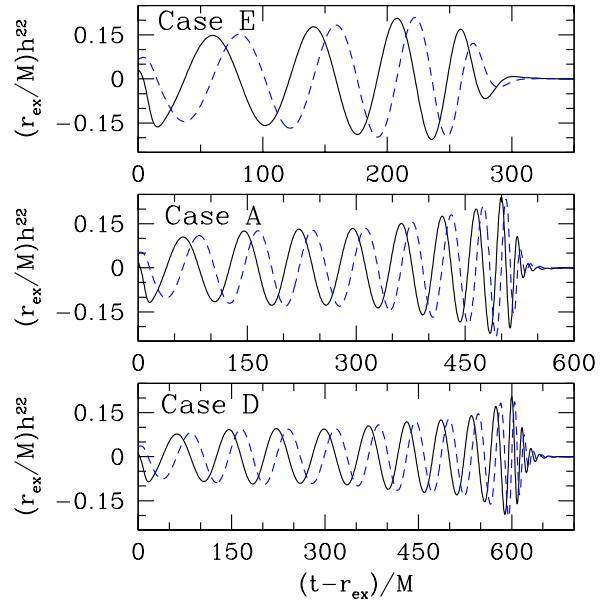}
\caption{Same as Fig.~\ref{fig:hphc_DAB} but for cases (top to bottom) 
\G, \A\ and \F.}
\label{fig:hphc_GAF}
\end{figure}

\begin{figure}
\epsfxsize=3.4in
\leavevmode
\epsffile{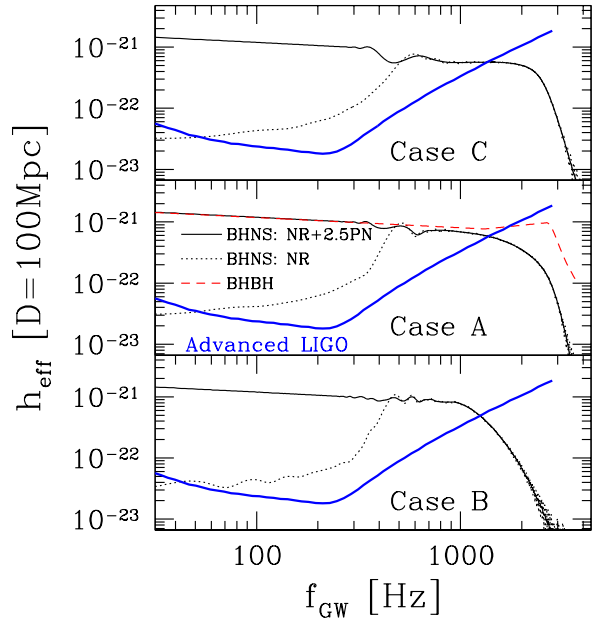}
\caption{Gravitational wave spectrum for the Cases \D, \A\ and \B,
  computed via the method summarized in Sec.~IIIF of Paper~I.  
  The solid curve shows the combined waveform found by attaching 
  the restricted 2.5 order PN waveform to our numerical signal
  (including only the dominant $(2,2)$ and $(2,-2)$ modes), while
  the dotted curve shows the contribution from the latter only. The
  dashed curve is the analytic fit derived by \protect\cite{Ajith}
  from analysis of multi-orbit nonspinning BHBH binaries with the 
  same mass ratios $q$ as the BHNS, which
  maintain significantly more power at higher frequencies.  The heavy
  solid curve is the effective strain of the Advanced LIGO detector,
  defined such that $h_{\rm LIGO}(f)\equiv \sqrt{fS_h(f)}$.  To set
  physical units, we assume a NS rest mass of $M_0=1.4M_\odot$ and a
  source distance of $D$=100Mpc.}
\label{fig:heff_DAB}
\end{figure}

\begin{figure}
\epsfxsize=3.4in
\leavevmode
\epsffile{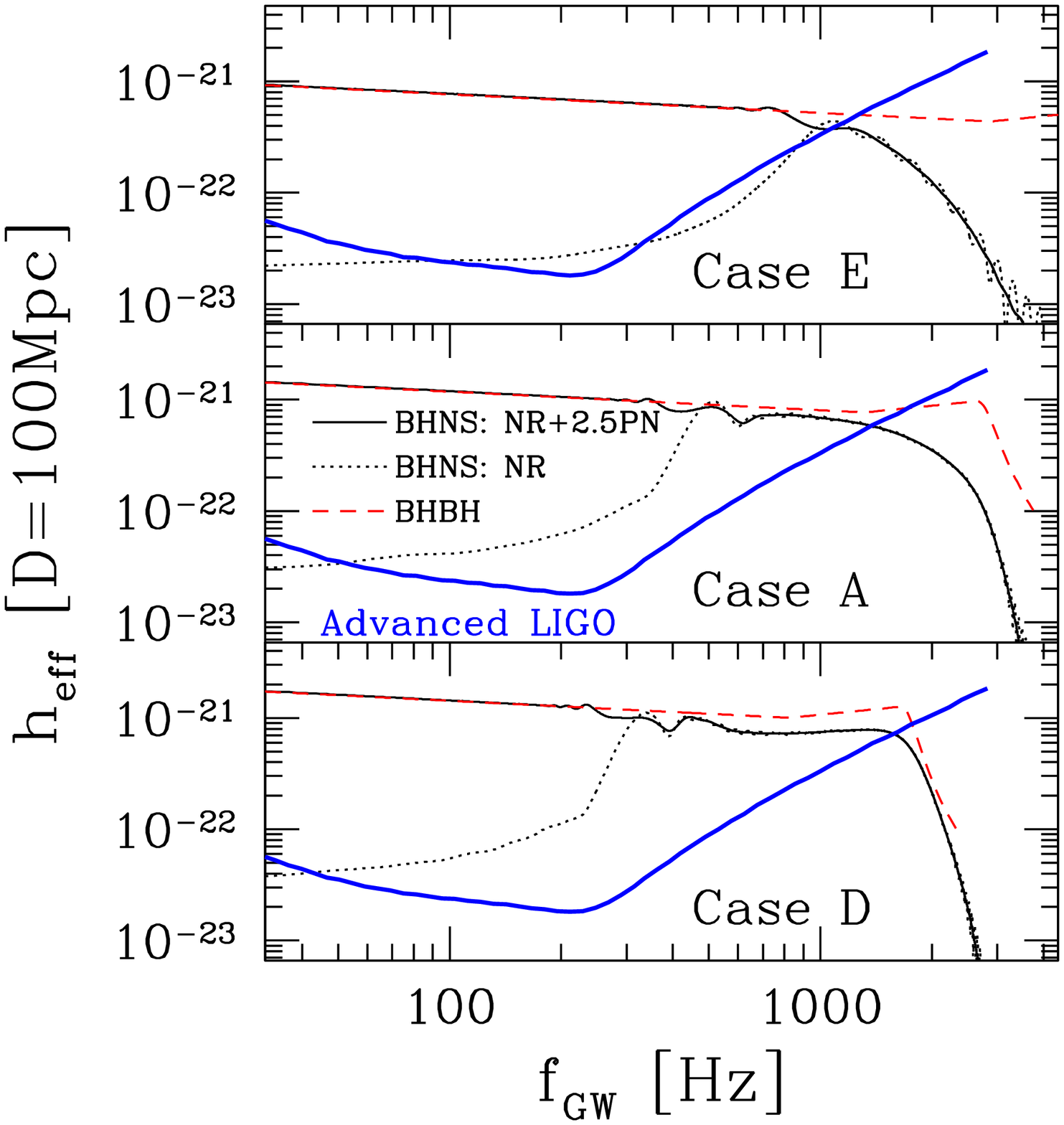}
\caption{Same as Fig.~\ref{fig:heff_DAB} but for Cases \G, \A\ and \F.}
\label{fig:heff_GAF}
\end{figure}

Following the literature, we decompose the gravitational 
waveform $h_+$ and $h_\times$ into $s=-2$ spin-weighted spherical 
harmonics ${}_{-2}Y_{lm}$ as follows
\beq
  h_+ - i h_\times = \sum_{l,m} (h^{lm}_+ - i h^{lm}_\times ) \, 
{}_{-2}Y_{lm} \ ,
\eeq
where $h^{lm}_+$ and $h^{lm}_\times$ are real functions. Each $(l,m)$ 
mode is a function of radius and time only. We extract gravitational waves 
using both the Regge-Wheeler-Zerilli-Moncrief formalism and the 
Newman-Penrose Weyl scalar $\psi_4$. We find good agreement between 
waveforms computed by both methods, as demonstrated in Paper~I. 
Here we only present the waveforms computed from $\psi_4$.

Gravitational waves are extracted at several different radii.  We find
that the extracted waveforms overlap extremely well 
provided the extraction radius $r_{\rm ex}
\gtrsim 40M$, as shown in Fig.~\ref{fig:hphc_A}.  

Figures~\ref{fig:hphc_DAB} and \ref{fig:hphc_GAF} demonstrate
that the maximum amplitude of the waves is similar in Cases \A, \B,
\D, \F, and \G.  With approximately the same initial $M\Omega$, we
observe more wave cycles as the BH spin is increased
(Fig.~\ref{fig:hphc_DAB}) and as the mass ratio $q$ is decreased
(Fig.~\ref{fig:hphc_GAF}), consistent with the discussion in
Secs.~\ref{sec:bhspin} and \ref{sec:q}.  Table~\ref{table:results}
specifies the radiated energy $\Delta E_{\rm GW}$ and angular momentum
$\Delta J_{\rm GW}$ in each case, as well as the kick velocity $v_{\rm
  kick}$ due to recoil.

For Case~\G, NS disruption occurs at relatively low orbital
frequency.  In fact, the total radiated angular momentum $\Delta J_{GW}$
reaches 90\% of its final value at $t_{10}$ (the time at which only 10\%
of the NS has been accreted), and 99\% of its final value at $t_{30}$.
Hence there is a large amount of angular momentum still remaining in the NS
matter before it is accreted, and during the accretion phase about 98\% 
of the NS rest mass is accreted. Thus the bulk
of the orbital angular momentum is used to spin up the BH.
The final disk mass is only $\sim 1$\% of the binary's ADM mass, so we
can reliably estimate the final value of the BH spin parameter using the
Kerr formula (Eq.~(\ref{eq:spinbh})).  We obtain $\tilde{a}=0.85$, which
is significantly larger than all other cases that result in a
similarly small disk.  Compared to the final spin parameter from
equal-mass BHBH mergers, $\tilde{a}=0.686$ (e.g.~\cite{bhbh-highacc} 
and Table~\ref{tab:bbh_results} in Appendix~\ref{app:BHBH}),
we find that matter accretion in BHNS binary mergers is much more
effective at spinning up black holes.  This is not unexpected, since
NS tidal disruption spreads the NS matter around the BH,
diminishing the multipole moments of the system, and the associated 
gravitational wave emission.

Figures~\ref{fig:heff_DAB} and \ref{fig:heff_GAF} show the effective
gravitational wave strains in frequency space for all cases, computed
via the method summarized in Sec.~III~F of Paper~I, and including 
only the dominant $(2,2)$ and $(2,-2)$ modes.  For
comparison against the Advanced LIGO sensitivity curve $h_{\rm
  LIGO}(f)\equiv \sqrt{fS_h(f)}$ (\cite{ADVLIGO}), we plot
BHNS wave strains for the cases we study, assuming the NS has a rest mass of
$1.4M_\odot$ and binary distance of 100Mpc.  This is the distance
required to reach one merger per year, assuming an overall rate of 10
mergers per Myr per Milky Way-equivalent galaxy (and a density of
$0.1~{\rm gal/Mpc}^3$)~\cite{BBR}.  This distance is roughly
that of the Coma cluster, and approximately five times the distance to
the Virgo cluster.  
The gravitational wave spectra of nonspinning
BHBH mergers (computed from Eqs.~(4.12)--(4.19) of~\cite{Ajith})
with the same mass ratio as in Cases~\G, \A\ and \F\ are plotted
in the same graphs for comparison. 
We see that the wave signal drops significantly at
and above the frequency corresponding to the onset of NS tidal
disruption.  The difference in wave signals between BHBH and BHNS
mergers is marginally observable in the Advanced LIGO frequency band
in most cases. Distinguishing BHNS from BHBH inspirals and mergers
may require narrow-band detection techniques with
advanced detectors. The observation of an accompanying SGRB 
would also serve to distinguish the events.

\section{Summary and Discussion}
\label{sec:discussion}

In this paper we perform a new set of fully general relativistic BHNS
simulations using our code upgraded with AMR capability.  We
vary the initial binary separations and confirm that the simulation
outcomes do not change when the initial BHNS separation is large
enough. In most cases, a toroidal disk forms around the black 
hole remnant after the merger.  This result
revises our finding of Paper~I, where we artificially imposed a
uniform pressure ceiling on all the matter present in the simulation.
This ceiling removed a substantial amount of angular momentum in the
NS matter during the merger phase, causing the matter to fall into the
BH prematurely, and suppressing disk formation.

Fixing the mass ratio $q=3$, we study the effects of BH spin by evolving 
initial data with BH's spin parameter $\tilde{a}=-0.5$ (count-rotating), 
0 and 0.75 with nearly the same initial orbital angular 
velocity $M\Omega \approx 0.033$.  Not surprisingly, we find that 
the BHNS inspiral phase lasts longer as $\tilde{a}$ increases, and 
the final disk mass increases from $< 1\%$ ($\tilde{a}=-0.5$) to 
$\approx 4\%$ ($\tilde{a}=0$) and $\approx 15\%$ ($\tilde{a}=0.75$) 
of the NS's rest mass.

Next, we study the effect of varying the mass ratio $q$ by evolving
BHNS initial data with $q=1$ and 5. For the $q=5$ case almost all the
NS matter plunges into the BH, leaving $< 1\%$ of the NS's rest mass to
form a disk at the end of the simulation. For the $q=1$ case, a
low-density, hot spiral region of disrupted NS matter winds around the
BH and smashes into the nascent NS tidal tail, causing strong shock
heating before it rapidly falls into the BH.  Most of the remaining
matter in the tail eventually falls into the BH, leaving about 2\% of
the NS's rest mass behind to form a disk. 

The disks formed after the merger of Cases~\A\ and \B\ have a
characteristic density of $3\times 10^{11} \mbox{g cm}^{-3}$ and a
temperature of $T \sim 5\times 10^{10}$K, assuming the NS's rest mass
is $M_0=1.4M_\odot$.  Applying the results of the simulations
of BH disks in~\cite{SRJ}, the disk may produce a gamma-ray energy of
$E\sim 10^{47}$--$10^{50}$erg due to neutrino-antineutrino
annihilation. Hence the merger of a BHNS binary is a promising central
engine for a SGRB. But self-consistent simulations, taking into
account the detailed microphysics and including the correct EOS, are
necessary to fully explore the viability of and the mechanism for
generating SGRBs from BHNS mergers.

We compute the BHNS gravitational waveforms and power spectra, and
find that the signals are attenuated at frequencies roughly equal to
double the orbital frequency at which tidal disruption begins. The 
resulting deviation of BHNS and BHBH waveforms is visible in the Advanced
LIGO high frequency band out to distances $\gtrsim 100~{\rm Mpc}$. Should the
chirp mass determination, combined with higher order PN waveform phase
effects, allow for an independent determination of the component masses
and spins of the binary companions, the measurement of the BHNS tidal 
disruption frequency should give a good estimate of the NS radius and, 
hence, insight into the nuclear EOS.

Carpet-based AMR has provided us with a great improvement in
efficiency and accuracy, and future work will continue in this
direction.  To improve code performance (which is usually
memory-bound), we also plan to employ new techniques (e.g.,
loop-tiling) for minimizing cache misses.  To improve accuracy, we
will implement the piecewise parabolic method (PPM) or weighted
essentially non-oscillatory (WENO) reconstruction schemes in
hydrodynamics/MHD and experiment with tapered grids to replace second
order temporal prolongation (see~\cite{taperedgrids}).  We plan to
study techniques for minimizing refinement boundary crossings, by
adding more refinement boxes around the disrupted NS.  This
might help reduce our angular momentum conservation violation to below
1\% for all cases.

Rich with the challenges of modeling extreme matter in the presence of
intense gravitational fields, 
BHNS binary mergers
will likely remain a topic at the forefront of theoretical
astrophysics and numerical relativity for years to come. 
The most exciting possibility remains the simultaneous detection 
of a gravitational wave signal and a GRB. Our theoretical work 
on BHNSs is partly in anticipation and preparation for that discovery.

\acknowledgments
We would like to thank K.~Taniguchi for generating the BHNS initial 
data. We would also like to thank M.~Ansorg for providing
the {\tt TwoPunctures} code for generating BHBH initial data, and
E.~Schnetter for useful discussions about {\tt Carpet}. 
This paper was supported in part by NSF
Grants PHY02-05155 and PHY06-50377 as well as NASA
Grants NNG04GK54G and NNX07AG96G to the University of Illinois at
Urbana-Champaign, and NSF Grant PHY07-56514 to Bowdoin College. 
Simulations were performed under a TeraGrid Grant TG-MCA99S008 
and on the Illinois Numerical Relativity Beowulf Cluster.

\appendix

\section{Details of BHBH test simulations}
\label{app:BHBH}

Simulating the late inspiral, merger and ringdown of a binary black
hole was for many years the ``holy grail'' of numerical relativity.
Dozens of researchers have spent many years attempting to formulate a
stable algorithm capable of solving this important problem. 
In this appendix we summarize two BHBH simulations in detail. 
Performing a simulation of binary black hole coalescence provides a
highly nontrivial and reasonably comprehensive laboratory for testing
3+1 codes that solve Einstein's vacuum field equations in the
strong-field regime of general relativity.  It tests the caliber of
not only the basic evolution scheme, but also the all-important
diagnostic routines.  These routines measure globally conserved
quantities like the mass and angular momentum of the system, the
location and measure of all black hole horizons, the asymptotic
gravitational waveforms, the recoil motion of the black hole remnant,
etc. Appreciable effort must be expended to implement reliable
diagnostic routines in order to extract useful physical information
from the numerical output.

\begin{figure*}
\vspace{-4mm}
\begin{center}
\epsfxsize=2.15in
\leavevmode
\epsffile{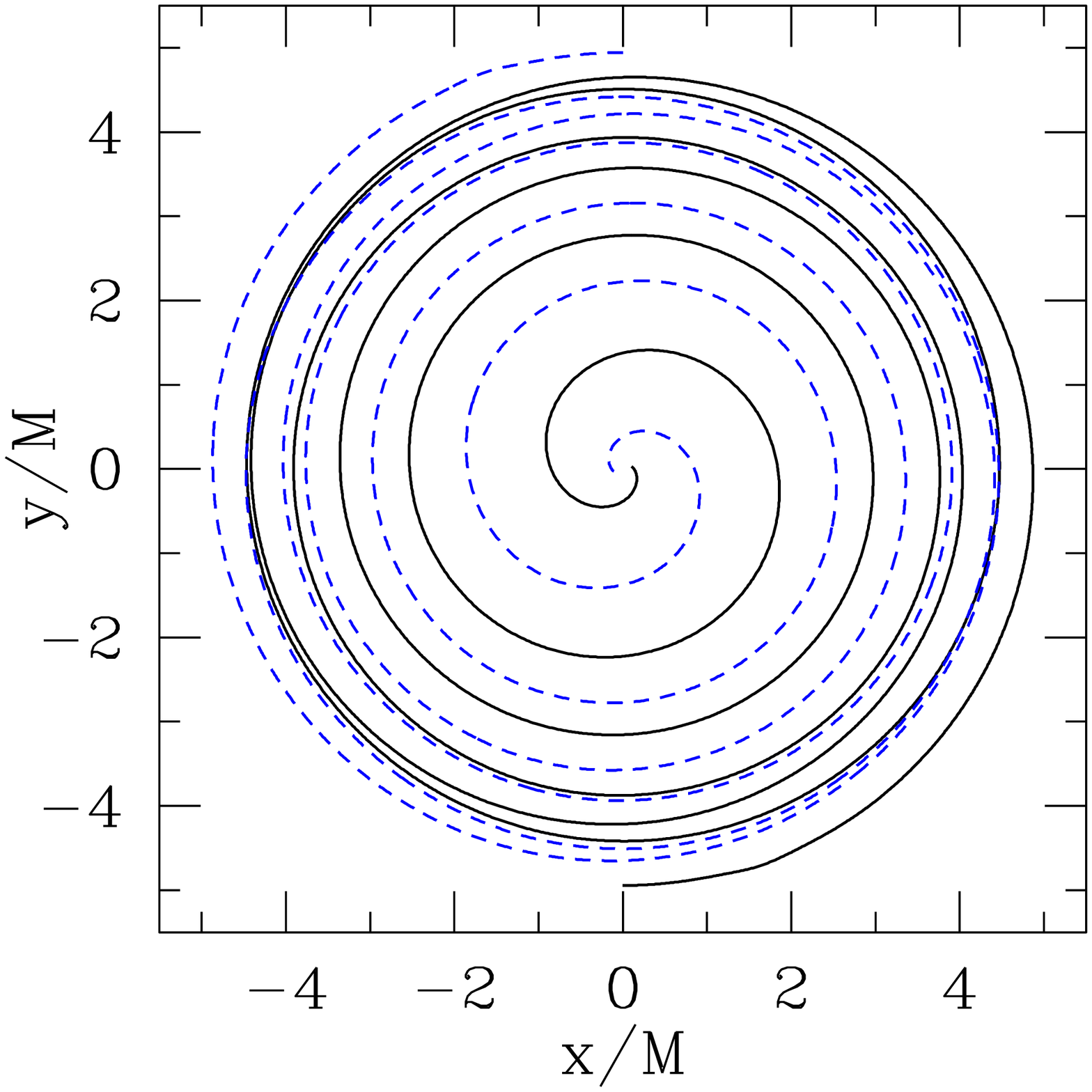}
\epsfxsize=2.15in
\leavevmode
\epsffile{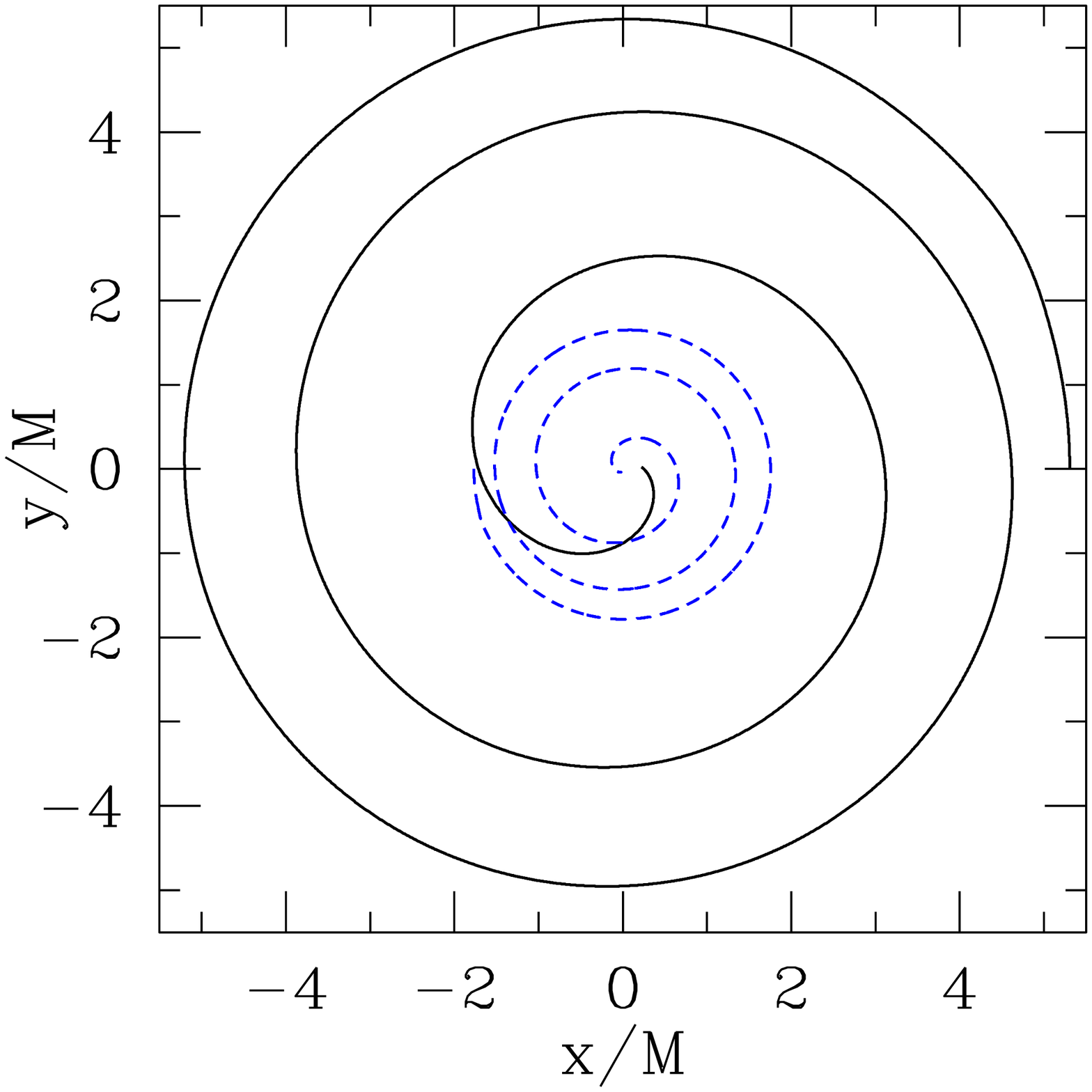}
\caption{Trajectories of the coordinate centroid of the ``+'' 
(black solid line) and ``-'' (blue dashed line) black hole 
from the equal-mass (left panel) and unequal-mass (right panel) BHBH simulations.} 
\label{fig:bh_centroids}
\end{center}
\end{figure*}

\begin{figure}
\epsfxsize=3.4in
\epsffile{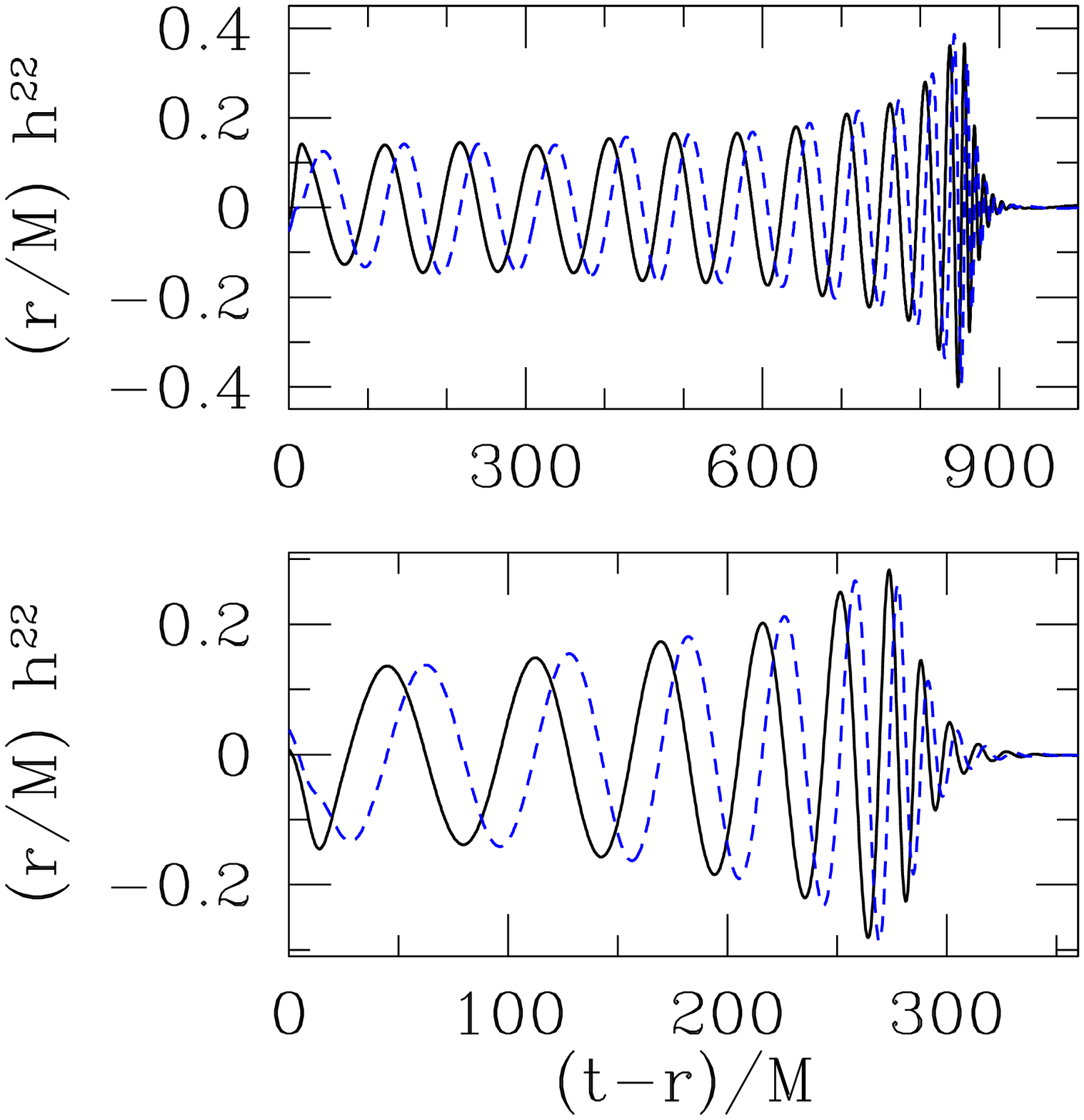}
\caption{Gravitational waveform from the equal-mass (upper panel) 
and unequal-mass (lower panel) BHBH merger. 
Shown here are $(r/M)h_+^{22}$ (black solid line) and 
$(r/M)h_\times^{22}$ (blue dashed line) as a function of the 
retarded time $t-r$ extracted at radius $r=50.53M$.}
\label{fig:hphc22_BHBH}
\end{figure}

We consider two cases of merging, nonspinning black holes to 
test our code: 
an equal-mass ($q=1$) and an unequal-mass ($q=3$) system.  Both cases are
constructed via the binary puncture 
technique discussed in~\cite{BB97}, with parameters as listed in
Table~\ref{tab:bbh_initial_data}.  Our code employs AMR with
equatorial symmetry using the moving-box Carpet infrastructure, so
that the boxes track the AH centroids throughout the simulations.  The
grid setup is described in Table~\ref{tab:bbh_grid_setup}.  The basic
evolution scheme and numerical techniques are described in
Secs.~\ref{sec:basic_eqns} and \ref{sec:numerical}.  A review
of the special algorithms and parameters chosen in these simulations is
summarized in Table~\ref{tab:bbh_technique}, and the results are
summarized in Table~\ref{tab:bbh_results}.  The coordinate
trajectories of the two black holes are plotted for the two cases in
Fig.~\ref{fig:bh_centroids}, and the gravitational waveforms are
plotted in Fig.~\ref{fig:hphc22_BHBH}.

The results found here are in good agreement with those reported by 
previous investigators~\cite{goddard06,goddard07,bcgshhb07,JenaQ}. 
Of special significance is the degree to which total energy and 
angular momentum are conserved when properly accounting for 
losses due to gravitational wave emission. The fractional errors 
are seen in Table~\ref{tab:bbh_results} to be a few times 
$10^{-4}$ for the energy and $10^{-3}$ for the angular momentum. 
Also, different measures of the mass and spin of the final 
(stationary) remnant Kerr black hole, such as 
the irreducible mass and the ADM mass, agree closely, given 
the adopted computational resources.

\begin{table*}
\caption{Initial data of our BHBH simulations}
\label{tab:bbh_initial_data}
\begin{tabular}{c|c|c}
\hline
 & Equal mass & Unequal mass \\
\hline
\hline
Location of punctures & $\ve{x}_+=(0,4.891,0)$, &
$\ve{x}_+=(5.25,0,0)$, \\
 & $\ve{x}_-=(0,-4.891,0)$ & $\ve{x}_-=(-1.75,0,0)$ \\
\hline
``Bare'' mass & $m_+=m_-=0.4856$ & $m_+=0.234$, $m_-=0.735$ \\
\hline
Spin & $\ve{S}_+=\ve{S}_-=0$ & $\ve{S}_+=\ve{S}_-=0$ \\
\hline
Momentum & $\ve{P}_+=(0.0969,0,0)$, &
$\ve{P}_+=(0,0.09407,0)$, \\
 & $\ve{P}_-(-0.0969,0,0)$ & $\ve{P}_-=(0,-0.09407,0)$ \\
\hline
ADM mass of the system & $M = 0.9894$ & $M = 0.9895$ \\
\hline
ADM angular momentum & $J = 0.9479$ & $J = 0.6587$ \\
\hline
Irreducible mass of the BHs & ${\cal M}_+={\cal M}_-=0.5000$ & ${\cal M}_+=0.2498$,
${\cal M}_-=0.7494$ \\
\hline
Mean coordinate radius of BHs' horizon & ${\cal R}_+ = {\cal R}_- = 0.2359$ &
${\cal R}_+ = 0.1097$, ${\cal R}_- = 0.3605$ \\
\hline
Binary coordinate separation ($|\ve{x}_+-\ve{x}_-|$) & $9.887M$ & $7.074M$ \\
\hline
\end{tabular}
\end{table*}

\begin{table*}
\caption{AMR grid setup in our BHBH simulations.}
\label{tab:bbh_grid_setup} 
\begin{tabular}{c|c|c}
\hline
  & equal-mass & unequal mass \\
\hline
\hline
 $L$ & 320 ($323.4M$) & 409.6 ($413.9M$)  \\
\hline
 $N_+$, $N_-$ & 8, 8 & 10, 9 \\
\hline
 $R_+$
& (160, 80, 140/3, 20, 5, 2.5, 1.25, 0.625)
& (256, 128, 64, 24, 12, 6, 3, 1.5, 0.75, 0.375) \\
\hline
 $\Delta_+$
& (4, 2, 1, 0.5, 0.25, 0.125, 0.0625, 0.03125)
& (5.12, 2.56, 1.28, 0.64, 0.32, 0.16, 0.08, 0.04, 0.02, 0.01) \\
\hline
 $R_-$
& (160, 80, 140/3, 20, 5, 2.5, 1.25, 0.625)
& (256, 128, 64, 32, 16, 8, 4, 2, 1) \\
\hline
 $\Delta_-$
& (4, 2, 1, 0.5, 0.25, 0.125, 0.0625, 0.03125)
& (5.12, 2.56, 1.28, 0.64, 0.32, 0.16, 0.08, 0.04, 0.02) \\
\hline
$N_{\rm AH}^+$, $N_{\rm AH}^-$ & $\approx 15, 15$ & $\approx 22, 36$ \\
\hline
\end{tabular}

\begin{flushleft}
$L$: Location of the outer boundary $x_{\rm max}=y_{\rm max}=z_{\rm max}=L$,
$x_{\rm min}=y_{\rm min}=-L$, $z_{\rm min}=0$.

$N_+$ ($N_-$): Number of refinement levels centered at the ``+'' (``-'')
puncture.

$R_+$ ($R_-$): Half side length of refinement boxes centered at the
``+'' (``-'') puncture (in code unit).

$\Delta_+$ ($\Delta_-$): Grid spacing in each refinement level centered
at the ``+'' (``-'') puncture (in code unit). Note that the grid spacing
doubles in each successive refinement level. The coarsest grid spacing is
twice of that of the outermost refinement level, which is
8 for the equal-mass case and 10.24 for the unequal-mass case.

$N_{\rm AH}^+$ ($N_{\rm AH}^-$): Number of grid points across
the mean diameter of the apparent horizon of the ``+'' (``-'')
BH at $t=0$.
\end{flushleft}
\end{table*}

\begin{table*}
\caption{Evolution technique used in our BHBH simulations} 
\label{tab:bbh_technique}
\begin{tabular}{c|l}
  \hline
   Formalism & BSSN-based \\
  \hline
  Lapse & 1+log slicing: $\partial_0 \alpha= -2\alpha K$,
  $\partial_0 \equiv \partial_t -\beta^j\partial_j$. \\
  \hline
  Shift & ``Gamma-freezing'': $\partial_0
  \beta^i =(3/4)B^i$,
$\partial_0 B^i =\partial_0 \tilde{\Gamma}^i-\eta
  B^i$, $\eta=0.25=0.2474/M$. \\
  \hline
  Conformal variable & Evolve $W=e^{-2\phi}$ instead of $\phi$. \\
  \hline
  Symmetry imposed & Equatorial (i.e., symmetry across equatorial plane). \\
  \hline
   & (1) Enforce auxiliary constraints
det$(\tilde{\gamma}_{ij}) = 1$ and Tr$(\tilde{A}_{ij}) = 0$ using the
      technique \\
   & described in Paper~I.\\
   & (2) Enforce auxiliary constraint $\tilde{\Gamma}^i=-\partial_j\tilde{\gamma}^{ij}$
using the technique as in, e.g.~\cite{marronetti07}. \\
   Other Numerical techniques  & (3) Add 5th order Kreiss-Oliger dissipation of the form \\
 & $(\epsilon/64) (\Delta x^5 \partial_x^6 + \Delta y^5 \partial_y^6
+ \Delta z^5 \partial_z^6 )f$ \\
    & for all BSSN, lapse and shift variables $f$. Here $\Delta x$,
$\Delta y$ and $\Delta z$ are grid spacings\\
    & and we set the strength $\epsilon=0.1$. \\
  \hline
  Grid-driver code & Carpet  \\
  \hline
  Spatial differencing & 4th order upwinded on shift advection terms, 4th
  order centered everywhere else  \\
\hline
  GW extraction radii & $10.11M$--$54.58M$ for the equal-mass case,
$30.32M$--$70.75M$ for the unequal-mass case.\\
  \hline
  Temporal differencing & MoL: 4th order Runge-Kutta (RK4) \\
  \hline
  Courant-Friedrichs-Lewy (CFL) factor & 0.25 \\
  \hline
  Prolongation & 5th order spatial prolongation, 2nd order temporal
prolongation \\
  \hline
\end{tabular}
\end{table*}

\begin{table}
\caption{Summary of our BHBH simulation results}
\label{tab:bbh_results}
\begin{tabular}{c|c|c}
\hline
 & equal mass & unequal mass ($q=3$) \\
\hline
\hline
 $M_{\rm BH}/M$ & 0.9617 & 0.9809 \\
\hline
 $J_{\rm BH}/M_{\rm BH}^2$ & 0.6852 & 0.5413 \\
\hline
 $\Delta E_{\rm GW}/M$ & 0.03794 & 0.01934 \\
\hline
 $\Delta J_{\rm GW}/M^2$ & 0.3306 & 0.1513 \\
\hline
 $v_{\rm kick}$ & --- & 174 km s$^{-1}$ \\
\hline
 $\delta E$ & $4\times 10^{-4}$ & $-2\times 10^{-4}$ \\
\hline
 $\delta J$ & $4\times 10^{-3}$ & $9\times 10^{-4}$ \\
\hline
\end{tabular}

\begin{flushleft}
$M_{\rm BH}$ and $J_{\rm BH}/M_{\rm BH}^2$: Mass and spin parameter
of the final BH, as determined by its irreducible mass and the ratio
of polar and equatorial circumferences (using the Kerr formula, given
in Eq.~(\ref{eq:spinbh})). Here $M$ is the initial ADM mass as listed 
in Table~\ref{tab:bbh_initial_data}

$\Delta E_{\rm GW}$ and $\Delta J_{\rm GW}$: Energy and angular momentum 
carried off by
gravitational radiation. GWs are extracted at $50.53M$ using $\psi_4$.

$v_{\rm kick}$: The kick velocity measured by GW. Following Gonz\'alez 
{\it et al.}~\cite{JenaQ}, GW data with $t-r< 50M$,
corresponding to the ``junk'' GW, for the unequal-mass case are removed before
computing the kick velocity.

$\delta E \equiv (M-M_{\rm BH}-\Delta E_{\rm GW})/M$ is a measure of violation 
of energy conservation.

$\delta J \equiv (J-J_{\rm BH}-\Delta J_{\rm GW})/J$ is a measure of violation of
angular momentum conservation.
\end{flushleft}
\end{table}

\section{Shock heating in low-density regions}
\label{app:shock-heating}

In this appendix, we estimate the amount of thermal energy generated
by strong shocks in low-density regions. Such shocks are generated 
when fluid elements of initially cold matter collide.
For simplicity, we assume
Newtonian hydrodynamics in our analysis, which is adequate for a rough
estimate, since the fluid velocity during the BHNS merger is only
mildly-relativistic ($v\sim \sqrt{M/r}\lesssim 0.4$ for $r\gtrsim 6M$) 
and $P\ll \rho_0$ in low-density regions. 

Let $\rho_1$, $P_1$, $v_1$, and $\kappa_1=P_1/\rho_1^\Gamma$ be 
the upstream rest-mass density, pressure, velocity, 
and polytropic constant, respectively as measured in the 
shock frame, and $\rho_2$, $P_2$, $v_2$ and $\kappa_2$ the
corresponding quantities in the downstream fluid. 
The upstream Mach number is given by
\beq
  {\cal M} \equiv \frac{v_1}{c_1} = 
v_1 \left( \frac{\Gamma P_1}{\rho_1}\right)^{-1/2} = \frac{v_1}
{\sqrt{\Gamma \kappa_1 \rho_1^{\Gamma-1}}} \ ,
\label{eq:mach}
\eeq
where $c_1$ is the sound speed in the upstream fluid. It 
can be shown that (see e.g.\ Sec.~89 of~\cite{LL})
\beqn
  \frac{\rho_2}{\rho_1} &=& \frac{(\Gamma+1){\cal M}^2}
{(\Gamma-1){\cal M}^2+2} \ , \label{eq:rho2orho1}\\ 
  \frac{P_2}{P_1} &=& \frac{2\Gamma {\cal M}^2}{\Gamma+1} 
 - \frac{\Gamma-1}{\Gamma+1} \ . 
\label{eq:P2oP1a}
\eeqn
In the shock frame, the flow always enters the front supersonically, 
(i.e.\ ${\cal M} > 1$), hence it follows from Eqs.~(\ref{eq:rho2orho1}) 
and (\ref{eq:P2oP1a}) that $\rho_2/\rho_1 > 1$ and 
$P_2/P_1>1$. Also, for adiabatic flow on either side of the shock, 
\beq
  \frac{P_2}{P_1} = K' \left( \frac{\rho_2}{\rho_1}\right)^\Gamma 
= K' \left[ \frac{(\Gamma+1){\cal M}^2}{(\Gamma-1){\cal M}^2+2}
\right]^\Gamma \ ,
\label{eq:P2oP1b}
\eeq
where $K'\equiv\kappa_2/\kappa_1$. 
Combining Eqs.~(\ref{eq:P2oP1a}) 
and (\ref{eq:P2oP1b}), we obtain 
\beq
  K' = \frac{2\Gamma {\cal M}^2 - (\Gamma-1)}{(\Gamma+1)^{\Gamma+1}} 
\left( \Gamma-1+\frac{2}{{\cal M}^2}\right)^\Gamma \ .
\label{eq:kp}
\eeq
As a shock generates entropy, we must have $K'>1$. 
In fact, it is easy to prove from Eq.~(\ref{eq:kp}) that 
$K' >1$ whenever ${\cal M}>1$ and $\Gamma>1$.
The thermal energy of the downstream fluid can be measured by 
$K\equiv P_2/\kappa \rho_2^\Gamma=(\kappa_1/\kappa)K'$, where $\kappa$ is 
the polytropic constant for a cold gas. Since we always 
have $\kappa_1 \geq \kappa$ (with the equality holding for cold, 
unshocked matter), we must always have $K>1$ when ${\cal M}>1$. 
In other words $K$ exceeds unity 
across a shock front. Using Eqs.~(\ref{eq:mach}), 
(\ref{eq:kp}) and $K=(\kappa_1/\kappa)K'$, we have 
\beq
  K \approx \frac{2v_1^2}{(\Gamma+1)\kappa \rho_1^{\Gamma-1}} 
\left( \frac{\Gamma-1}{\Gamma+1}\right)^\Gamma 
\label{eq:kestimate}
\eeq
in the strong shock limit (${\cal M} \gg 1$). 
Equation~(\ref{eq:kestimate}) shows that when the density $\rho_1$ 
of the upstream matter is low, the amount of the shock heating 
downstream can be substantial, i.e.\ $K\gg 1$. 
The initial maximum rest-mass density of the NSs in all of our BHNS
models is $\kappa \rho_{0,{\rm max}}=0.126$. Setting $\Gamma=2$, we
have
\beq
  K \approx 9 \left( \frac{v_1}{0.4}\right)^2 \left( 
\frac{0.01\rho_{0,{\rm max}}}{\rho_1}\right) 
\eeq 
in the strong shock limit.
We see that imposing the pressure ceiling $P<10\kappa \rho_0^\Gamma$ 
(i.e.\ $K<10$) may spuriously remove a substantial amount of thermal
energy and pressure generated by strong shocks when the density is below
$0.01\rho_{0,{\rm max}}$. Such low-density regions contribute a
substantial fraction of the NS rest mass during the merger phase
following tidal disruption. The pressure ceiling then serves to remove
a substantial amount of torque and angular momentum from the fluid, causing 
a large violation of angular momentum conservation.

\bibliography{bhns2}

\begin{thebibliography}{94}
\bibitem[{ft0()}]{ft01}
\bibinfo{note}{We find that we incorrectly calculated the extrinsic curvature
  on our numerical grid in Paper~I, which resulted in the BHNS inspiral phase
  lasting about 0.5--1 orbit shorter.}
\bibitem[{ft2()}]{ft2}
\bibinfo{note}{The radius of ISCO, $r_{\rm ISCO}$, in Boyer-Lindquist
  coordinates is estimated by assuming that the central BH can be described by
  the Kerr metric with spin parameter $\tilde{a}$ given by
  Table~\ref{table:results}. Given a circle of coordinate radius $R$ centered
  at the BH in the equatorial plane, the proper circumference $C(R)$ can be
  computed from the spatial metric. The coordinate radius of the ISCO in the
  adopted gauge can then be estimated by finding $R_{\rm ISCO}$ such that
  $C(R_{\rm ISCO})=2\pi \sqrt{r_{\rm ISCO}^2+a^2+2M_{\rm BH}a^2/r_{\rm
  ISCO}}$.}
\end{thebibliography}

\end{document}